\documentclass{mn}
\usepackage{amsfonts}
\usepackage{amsmath}
\usepackage{graphicx}

\def\gsim{\;\lower4pt\hbox{${\buildrel\displaystyle >\over\sim}$}\;}
\def\lsim{\;\lower4pt\hbox{${\buildrel\displaystyle <\over\sim}$}\;}
\def\grls{\;\lower4pt\hbox{${\buildrel\displaystyle >\over <}$}\;}
\begin{document}

\title[Configurations in a Composite MSID System]
{Global Perturbation Configurations in a Composite\\
Disc System with an Isopedic Magnetic Field}
\author[Y.-Q. Lou \& Y. Wu]{Yu-Qing Lou$^{1,2,3,4}$
and Yue Wu$^1$\\$^1$Physics Department and Tsinghua Center for
Astrophysics (THCA), Tsinghua University, Beijing 100084, China;
\\$^2$Centre de Physique des Particules de Marseille (CPPM)
/Centre National de la Recherche Scientifique (CNRS)\\
\qquad\quad
/Institut National de Physique Nucl\'eaire
et de Physique des Particules (IN2P3) et\\ \qquad\quad
Universit\'e de la M\'editerran\'ee Aix-Marseille II,
163, Avenue de Luminy F-13288 Marseille, Cedex 09, France;
\\$^3$Department of Astronomy and Astrophysics, The University
of Chicago, 5640 S. Ellis Ave, Chicago, IL 60637, USA;
\\$^4$National Astronomical Observatories, Chinese Academy
of Sciences, A20, Datun Road, Beijing 100012, China.
\\}
\date{Accepted 2004... Received 2003...;
in original form 2003}\date{Accepted .
      Received ;
      in original form }
\maketitle

\begin{abstract}
We construct stationary global configurations of both aligned
and unaligned logarithmic spiral perturbations in a composite
disc system of stellar and isopedically magnetized gaseous
singular isothermal discs (SIDs) coupled by gravity.
Earlier models are generalized to a more general theoretical framework.
The thin gaseous SID is threaded across by a vertical magnetic field
$B_z$ with a constant ratio of the surface gas mass density to $B_z$.
In reference to SID models of Shu \& Li, Shu et al., Lou \& Shen,
Lou \& Zou, Shen, Liu \& Lou, there exist two classes of stationary
magnetohydrodynamic (MHD) solutions with in-phase and out-of-phase
density perturbations here. Relevant parameter regimes are explored
numerically.
For both aligned and unaligned cases with azimuthal periodicities
$|m|\geq 2$ ($m$ is an integer), there may be two, one, and no
solution situations, depending on the chosen parameters. For the
transition criteria from an axisymmetric equilibrium to aligned secular
bar-like instabilities, the corresponding $\mathcal T/|\mathcal W|$ ratio
can be much lower than the oft-quoted value of $\mathcal T/|\mathcal W|
\sim 0.14$, where $\mathcal T$ is the total rotational kinetic energy
and $\mathcal W$ is the total gravitational potential energy plus the
magnetic energy. The $\mathcal T/|\mathcal W|$ ratios for the two sets
of solutions in different ranges are separated by $m/(4m+4)$. For the
unaligned cases, we study marginal stabilities for axisymmetric ($m=0$)
and non-axisymmetric ($m\neq 0$) disturbances. By including the
gravitational influence of an axisymmetric dark matter halo on the
background, the case of a composite partial magnetized SID system is
also examined. The global analytical solutions and their properties
are valuable for testing and benchmarking numerical MHD codes. For
astrophysical applications to large-scale galactic dynamics, our
model analysis contains more realistic elements and offers useful
insights into the structures and dynamics of disc galaxies consisting
of stars and magnetized interstellar medium. In particular, in the
presence of star burst activities, supernovae, hypernovae, superbubbles
etc., our open magnetic field geometry in disc galaxies bears strong
implications on circumnuclear and spiral galactic winds.
\end{abstract}

\begin{keywords}
galaxies: kinematics and dynamics --- galaxies: spiral ---
galaxies: structure --- ISM: general --- MHD --- waves.

\end{keywords}

\section{Introduction}

The theoretical magnetohydrodynamic (MHD) disc model we set out to
formulate in this paper is to explore possible large-scale global
perturbation structures and stationary MHD density waves (Fan \& Lou
1996; Lou \& Fan 1998) in a composite system of a stellar disc and
an isopedicallly magnetized gaseous disc intended for the interstellar
medium (ISM). These two gravitationally coupled discs are approximately
treated as `fluid' and `magnetofluid' respectively, and are both
geometrically idealized as razor-thin singular isothermal discs (SIDs)
with the gaseous SID being threaded across by an almost vertical
magnetic field throughout. In astrophysical contexts of large-scale
structures in disc galaxies, we also include gravitational effects
of a massive axisymmetric dark matter halo and adopt a background
composite system of two coupled partial SIDs (Syer \& Tremaine 1996;
Shu et al. 2000; Lou 2002; Lou \& Shen 2003; Shen \& Lou 2003, 2004a,
b; Lou \& Zou 2004, 2005; Shen, Liu \& Lou 2005). Our motivation is to
construct solutions with combined analytical and numerical techniques,
to understand their basic properties, to provide observational
diagnostics, and to reveal or speculate physical implications.

Chakrabarty, Laughlin \& Shu (2003) studied substructures in
grand-design spiral galaxies, such as branches, spurs and feathers,
using a two-component disc model in which the gas component responds
passively and nonlinearly to the potential of a rigidly rotating
spiral structure involving old stars and halos. We here treat a
dynamically coupled two-component disc system without or with a
massive dark matter halo and focus on stationary global MHD density
wave configurations. Specifically, we construct stationary MHD
configurations for aligned and unaligned logarithmic spiral
perturbations in a composite disc system of two SIDs with flat
rotation curves. For observational diagnostics of nearby disc
galaxies, we also examine phase relationships among perturbation
patterns of the stellar surface mass density, the ISM surface mass
density and the isopedic magnetic field.

We now proceed to provide the more general background
information relevant to the idealized MHD composite
disc problem to be formulated and investigated here.

In a pioneering work on a composite disc system of stellar and
gaseous discs dynamically coupled by gravity, Lin \& Shu (1966,
1968) proposed a combined approach involving a distribution
function for the stellar disc and a fluid description for the
gas disc to derive and analyze the local dispersion relation
of coplanar galactic spiral density waves in the WKBJ
approximation. The basic physical scenario is that stars form
out of gas clouds in the ISM disc, leading to the coexistence
of a stellar disc and a magnetized ISM disc at a later evolution
stage of disc galaxy. Since the seminal work of Lin \& Shu, there
have been extensive theoretical researches on density wave
oscillations, perturbation configurations and stability properties
of a rotating composite disc system, mainly in the galactic context.
Kato (1972) studied oscillations and overstabilites of density waves
using a  formalism similar to that of Lin \& Shu (1966, 1968). Jog
\& Solomon (1984a, b) discussed the growth of local axisymmetric
perturbations using a two-fluid formalism in a composite disc
system. Bertin \& Romeo (1988) investigated the role of a gas disc
for spiral modes in a two-fluid model framework. The influence of
interstellar gas on oscillations and stabilities of spheroidal
galaxies was studied by Vandervoort (1991a, b). In order to account
for the effects of the disc thickness, Romeo (1992) adopted a
two-fluid approach to investigate a two-component disc system with
finite disc thickness. Lowe et al. (1994) performed an extensive
analysis for morphologies of disc galaxies. Different effective
$Q_{eff}$ parameters (Safronov 1960; Toomre 1964) have been
suggested for the axisymmetric stability of a composite disc system
in a two-fluid formalism by Elmegreen (1995) and Jog (1996). Lou \&
Fan (1998b) recently used the two-fluid formalism to study properties
of open and tight-winding spiral density-wave modes in a composite
disc system. Lou \& Shen (2003) discussed stationary global
perturbation structures in a two-fluid formalism and offered a more
straightforward $D-$criterion for the axisymmetric instability in a
composite SID system instead of a redefinition of a new $Q_{eff}$
parameter (Shen \& Lou 2003). Considering the magnetic field embedded
in the gaseous disc, Lou \& Zou (2004, 2005) studied the stationary
global coplanar MHD perturbation structures and axisymmetric stability
in a composite SID system.

A series of astrophysical disc problems involves stability analysis
of a SID. Since the pioneer work of Mestel (1963) more than four
decades ago, numerous theoretical and numerical studies have been
carried out in this subject area (e.g. Zang 1976; Toomre 1977; Lemos,
Kalnajs \& Lynden-Bell 1991; Lynden-Bell \& Lemos 1999; Goodman \&
Evans 1999; Chakrabarti, Laughlin \& Shu 2003). An important
breakthrough by Syer \& Tremaine (1996) was to derive the
semi-analytic solutions for stationary coplanar perturbation
configurations in a class of power-law discs (i.e., SID is only a
special case). Shu et al. (2000) obtained stationary solutions for
global perturbation configurations in an isopedically magnetized
SID with a flat rotation curve. Based on extensive numerical
calculations, they interpreted these stationary aligned and
unaligned logarithmic spiral configurations as the onsets of
bar-type and barred-spiral instabilities (see also Galli et al.
2001). As a different yet complementary work to the analysis of
Shu et al. (2000), Lou (2002) studied a coplanar MHD perturbation
analysis in a single background SID embedded with an azimuthal
magnetic field, from the perspective of stationary fast and slow
MHD density waves (FMDWs and SMDWs; Fan \& Lou 1996; Lou \& Fan
1998a). Lou (2002) also derived a form of magnetic virial theorem
for an MSID and suggested by analogy that the ratio of the disc
rotational kinetic energy to the sum of the gravitational and
magnetic energies may be crucial for the onset of bar-like
instability in an MSID system. To model large-scale effects of
a galactic magnetic field, it would be more realistic to treat
large-scale structures and dynamics as a composite disc system
of stars and a magnetized ISM. Lou \& Shen (2003) made an foreray
into this model problem, constructed stationary aligned and
unaligned logarithmic spiral configurations in such a composite
SID system and further examined axisymmetric instability properties
(Shen \& Lou 2003). Meanwhile, Lou \& Zou (2004, 2005) incorporated
a coplanar magnetic field into the composite SID system for a more
specific analysis.

The geometry of an isopedic magnetic field proposed by Shu \& Li (1997)
is an open magnetic field configuration across the disc plane. For disc
galaxies, open magnetic fields are most likely interlaced or intermingled
with `closed' coplanar magnetic fields in a stochastic manner.\footnote{The
closed and open magnetic fields over the solar surface into the lower solar
corona serve as an empirical analogy.} For a much simplified theoretical
model analysis at this stage, we take an isopedic magnetic field geometry
all over a composite SID system. Together with other physical mechanisms
(e.g., massive star formation activities, supernovae, hypernovae, hot
bubbles and superbubbles etc.), such an open magnetic field geometry can
generate and channel hot gas outflows, often referred to as `galactic
winds' under increasing scrutiny of multi-band observations and the X-ray
band in particular. Along with observational and theoretical studies of
stellar winds (e.g., Burke 1968; Holzer \& Axford 1970), an initial
impetus was made to formulate the seminal concept of galactic-scale winds
blowing away from both sides of the disc plane (e.g., Johnson \& Axford
1971). As galactic winds are expected to be a relatively weak phenomenon
in terms of radiative signatures, we start to accumulate some substantial
observational evidence in recent years. It is generally conceived
that supernova explosions and stellar radiations drive galactic
outflows. Strickland, Ponman \& Stevens (1997) have reported X-ray
observations of galactic outflows from the starburst galaxy M82,
which may imply an overall configuration of a galactic wind. On
both observational and theoretical grounds, magnetic field
configurations away from a galactic plane should play dynamical,
diagnostic and geometric roles in probing galactic winds. As a
result of such open magnetic field, the outflow of hot gas can
lead to large-scale observable effects.

For such a magnetized composite disc system, MHD will play an
indispensable role in the gaseous MSID and reveal more realistic
aspects of coupled large-scale dynamics as well as diagnostic
information. These important MHD disc problems (Shu et al. 2000;
Lou \& Shen 2003; Lou \& Zou 2004, 2005; Shen, Liu \& Lou 2005)
are not only interesting for their own sakes, but also serve as
necessary steps for developing more and more realistic physical
models. We construct here stationary perturbation configurations
for aligned and unaligned logarithmic spiral cases in a composite
system of a stellar SID and an isopedically magnetized gaseous
SID, and discuss their stability properties. For numerical MHD
simulations into the nonlinear regime, the global perturbation
solutions here are valuable for initializing, benchmarking and
understanding the MHD code development.

We adopt a relatively simple formalism for a composite disc system
containing infinitely thin fluid (for stars) and magnetofluid (for
the magnetized ISM) discs dynamically coupled by gravity. We describe
in Section 2 the isopedic magnetic field and a few important theorems
for the gaseous MSID. In Section 3, we determine conditions for the
background axisymmetric equilibrium state for both stellar SID and
gaseous MSID, and derive linearized equations for MHD perturbations.
The global MHD perturbation solutions can be classified as aligned
perturbations and unaligned logarithmic spiral perturbations; these
solutions are analyzed in detail in Section 4. The exact solutions
of stationary MHD perturbations, their properties and their
corresponding phase relationships among perturbation variables are
examined and summarized in Section 5. Several details of mathematical
analysis are included in Appendices A and B for the convenient of
references.

\section[]{Isopedic Magnetic Fields}

Magnetic fields may be generated and sustained through nonlinear
MHD dynamo processes in an electrically conducting gas medium in
motion (Parker 1979 and extensive references therein). Several
numerical simulation results and some theories for galaxy
formation lend supports to the basic assumption that the
mass-to-magnetic flux ratio $\Lambda$ may be constant in space
and time. For example, model calculations and numerical simulations
of cloud core formation by ambipolar diffusion tend to produce
gravitationally unstable inner regions with approximately constant
values of $\Lambda$ (e.g., Nakano 1979; Lizano \& Shu 1989). In
the models of Basu \& Mouschovias (1994), the mass-to-magnetic
flux ratio $\Lambda$ remains almost constant when the variation
ranges of density and magnetic field flux are large.
%
In the theoretical formulation of Shu \& Li (1997) and Shu et al.
(2000), it was thus {\it presumed} that $\Lambda$ is a constant;
in particular, they referred to these earlier results to justify
this {\it assumption}. Here, we emphasize the theoretical fact
that $\Lambda$ remains constant is a natural consequence of the
standard ideal MHD equations. In cylindrical coordinates
$(r,\varphi,z)$ and by using the gas mass conservation equation
(\ref{basic4}) below and the $\hat z-$component of the magnetic
induction equation
$$
{\partial B_z\over\partial t}=-{1\over r}{\partial\over\partial r}
(rv_rB_z)-{1\over r}{\partial\over\partial\varphi}(v_{\varphi}B_z)
$$
in the thin disc geometry, one can readily show that
$$
{\partial\over\partial t}\bigg({\Sigma^g\over B_z}\bigg)
+v_r{\partial\over\partial r}\bigg({\Sigma^g\over B_z}\bigg)
+{v_{\varphi}\over r}{\partial\over\partial\varphi}
\bigg({\Sigma^g\over B_z}\bigg)=0\ ,
$$
where the relevant notations bear the conventional meanings
in MHD. In reference to the works of Mouschovias (1994) and Li
\& Shu (1996) as well as the earlier numerical results (Nakano
1979; Lizano \& Shu 1989), it is fairly clear that the above
equation provides a natural explanation for the fact that
$\Lambda$ remains constant in the evolution of an MHD disc.

For the model problem under consideration, we ignore vertical
structures within the disc thickness as the horizontal scale
of the problem is very much larger. To analyze properties of
a drastically flattened disc, we adopt the cylindrical coordinates
$(r,\varphi,z)$ in our mathematical model prescription. We assume
that the mass density $\rho$ and the electric current density
${\bf \sl j_e}$ in our model to have nonzero values only within
a narrow range $\Delta z\ll r$ about the disc midplane $z=0$.
We are justified to define a surface mass density $\Sigma$ and
a surface electric current density ${\bf\sl J}$ by
$$
\Sigma(r,\varphi,t)\equiv\int_{\Delta z}\rho(r,\varphi,z,t)dz\
$$
and
$$
{\bf \sl J}(r,\varphi,t)\equiv
\int_{\Delta z}{\bf \sl j_e}(r,\varphi,z,t)dz\ ,
$$
with ${\bf\sl J}={\bf\sl J_r\hat{e_r}}
+{\bf\sl J_\varphi\hat{e_\varphi}}$ having only two orthogonal
components coplanar with the disc and the $z-$component
of ${\bf\sl J}$ within the disc thickness is omitted.

As for the two SIDs, one is a stellar SID approximated as a
fluid for large-scale dynamics and the other is a magnetized
gas SID treated as a magnetofluid. Over large scales, we
ignore the interaction between the magnetic field embedded
in the ISM and the stellar disc. It then comes the physical
scenario that the stellar and magnetized gas discs are
dynamically coupled by gravity, while the magnetic field
directly interacts with the gaseous ISM disc.

On the basis of the Poisson equation relating the gravitational
potential and the mass density, there will also be a jump in the
vertical gravitational field given by $4\pi G\Sigma$ across a
thin disc. In other words, the vertical gravitational field just
above and below the disc point towards the midplane $z=0$ and
have the same strength of $2\pi G\Sigma$. We choose to define
the ratio $\eta$ of the horizontal gravitational acceleration
$f_\parallel$ (continuous across the thin disc in vertical
direction) to the vertical gravitational acceleration just
above the disc as a dimensionless parameter, namely
$$
\eta\equiv\frac{f_\parallel}{(2\pi G\Sigma )}
$$
and the local value of the mass-to-magnetic flux ratio as
$$
\Lambda\equiv\frac{\Sigma^g(r,\varphi,t)}{B_z(r,\varphi,0,t)}\ .
$$
We then take $\Lambda$ to be constant in both space and time,
when computed at the footpoint of a magnetic field line anchored
through the disc (the adjective `isopedic' comes from the Latin
word `ped' for foot). For ideal MHD with the above frozen-in
condition, an initially isopedic disc retains the same value of
$\Lambda$ throughout the entire MHD evolution.

It is convenient and advantageous to nondimensionalize
this ratio in the form of
$$
\lambda\equiv 2\pi G^{\frac{1}{2}}\Lambda\ .
$$
We now discuss the two MHD theorems proven by Shu \&
Li (1997) that form the fundamental part of the MHD
equations to come in our theoretical model analysis.

THEOREM 1: {\it For an arbitrary distribution of the surface
mass density $\Sigma(r,\varphi, t)$, which may or may not be
in mechanical equilibrium in the horizontal directions, the
magnetic tension force in the plane of a thin isopedic disc
equals $-1/\lambda^2$ times the horizontal self-gravitational
force $f_{\parallel}$.}

THEOREM 2: {\it If we also assume that the disc remains
in a vertical magnetostatic equilibrium, then the
magnetic pressure integrated over the disc thickness
may be approximated as $(1+\eta^2)/(\lambda^2+\eta^2)$
times the thermal gas pressure integrated over $z$.   }

By the first theorem, the magnetic tension force
$f_{ten}$ tangential to the disc plane is simply
$$
f_{ten}=-f_{\parallel}/{\lambda^2}
$$
and the combined forces of magnetic tension and
gravitational field act in a form of a diluted
horizontal gravity
$$
f\equiv f_{ten}+f_\parallel=\epsilon f_\parallel\ ,
$$
where $\epsilon\equiv 1-1/\lambda^2$ is the reduction factor.

By the second theorem, it follows approximately that
$$
\Pi_m=\Pi_g\bigg(\frac{1+\eta^2}{\lambda^2+\eta^2}\bigg)
$$
where $\Pi_m$ is the projected magnetic pressure acting on
the gas material and $\Pi_g$ is the vertically integrated
thermal gas pressure. Therefore, the total projected
pressure is
$$
\Pi\equiv\Pi_m+\Pi_g=\Theta\Pi_g=\Theta a_g^2\Sigma^g\ ,
$$
where the enhancement factor $\Theta$ is given by
$$
\Theta\equiv\frac{(1+\lambda^2+2\eta^2)}
{(\lambda^2+\eta^2)}\ .
$$
Coming to a composite disc system of two gravitationally coupled
discs with the gas SID being isopedically magnetized, the above
two theorems should be modified accordingly. By carefully examining
the two theorems in details, we note that as the stellar disc does
not interact with the magnetic field, the first theorem of Shu \&
Li should retain its basic form with some relevant parameters being
properly adjusted. For example, $\Lambda$ should now stand for
$\Sigma^g/B_z$ where $\Sigma^g$ is the surface mass density of the
gas disc and the magnetic tension force $f_{ten}$ acts on the
magnetized gas SID. Naturally, we introduce an appropriate $\eta$
parameter for convenience, namely
$\eta\equiv f_{\parallel g}/(2\pi G\Sigma^g)$, where
$f_{\parallel g}$ is the contribution to the total $f_{\parallel}$
from the gaseous MSID. For the two SIDs having the same structure
and density profile, we further obtain
\begin{equation}\label{lambda}
\eta\equiv\frac{f_{\parallel g}}{2\pi G\Sigma^g}
=\frac{(f_{\parallel g}+f_{\parallel s})}{2\pi G
(\Sigma^g+\Sigma^s)}=\frac{f_{\parallel}}{2\pi G\Sigma}\ ,
\end{equation}
where $\Sigma^s$ is the surface mass density of the stellar SID,
$\Sigma$ is the total surface mass density and $f_{\parallel s}$ is
the contribution to the total $f_{\parallel}$ from the stellar SID.

For the second theorem of Shu \& Li, the generalization and extension
are somewhat different. We now examine the procedure of Shu \& Li to
generalize the second theorem for a composite disc system with one
stellar SID and one isopedically magnetized SID.

The force per unit volume associated with the magnetic
pressure gradient is given by $-\nabla[B^2/(8\pi)]$, where
\begin{equation}\label{B}
B^2=B_r^2+B_\varphi^2+B_z^2\ .
\end{equation}
By integrating $-\nabla[B^2/(8\pi)]$ from $z=-\infty$ to
$z=+\infty$, the contribution from the $z-$component of the
magnetic pressure gradient vanishes for $B^2\rightarrow 0$
as $|z|\rightarrow +\infty$. We thus obtain the parallel
magnetic force per unit projected surface area as
\begin{equation}\label{surfacetension}
-\nabla_{\parallel}\int_{-\infty}^{+\infty}
\frac{B^2}{(8\pi )}dz\ ,
\end{equation}
where the parallel gradient operation
within the disc plane is defined by
$$
\nabla_\parallel\equiv\hat{e_r}\frac{\partial}{\partial r}
+\frac{\hat{e_\varphi}}{r}\frac{\partial}{\partial\varphi}\ .
$$

Away from the disc, the total integral of equation
~(\ref{surfacetension}) does not act on materials contained
in the gaseous MSID. In the vacuum regions above and below
the MSID, the horizontal force of magnetic pressure gradient
does not vanish but counteracts instead a nonzero force of
magnetic tension, such that the sum produces a force-free
environment in vacuum (i.e., no electric current flowing
around). We specifically denote the part of the projected
magnetic pressure acting on the gaesous MSID as $\Pi_m$. For
a magnetized gas disc of a thin but nonvanishing thickness,
we estimate $\Pi_m$ by
\begin{equation}\label{Pim}
\Pi_m\equiv\int_{\Delta z}\frac{B^2}{8\pi}dz
=\frac{z_0B_+^2}{(4\pi)}\ ,
\end{equation}
where $z_0$ is the effective half-thickness of the gaseous MSID
and $B_+^2$ is the value of $B^2$ evaluated just above the upper
disc surface. While our estimate for $\Pi_m$ is plausible on the
dimensional ground, the possibility remains for a multiplicative
factor of order unity whose exact value would depend on the
vertical structure within a thin MSID.

In order to determine the two expressions of $z_0$ and $B_+^2$,
we begin with a generalization which states that a quasi vertical
magnetostatic equilibrium in a thin gas disc requires the sum of
the gas, magnetic and gravitational pressures to be independent
of the vertical height $z$, namely,
\begin{eqnarray}\label{K}
P_g+\frac{B^2}{8\pi}+\frac{\pi}{2}G\sigma^2=K\ ,
\end{eqnarray}
where $P_g$ is the thermal pressure of the gaseous MSID
and $K$ is an integration constant; and by assuming that
the gaseous MSID and the stellar SID interact only via
the mutual gravity, we have the total effective surface
mass density in equilibrium as
\begin{equation}\label{sigma}
\sigma\equiv 2\int_0^z(\rho_s+\rho_g)dz\ .
\end{equation}
In the midplane $z=0$, we have by symmetry $\sigma=0$,
$B_r=B_{\varphi}=0$, $B_z=\Sigma^g/\Lambda$, while $P_g=\Pi_g/(2z_0)$
with $\Pi_g$ being the vertically integrated thermal gas pressure and
$z_0$ is the half-scaleheight in the vertical direction. Thus equation
~(\ref{K}) yields at $z=0$ the following expression for constant $K$,
\begin{eqnarray}\label{K0}
K=\frac{\Pi_g}{2z_0}+\frac{(\Sigma^g)^2}{8\pi\Lambda^2}\ .
\end{eqnarray}
Meanwhile for $r\gg z\gg z_0$, we have $P_g=0$, $P_s=0$,
$\sigma=\Sigma^s+\Sigma^g$, and ${\mathbf B}=
(\Sigma^g/\Lambda)\hat{\mathbf e_z}-{\bf\sl f_{\parallel g}}
/(2\pi G\Lambda)$ [see derivations of Shu \& Li (1997) that
lead to their equations (2.14) and (2.15)]. In these spatial
regions, equation (\ref{K}) leads to
\begin{eqnarray}\label{K1}
\label{K01} K=\frac{(\Sigma^g)^2}{8\pi\Lambda^2}
+\frac{f_{\parallel g}^2}{32\pi^3G^2\Lambda^2}
+\frac{\pi}{2}G(\Sigma^g+\Sigma^s)^2\ .
\end{eqnarray}
By introducing the gas-to-stellar surface mass density ratio
$\delta\equiv\Sigma^g/\Sigma^s$ from the two SIDs, we derive
from equations (\ref{K0}) and (\ref{K1}) the following relation
\begin{eqnarray}
\label{Piz}\frac{\Pi_g}{z_0}=\pi
G(\Sigma^g)^2\bigg(1+\frac{1}{\delta}\bigg)^2
+\frac{f_{\parallel g}^2}{(4\pi G\lambda^2)}
\nonumber \\ \qquad\qquad
=\pi G(\Sigma^g)^2\frac{[(1+\delta)^2\lambda^2
+\delta^2\eta^2]}{\delta^2\lambda^2}\ .
\end{eqnarray}
It then follows immediately that
\begin{equation}
\label{z0} z_0=\frac{\Pi_g}{\pi
G(\Sigma^g)^2}\frac{\delta^2\lambda^2}
{[(1+\delta)^2\lambda^2+\delta^2\eta^2]}\ .
\end{equation}
Just above the disc, equation (\ref{K}) leads to
\begin{equation}\label{B2}
\frac{B_+^2}{8\pi}=K-\frac{\pi}{2}G
(\Sigma^g)^2\bigg(1+\frac{1}{\delta}\bigg)^2\ .
\end{equation}
Using expression (\ref{K0}) for
constant $K$, we then obtain
\begin{equation}\label{B2+}
\frac{B_+^2}{8\pi}=\frac{\Pi_g}{2z_0} +\frac{\pi
G(\Sigma^g)^2}{2\lambda^2} -\frac{\pi G(\Sigma^g)^2}{2}
\bigg(1+\frac{1}{\delta}\bigg)^2\ .
\end{equation}
Multiplying equation (\ref{B2+}) by $2z_0$ with $z_0$
being given explicitly by expression (\ref{z0}), we
derive from definition (\ref{Pim})
\begin{equation}\label{pimpig}
\Pi_m=\frac{z_0B_+^2}{(4\pi )}
=\frac{(1+\eta^2)\Pi_g}
{[(1+\delta^{-1})^2\lambda^2+\eta^2]}\ .
\end{equation}
It is convenient to introduce a modified $\lambda$
parameter $\hat{\lambda}\equiv(1+\delta^{-1})\lambda$
such that expression (\ref{pimpig}) becomes
\begin{equation}\label{Pim2}
\Pi_m=\Pi_g\frac{(1+\eta^2)}{(\hat{\lambda}^2+\eta^2)}\ .
\end{equation}
By summing up the gas and magnetic pressure contributions,
we obtain the total pressure in the form of
\begin{equation}\label{Pi}
\Pi=\Pi_g+\Pi_m\equiv\Theta\Pi_g\ ,
\end{equation}
where the modified enhancement factor $\Theta$ is given by
\begin{equation}
\label{finalT}
\Theta\equiv\frac{(1+\hat{\lambda}^2+2\eta^2)}
{(\hat{\lambda}^2+\eta^2)}\ .
\end{equation}
This completes our extension and generalization of the second theorem
of Shu \& Li as applied to the MHD problem of a composite disc system
consisting of one stellar SID and one gaseous isopedic MSID.

At the end of this section, we summarize the two extended
theorems for a composite MSID system with an isopedic magnetic
field as follows. In the presence of an isopedic magnetic field,
the horizontal self-gravity force $\mathbf f_{\parallel g}$ and
the gas pressure integral $\Pi_g (=a_g^2\Sigma^g)$ of the
gaseous MSID will be modified as
\begin{eqnarray*}
{\mathbf f}_{\parallel g}\rightarrow\epsilon
{\mathbf f}_{\parallel g},
\qquad\qquad \Pi_g\rightarrow\Theta\Pi_g\ ,
\end{eqnarray*}
where $\epsilon\equiv 1-\lambda^{-2}$ and $\Theta$ is defined
above by equation (\ref{finalT}) that involves essentially
three dimensionless parameters $\delta\equiv\Sigma^g/\Sigma^s$,
$\lambda\equiv 2\pi G^{1/2}\Sigma^g/B_z$, $\eta\equiv
f_{\parallel g}/(2\pi G\Sigma^g)=f_{\parallel}/(2\pi G\Sigma)$
and $\hat{\lambda}\equiv(1+\delta^{-1})\lambda$.

\section{FLUID-MAGNETOFLUID FORMALISM\\
\quad\ \ FOR TWO THIN COUPLED SIDS}

\subsection{Basic Equations for a Composite
MSID System and the Background Equilibrium}

The two coupled SIDs located at $z=0$ are both approximated
as infinitesimally thin for expediency and they are coupled
through the mutual gravity. For large-scale stationary MHD
perturbations, diffusive processes such as viscosity,
ambipolar diffusion and thermal diffusion etc. are all
ignored in our formulation for simplicity. For physical
variables under consideration, we use either superscript
or subscript $s$ to indicate the association with the
stellar SID and either superscript or subscript $g$ to
indicate the association with the gaseous MSID. In the
cylindrical coordinates ($r$, $\varphi$, $z$), the ideal
fully nonlinear fluid-magnetofluid equations for a
composite MSID system can be readily written out.

For the stellar SID in the `fluid' approximation, we have
\begin{equation}\label{basic1}
\frac{\partial \Sigma^s}{\partial t}+
\frac{1}{r}\frac{\partial}{\partial r}(r\Sigma^s u^s)
+\frac{1}{r^2}\frac{\partial}{\partial \varphi}(\Sigma^sj^s)=0\
\end{equation}
for the two-dimensional stellar mass conservation,
\begin{equation}\label{basic2}
\frac{\partial u^s}{\partial t}+
u^s\frac{\partial u^s}{\partial r}
+\frac{j^s}{r^2}\frac{\partial u^s}{\partial \varphi}
-\frac{j^{s2}}{r^3}
=-\frac{1}{\Sigma^s}\frac{\partial}{\partial r}(a^2_s\Sigma^s)
-\frac{\partial (\phi^s+\phi^g)}{\partial r}
\end{equation}
for the radial momentum equation, and
\begin{equation}\label{basic3}
\frac{\partial j^s}{\partial t}+
u^s\frac{\partial j^s}{\partial r}
+\frac{j^s}{r^2}\frac{\partial j^s}{\partial\varphi}
=-\frac{1}{\Sigma^s}\frac{\partial}{\partial\varphi}(a^2_s\Sigma^s)
-\frac{\partial (\phi^s+\phi^g)}{\partial\varphi}
\end{equation}
for the azimuthal momentum equation, where the
`isothermal' approximation has been invoked.

In parallel, for the gaseous isopedic MSID
in the magnetofluid approximation, we have
\begin{equation}\label{basic4}
\frac{\partial \Sigma^g}{\partial t}+
\frac{1}{r}\frac{\partial}{\partial r}(r\Sigma^g u^g)
+\frac{1}{r^2}\frac{\partial}{\partial\varphi}(\Sigma^gj^g)=0
\end{equation}
for the two-dimensional gas mass conservation,
\begin{equation}\label{basic5}
\frac{\partial u^g}{\partial t}+
u^g\frac{\partial u^g}{\partial r}
+\frac{j^g}{r^2}\frac{\partial u^g}{\partial\varphi}
-\frac{j^{g2}}{r^3}=-\frac{1}{\Sigma^g}
\frac{\partial}{\partial r}(\Theta a^2_g\Sigma^g)
-\frac{\partial (\phi^s+\epsilon\phi^g)}{\partial r}
\end{equation}
for the radial momentum equation, and
\begin{equation}\label{basic6}
\frac{\partial j^g}{\partial t}+
u^g\frac{\partial j^g}{\partial r}
+\frac{j^g}{r^2}\frac{\partial j^g}{\partial\varphi}
=-\frac{1}{\Sigma^g}\frac{\partial}
{\partial\varphi}(\Theta a^2_g\Sigma^g)
-\frac{\partial (\phi^s+\epsilon\phi^g)}{\partial\varphi}
\end{equation}
for the azimuthal momentum equation, where
$\epsilon\equiv 1-1/\lambda^2$ and the modified $\Theta$
here is defined by equation (\ref{finalT}). The dynamical
coupling between the two sets of fluid and magnetofluid
equations is caused by the gravitational potential
through the Poisson integral relations
\begin{equation}\label{basic7}
\phi^s(r,\varphi,t)=\oint d\psi\int_0^{\infty}
\frac{-G\Sigma^s(r^\prime,\psi,t)r^\prime dr^\prime}
{[r^{\prime2}+r^2-2rr^{\prime}
\cos(\psi-\varphi)]^{1/2}}\ ,
\end{equation}
and
\begin{equation}\label{basic8}
\phi^g(r,\varphi,t)=\oint d\psi\int_0^{\infty}
\frac{-G\Sigma^g(r^\prime,\psi,t)r^\prime dr^\prime}
{[r^{\prime2}+r^2-2rr^{\prime}
\cos(\psi-\varphi)]^{1/2}}\ ,
\end{equation}
as well as through the effects of $\delta$ and $\eta$ in the $\Pi_m$
term. In equations (\ref{basic1})$-$(\ref{basic8}), $\Sigma^s$ is
the stellar surface mass density, $u^s$ is the radial component of
the bulk stellar `fluid' velocity, $j^s$ is the $z-$component of
the specific angular momentum of the stellar `fluid', and $a_s$ is
the stellar velocity dispersion presumed to be constant (or an
equivalent `isothermal sound speed' for the stellar SID),
$a^2_s\Sigma^s$ stands for an effective pressure in the isothermal
approximation. For physical variables of the gaseous isopedic MSID,
we simply replace the superscript $s$ by $g$ systematically. Here,
we assume that the stellar and magnetized gaseous SIDs interact
mainly through the mutual gravitational coupling on large scales.
As the isopedic magnetic field only interacts with the gas disc,
$\Theta$ and $\epsilon$ only appear in the MHD equations for the
gaseous MSID.

Before an MHD perturbation analysis, one needs to establish a
background magneto-rotational equilibrium for the composite
MSID system that is dynamically consistent with equations
(\ref{basic1})$-$(\ref{basic8}). For that purpose, we presume
an axisymmetric background for both stellar and magnetized
gaseous SIDs, with the same form of power-law surface mass
densities (i.e., $\Sigma \propto r^{-1}$) yet with different
flat rotation curves in general.

Using the radial momentum equations (\ref{basic2}) and
(\ref{basic5}) for the background equilibrium of
axisymmetry with $u_0^s=u_0^g=0$, $\Omega_s=j_0^s/r^2$,
and $\Omega_g=j_0^g/r^2$, we readily obtain
\begin{equation}\label{backs}
-\Omega^2_sr=\frac{a_s^2}{r}-2\pi G(\Sigma_0^s+\Sigma_0^g)\
\end{equation}
and
\begin{equation}\label{backg}
-\Omega^2_gr=\Theta\frac{a_g^2}{r}
-2\pi G(\Sigma_0^s+\epsilon\Sigma_0^g)\ ,
\end{equation}
where $\delta\equiv\Sigma^g_0/\Sigma^s_0$, and
$\Omega_s\equiv a_sD_s/r$ and $\Omega_g\equiv
\Theta^{\frac{1}{2}}a_gD_g/r$ for the two rotational
Mach numbers $D_s$ and $D_g$, respectively. The two
equilibrium surface mass densities $\Sigma^s_0$ and
$\Sigma^g_0$ can be immediately determined as
\begin{equation}\label{sigmas0}
\Sigma^s_0=\frac{a_s^2}{2\pi Gr}\frac{(1+D^2_s)}{(1+\delta )}\
\end{equation}
and
\begin{equation}\label{sigmag0}
\Sigma^g_0=\frac{\Theta a_g^2}{2\pi Gr}
\frac{\delta (1+D^2_g)}{(1+\epsilon\delta )}\ .
\end{equation}
For stationary global MHD perturbation configurations to persist,
$D_s$ and $D_g$ can only take on specific values (e.g. Lou \&
Shen 2003; Shen \& Lou 2003, 2004a, b; Lou \& Zou 2004, 2005). In
the above equations, the two parameters $\delta$ and $\epsilon$
are constant, while $\Theta$ is a variable in general. However,
in our MHD perturbation analysis, $\Theta$ variation will be
ignored merely for simplicity without losing essential physics
(see Shu \& Li 1997; Shu et al. 2000). We now further introduce
a new dimensionless parameter $\beta\equiv a_s^2/a_g^2$ for the
square of the `sound speed ratio'.

Because $|f_{\parallel g}|=|d\phi^g_0/dr|=2\pi G\Sigma^g_0$
in an axisymmetric equilibrium state, we take $\eta=1$ and have
$\Theta>1$ in general. For a positive $\Sigma^g_0$ as a necessary
physical requirement, the equilibrium solution (\ref{sigmag0})
implies the following three inequalities in sequence
\begin{eqnarray}\nonumber
\qquad
1+\epsilon\delta=1+\bigg(1-\frac{1}{\lambda^2}\bigg)\delta>0\ ,
\\ \nonumber
\delta+1>\frac{\delta}{\lambda^2}\ ,
\\ \nonumber
\lambda(1+\delta^{-1})>\frac{1}{\lambda}\ .
\end{eqnarray}
As $\delta>0$ by definition, it is obvious that
$\lambda(\delta^{-1}+1)>\lambda$. A multiplication of this
inequality with the third inequality above on both sides leads to
$\hat{\lambda}\equiv\lambda(1+\delta^{-1})>1$ for a physical
magneto-rotational background equilibrium. It then follows that
the enhancement factor $\Theta$ defined by equation (\ref{finalT})
should be constrained within a finite range of {$1<\Theta < 2$}.

We now discuss the parameter $\epsilon$. In Shu \& Li (1997), they
require $\epsilon\geq 0$ which is the same as $\lambda\geq 1$ for
a gravitationally bound disc. In the case of a single SID, the
physical explanation of this condition is that a rotating disc
with self-gravity must be held together in a radial force balance.
As a result of rotation and singular isothermal density distribution,
the centrifugal force and the effective and thermal pressure forces
all point radially outward. The only counterforce to keep the disc
in equilibrium is the radially inward gravity force. For $\epsilon<0$,
the magnetic tension force will be stronger than the horizontal
gravitational force; in this case, there will be no rotational
equilibrium for the disc system at all to begin with. In our
composite model containing one SID and one isopedic MSID, the
necessary condition for an equilibrium to exist
is $1+\epsilon\delta>0$ in comparison with the condition of
$\epsilon>0$ of Shu \& Li.
More precisely, we should have $0<\epsilon\leq 1$
for a single SID and $0<(1+\epsilon\delta)\leq (1+\delta )$
for a composite system of two gravity-coupled SIDs.

\subsection{MHD Perturbation Equations}

We now introduce nonaxisymmetric MHD perturbations to the
disc system, denoted by physical variables with subscript
1, in both stellar SID and gaseous isopedic MSID, namely
$$
\Sigma^s=\Sigma^s_0+\Sigma^s_1\ ,\qquad
\Sigma^g=\Sigma^g_0+\Sigma^g_1\ ,
$$
$$
u^s=u^s_0+u^s_1=u^s_1\ ,\qquad u^g=u^g_0+u^g_1=u^g_1\ ,
$$
$$
j^s=j^s_0+j^s_1\ ,\qquad j^g=j^g_0+j^g_1\ .
$$
It is then straightforward to write down the linearized
MHD perturbation equations in the forms of
\begin{equation}\label{liner1}
\frac{\partial \Sigma_1^s}{\partial t}
+\frac{1}{r}\frac{\partial}{\partial r}(r\Sigma_0^s u_1^s)
+\Omega_s \frac{\partial \Sigma_1^s}{\partial\varphi}
+\frac{\Sigma_0^s}{r^2}\frac{\partial j_1^s}{\partial\varphi}=0\ ,
\end{equation}
\begin{equation}\label{liner2}
\frac{\partial u_1^s}{\partial t}
+\Omega_s\frac{\partial u_1^s}{\partial\varphi}
-2\Omega_s\frac{j_1^s}{r}=-\frac{\partial}{\partial r}
\bigg(a_s^2\frac{\Sigma_1^s}{\Sigma_0^s}+\phi_1^s+\phi_1^g\bigg)\ ,
\end{equation}
\begin{equation}\label{liner3}
\frac{\partial j_1^s}{\partial t}+r\Omega_su_1^s
+\Omega_s\frac{\partial j_1^s}{\partial\varphi}
=-\frac{\partial}{\partial\varphi}
\bigg(a_s^2\frac{\Sigma_1^s}{\Sigma_0^s}+\phi_1^s+\phi_1^g\bigg)
\end{equation}
for the stellar SID, and
\begin{equation}\label{liner4}
\frac{\partial\Sigma_1^g}{\partial t}
+\frac{1}{r}\frac{\partial}{\partial r}(r\Sigma_0^g u_1^g)
+\Omega_g\frac{\partial\Sigma_1^g}{\partial\varphi}
+\frac{\Sigma_0^g}{r^2}\frac{\partial j_1^g}{\partial\varphi}=0\ ,
\end{equation}
\begin{equation}\label{liner5}
\frac{\partial u_1^g}{\partial t}
+\Omega_g\frac{\partial u_1^g}{\partial\varphi}
-2\Omega_g\frac{j_1^g}{r}=-\frac{\partial}{\partial r}
\bigg(\Theta a_g^2\frac{\Sigma_1^g}{\Sigma_0^g}
+\phi_1^s+\epsilon\phi_1^g\bigg)\ ,
\end{equation}
\begin{equation}\label{liner6u}
\frac{\partial j_1^g}{\partial t}+r\Omega_gu_1^g
+\Omega_g\frac{\partial j_1^g}{\partial\varphi}
=-\frac{\partial}{\partial\varphi}
\bigg(\Theta a_g^2\frac{\Sigma_1^g}{\Sigma_0^g}
+\phi_1^s+\epsilon\phi_1^g\bigg)
\end{equation}
for the gaseous isopedic MSID, with the
total gravitational potential perturbation
$\phi_1=\phi^s_1+\phi^g_1$ given by the sum of
\begin{equation}\label{liner7}
\phi^s_1(r,\varphi,t)=\oint\!\! d\psi\!\int_0^{\infty}\!\!
\frac{-G\Sigma^s_1(r^\prime\ ,\psi\ ,t)r^\prime
dr^\prime}{[r^{\prime2}+r^2-2rr^{\prime}
\cos(\psi-\varphi)]^{1/2}}
\end{equation}
and
\begin{equation}\label{liner8}
\phi^g_1(r,\varphi,t)=\oint\!\! d\psi\!\int_0^{\infty}\!\!
\frac{-G\Sigma^g_1(r^\prime\ ,\psi\ ,t)r^\prime
dr^\prime}{[r^{\prime2}+r^2-2rr^{\prime}
\cos(\psi-\varphi)]^{1/2}}\ .
\end{equation}
Assuming a Fourier harmonics in the periodic form of
$\exp[i(\omega t-m\varphi)]$ for MHD perturbation
solutions in general (after taking the real part),
we write for perturbations
\begin{equation}\label{perturb1}
\Sigma_1^l=\mu^l(r)\exp[i(\omega t-m\varphi)]\ ,
\end{equation}
\begin{equation}\label{perturb2}
u_1^l=U^l(r)\exp[i(\omega t-m\varphi)]\ ,
\end{equation}
\begin{equation}\label{perturb3}
j_1^l=J^l(r)\exp[i(\omega t-m\varphi)]\ ,
\end{equation}
\begin{equation}\label{perturb4}
\phi_1^l=V^l(r)\exp[i(\omega t-m\varphi)]\ ,
\end{equation}
where $l=s$ or $g$ denotes associations with stellar or
gaseous discs respectively. By substituting expressions
(\ref{perturb1})$-$(\ref{perturb4}) into equations
(\ref{liner1})$-$(\ref{liner8}), we derive for the
stellar SID
\begin{equation}\label{rel1}
i(\omega-m\Omega_s)\mu^s+\frac{1}{r}\frac{d}{dr}(r\Sigma_0^sU^s)
-im\Sigma_0^s\frac{J^s}{r^2}=0\ ,
\end{equation}
\begin{equation}\label{rel2}
i(\omega-m\Omega^s)U^s-2\Omega_s\frac{J^s}{r}
=-\frac{d}{dr}\bigg(a_s^2\frac{\mu^s}{\Sigma_0^s}+V^s+V^g\bigg)\ ,
\end{equation}
\begin{equation}\label{rel3}
i(\omega-m\Omega^s)J^s+r\Omega_sU^s
=im\bigg(a_s^2\frac{\mu^s}{\Sigma_0^s}+V^s+V^g\bigg)\ ,
\end{equation}
for the gaseous isopedic MSID
\begin{equation}\label{rel4}
i(\omega-m\Omega_g)\mu^g+\frac{1}{r}
\frac{d}{dr}(r\Sigma_0^gU^g)-im\Sigma_0^g\frac{J^g}{r^2}=0\ ,
\end{equation}
\begin{equation}\label{rel5}
i(\omega-m\Omega^g)U^g-2\Omega_g\frac{J^g}{r}=-\frac{d}{dr}
\bigg(\Theta
a_g^2\frac{\mu^g}{\Sigma_0^g}+V^s+\epsilon V^g\bigg)\ ,
\end{equation}
\begin{equation}\label{rel6}
i(\omega-m\Omega^g)J^g+r\Omega_gU^g=im\bigg(\Theta
a_g^2\frac{\mu^g}{\Sigma_0^g}+V^s+\epsilon V^g\bigg)\ ,
\end{equation}
and for the two gravitational potential perturbations
\begin{equation}\label{rel7}
V^s(r)=\oint d\psi\int_0^{\infty}
\frac{-G\mu^s(r^\prime)\cos{(m\psi)}r^\prime
dr^\prime}{[r^{\prime2}+r^2-2rr^{\prime} \cos\psi]^{1/2}}\ ,
\end{equation}
\begin{equation}\label{rel8}
V^g(r)=\oint d\psi\int_0^{\infty}
\frac{-G\mu^g(r^\prime)\cos{(m\psi)}r^\prime
dr^\prime}{[r^{\prime2}+r^2-2rr^{\prime} \cos\psi]^{1/2}}\ .
\end{equation}
We use equations (\ref{rel2}) and (\ref{rel3}) to express $U^s$
and $J^s$ in terms of $\Psi^s\equiv (a^2_s\mu^s/\Sigma_0^s)+V^s+V^g$
for the stellar SID and similarly, use equations (\ref{rel5}) and
(\ref{rel6}) to express $U^g$ and $J^g$ in terms of $\Psi^g\equiv
(\Theta a^2_g\mu^g/\Sigma_0^g)+V^s+\epsilon V^g$ for the gaseous MSID.
In the following analysis, we shall assume for simplicity that
$\Theta$ remains constant in the presence of MHD perturbations
in a composite MSID system (see Shu et al. 2000). The resulting
expressions then become
\begin{equation}\label{Us}
U^s=\frac{-i(\omega-m\Omega_s)}{(\omega-m\Omega_s)^2-2\Omega_s^2}
\bigg[\frac{2\Omega_s m}{r(\omega-m\Omega_s)}
-\frac{d}{dr}\bigg]\Psi^s
\end{equation}
and
\begin{equation}\label{Js}
\frac{J^s}{r}=\frac{(\omega-m\Omega_s)}{(\omega-m\Omega_s)^2
-2\Omega_s^2}\bigg[\frac{m}{r}-\frac{\Omega_s}{(\omega-m\Omega_s)}
\frac{d}{dr}\bigg]\Psi^s
\end{equation}
for the stellar SID, and
\begin{equation}\label{Ug}
U^g=\frac{-i(\omega-m\Omega_g)}{(\omega-m\Omega_g)^2-2\Omega_g^2}
\bigg[\frac{2\Omega_gm}{r(\omega-m\Omega_g)}-\frac{d}{dr}\bigg]\Psi^g
\end{equation}
and
\begin{equation}\label{Jg}
\frac{J^g}{r}=\frac{(\omega-m\Omega_g)}{(\omega-m\Omega_g)^2-2\Omega_g^2}
\bigg[\frac{m}{r}-\frac{\Omega_g}{(\omega-m\Omega_g)}
\frac{d}{dr}\bigg]\Psi^g
\end{equation}
for the gaseous isopedic MSID, respectively.

A substitution of expressions (\ref{Us}) and
(\ref{Js}) into equation (\ref{rel1}) leads to
\begin{eqnarray}
0=(\omega-m\Omega_s)\mu^s
\qquad\qquad\qquad\qquad\qquad\qquad\qquad\nonumber \\
-\frac{1}{r}\frac{d}{dr}
\bigg\{\frac{r\Sigma_0^s(\omega-m\Omega_s)}
{(\omega-m\Omega_s)^2-2\Omega_s^2}
\bigg[\frac{2\Omega_sm}{r(\omega-m\Omega_s)}-
\frac{d}{dr}\bigg]\Psi^s\bigg\}
\nonumber \\ \label{final1}
-\frac{m\Sigma_0^s (\omega-m\Omega_s)}
{r[(\omega-m\Omega_s)^2-2\Omega_s^2]}
\bigg[\frac{m}{r}-\frac{\Omega_s}{(\omega-m\Omega_s)}
\frac{d}{dr}\bigg]\Psi^s
\end{eqnarray}
for the stellar SID. Likewise, a substitution of expressions
(\ref{Ug}) and (\ref{Jg}) into equation (\ref{rel4}) leads to
\begin{eqnarray}
0=(\omega-m\Omega_g)\mu^g
\qquad\qquad\qquad\qquad\qquad\qquad\qquad\nonumber \\
-\frac{1}{r}\frac{d}{dr}
\bigg\{\frac{r\Sigma_0^g(\omega-m\Omega_g)}
{(\omega-m\Omega_g)^2-2\Omega_g^2}
\bigg[\frac{2\Omega_gm}{r(\omega-m\Omega_g)}
-\frac{d}{dr}\bigg]\Psi^g\bigg\}
\nonumber \\ \label{final2}
-\frac{m\Sigma_0^g(\omega-m\Omega_g)}
{r[(\omega-m\Omega_g)^2-2\Omega_g^2]}
\bigg[\frac{m}{r}-\frac{\Omega_g}{(\omega-m\Omega_g)}
\frac{d}{dr}\bigg]\Psi^g
\end{eqnarray}
for the gaseous isopedic MSID. Equations (\ref{final1})
and (\ref{final2}) should be solved together with
Poisson integral relations (\ref{rel7}) and (\ref{rel8}).

For stationary MHD perturbation solutions ($\omega=0$)
with zero pattern speed, we readily have
\begin{eqnarray}
m\Omega_s\mu^s=-\frac{1}{r}\frac{d}{dr}
\bigg\{\frac{m\Sigma_0^s}{(m^2-2)\Omega_s}
\bigg[2+r\frac{d}{dr}\bigg]\Psi^s\bigg\}
\nonumber \\ \label{Sstation}
+\frac{m\Sigma_0^s}{r(m^2-2)\Omega_s}
\bigg[\frac{m^2}{r}+\frac{d}{dr}\bigg]\Psi^s
\end{eqnarray}
for the stellar SID, and
\begin{eqnarray}
m\Omega_g\mu^g=-\frac{1}{r}\frac{d}{dr}
\bigg\{\frac{m\Sigma_0^g}{(m^2-2)\Omega_g}
\bigg[2+r\frac{d}{dr}\bigg]\Psi^g\bigg\}
\nonumber \\ \label{Gstation}
\frac{m\Sigma_0^g}{r(m^2-2)\Omega_g}
\bigg[\frac{m^2}{r}+\frac{d}{dr}\bigg]\Psi^g
\end{eqnarray}
for the gaseous isopedic MSID, respectively.

Using conditions and relationships
(\ref{backs})$-$(\ref{sigmag0}), we further
reduce the above two equations to the forms of
\begin{equation}\label{trans}
m\Omega_s\bigg[-\mu^s+\frac{\Sigma_0^s}{(m^2-2)
\Omega_s^2r}\bigg(\frac{m^2}{r}-2\frac{d}{dr}
-r\frac{d^2}{dr^2}\bigg)\Psi^s\bigg]=0
\end{equation}
and
\begin{equation}\label{trang}
m\Omega_g\bigg[-\mu^g+\frac{\Sigma_0^g}{(m^2-2)
\Omega_g^2r}\bigg(\frac{m^2}{r}-2\frac{d}{dr}
-r\frac{d^2}{dr^2}\bigg)\Psi^g\bigg]=0
\end{equation}
for the stellar SID and the gaseous isopedic MSID,
respectively. For nonzero $\Omega_s$ and $\Omega_g$,
we have
\begin{eqnarray}
\qquad m\bigg[-\mu^s+\frac{1}{D^2_s(m^2-2)}\bigg(\frac{m^2}{r}
-2\frac{d}{dr}-r\frac{d^2}{dr^2}\bigg)
\nonumber\\ \label{starel1}
\times\bigg(r\mu^s+\frac{1+D^2_s}
{2\pi G}\frac{V^s+V^g}{1+\delta}\bigg)\bigg]=0
\end{eqnarray}
for the stellar SID and
\begin{eqnarray}
\qquad m\bigg\{-\mu^g+\frac{1}{D^2_g(m^2-2)}
\bigg(\frac{m^2}{r}-2\frac{d}{dr}
-r\frac{d^2}{dr^2}\bigg)
\nonumber\\ \label{starel2}
\times\bigg[r\mu^g+\frac{(1+D^2_g)}{2\pi G}
\frac{\delta(V^s+\epsilon V^g)}
{(1+\epsilon\delta )}\bigg]\bigg\}=0
\end{eqnarray}
for the gaseous isopedic MSID, respectively. Equations
(\ref{starel1}) and (\ref{starel2}) are to be solved
simultaneously with Poisson integrals (\ref{rel7}) and
(\ref{rel8}) in order to determine the two rotational
Mach numbers $D_s^{2}$ and $D_g^{2}$.

\subsection{A Composite System of Partial (M)SIDs}

Since the early pioneer work of Zwicky (1933) and Smith (1936),
evidence has gradually emerged for the presence of dark matters
in clusters of galaxies. In the 1970s, systematic observations
of field spiral galaxies further revealed the generic presence
of dark matters in isolated spiral galaxies. By careful and
independent measurements of galactic rotations (e.g., Rubin
1987; Kent 1986, 1987, 1988), the more or less flat rotation
curves of galaxies determined by different methods clearly
imply the presence of dark matter halos in spiral galaxies.
To make our formulation more general and realistic, we here propose
a theoretical model of a composite {\it partial} MSID system to take
into account of the Newtonian gravitational effect of a presumed
axisymmetric dark matter halo (e.g., Binney \& Tremaine 1987; Bertin
\& Lin 1996; Syer \& Tremaine 1996; Shu et al. 2000; Lou 2002; Lou
\& Shen 2003; Shen \& Lou 2003, 2004a, b; Lou \& Zou 2004, 2005;
Shen, Liu \& Lou 2005). For a composite {\it partial} MSID system,
the preceding mathematical formulation for a composite MSID system
remains to be its essential part yet with an important distinctive
feature of adding an axisymmetric gravitational potential in the
background magneto-rotational equilibrium. In numerical simulations
designed for galaxy formation, velocity dispersions of dark matter
`particles' in the halo are typically high
(e.g., Hoeft et al. 2004). For simplicity, we may therefore ignore
dynamical perturbation responses of this massive dark matter halo
to MHD perturbations in the composite partial MSID system.
More specifically, we introduce an additive gravity term
$-\partial\Phi/\partial r$ to the basic equations ~(\ref{basic2}),
~(\ref{basic3}) and ~(\ref{basic5}), ~(\ref{basic6}), where $\Phi$
represents the gravitational potential contribution from the
background dark matter halo of axisymmetry. In the two components
of the momentum equation for the stellar disc, we then have
\begin{eqnarray}
\frac{\partial u^s}{\partial t}+
u^s\frac{\partial u^s}{\partial r}
+\frac{j^s}{r^2}\frac{\partial u^s}{\partial\varphi}
-\frac{j^{s2}}{r^3}
\qquad\qquad\qquad \nonumber\\ \qquad\qquad
\label{Dbasic2}=-\frac{1}{\Sigma^s}
\frac{\partial}{\partial r}(a^2_s\Sigma^s)
-\frac{\partial (\phi^s+\phi^g+\Phi)}{\partial r}\ ,
\end{eqnarray}
\begin{eqnarray}
\frac{\partial j^s}{\partial t}+
u^s\frac{\partial j^s}{\partial r}
+\frac{j^s}{r^2}\frac{\partial j^s}{\partial\varphi}
\qquad\qquad\qquad\qquad\qquad \nonumber\\ \qquad\qquad
\label{Dbasic3}=-\frac{1}{\Sigma^s}
\frac{\partial}{\partial\varphi}(a^2_s\Sigma^s)
-\frac{\partial (\phi^s+\phi^g+\Phi)}{\partial\varphi}\ ;
\end{eqnarray}
meanwhile, for the two components of the momentum
equations in the gaseous isopedic MSID, we have
\begin{eqnarray}
\frac{\partial u^g}{\partial t}+
u^g\frac{\partial u^g}{\partial r}
+\frac{j^g}{r^2}\frac{\partial u^g}{\partial\varphi}
-\frac{j^{g2}}{r^3}
\qquad\qquad\qquad\quad
\nonumber\\\label{Dbasic5}
\qquad\qquad
=-\frac{1}{\Sigma^g}\frac{\partial}{\partial r}(\Theta a^2_g\Sigma^g)
-\frac{\partial (\phi^s+\epsilon\phi^g+\Phi)}{\partial r}\ ,
\end{eqnarray}
\begin{eqnarray}
\frac{\partial j^g}{\partial t}+
u^g\frac{\partial j^g}{\partial r}
+\frac{j^g}{r^2}\frac{\partial j^g}{\partial\varphi}
\qquad\qquad\qquad\qquad\qquad\quad
\nonumber\\ \qquad\qquad\label{Dbasic6}
=-\frac{1}{\Sigma^g}
\frac{\partial}{\partial\varphi}(\Theta a^2_g\Sigma^g)
-\frac{\partial (\phi^s+\epsilon\phi^g+\Phi)}{\partial\varphi}\ .
\end{eqnarray}
For such a composite partial MSID system, we now introduce
a dimensionless ratio ${\mathcal F}$ parameter defined by
$\mathcal{F}$$\equiv(\phi^g+\phi^s)/(\phi^g+\phi^s+\Phi)$ (Lou
\& Shen 2003; Shen \& Lou 2003; Lou \& Zou 2004, 2005). The
background magneto-rotational equilibrium should be modified
accordingly. As before, we still write $\Omega_s=a_sD_s/r$,
$\Omega_g=a_gD_g/r$, and $u_0^s=u_0^g=0$ for the background.
The background equilibrium equations then become
\begin{equation}\label{Dstationary1}
-\Omega^2_sr=\frac{a_s^2}{r}-2\pi G
\frac{(\Sigma_0^s+\Sigma_0^g)}{\mathcal{F}}\
\end{equation}
and
\begin{equation}\label{Dstationary2}
-\Omega^2_gr=\Theta\frac{a_g^2}{r}-2\pi G
\bigg[\frac{(\Sigma_0^s+\Sigma_0^g)}{\mathcal{F}}
-(1-\epsilon)\Sigma_0^g\bigg]\ .
\end{equation}
It follows immediately that the two modified equilibrium
surface mass density profiles are now given by
\begin{equation}\label{Dsigmas0}
\Sigma^s_0=\frac{a_s^2}{2\pi Gr}
\frac{\mathcal{F}(1+D^2_s)}{(1+\delta)}
\end{equation}
and
\begin{equation}\label{Dsigmag0}
\Sigma^g_0=\frac{\Theta a_g^2}{2\pi Gr}
\frac{(1+D^2_g)\mathcal{F}\delta}
{1+[1-{\mathcal{F}}(1-\epsilon)]\delta}\ ,
\end{equation}
respectively. From these two radial equilibrium
conditions, we readily derive the requirement
\begin{equation}\label{condition}
a_s^2\{1+[1-{\mathcal{F}}(1-\epsilon)]\delta\}
(1+D_s^2)=\Theta a_g^2(1+\delta)(1+D_g^2)\
\end{equation}
resulting from the same total gravitational potential
that couples the two SIDs. We refer to such a system
as a composite system of partial (M)SIDs for
$0<{\mathcal{F}}<1$, in contrast to a composite
system of full (M)SIDs with ${\mathcal{F}}=1$.

The linearized MHD perturbation equations in a composite
system of partial (M)SIDs should take the same forms as
those in a composite system of full (M)SIDs explicitly
written down earlier in this section. For stationary
perturbation configurations in the modified equilibrium
with $\omega=0$, we then have for the stellar partial SID
\begin{eqnarray}
\quad m\bigg\{-\mu^s+\frac{1}{D^2_s(m^2-2)}
\bigg(\frac{m^2}{r}-2\frac{d}{dr}-r\frac{d^2}{dr^2}\bigg)
\qquad\nonumber\\ \quad \label{pstarel1}
\times\bigg[r\mu^s+\frac{(1+D^2_s)}{2\pi G}
\frac{{\mathcal{F}}(V^s+V^g)}{(1+\delta )}\bigg]\bigg\}=0\ ,
\end{eqnarray}
and simultaneously for the gaseous partial MSID
\begin{eqnarray}
\quad m\bigg\{-\mu^g+\frac{1}{D^2_g(m^2-2)}
\bigg(\frac{m^2}{r}-2\frac{d}{dr}-r\frac{d^2}{dr^2}\bigg)
\qquad\quad \nonumber\\ \label{pstarel2}
\times\bigg[r\mu^g+\frac{(1+D^2_g)}{2\pi G}
\frac{{\mathcal{F}}\delta(V^s+\epsilon V^g)}
{1+[1-{\mathcal{F}}(1-\epsilon)]\delta}\bigg]\bigg\}=0\ ,
\end{eqnarray}
where $\mathcal{F}$ factor appears
explicitly in both relations.

\subsection{Two Different Limiting Procedures}

Here, we discuss specifically the axisymmetric case of $m=0$
and start from the very beginning with $\omega\neq 0$ in
Fourier harmonics (\ref{perturb1})$-$(\ref{perturb4}).
%
By substituting expressions (\ref{perturb1})$-$(\ref{perturb4})
with $m=0$ into equations (\ref{liner1})$-$(\ref{liner8}), we
derive
\begin{equation}\label{rel1o}
i\omega\mu^s+\frac{1}{r}\frac{d}{dr}(r\Sigma_0^sU^s)=0\ ,
\end{equation}
\begin{equation}\label{rel2o}
i\omega U^s-2\Omega_s\frac{J^s}{r}
=-\frac{d}{dr}\bigg(a_s^2\frac{\mu^s}{\Sigma_0^s}+V^s+V^g\bigg)\ ,
\end{equation}
\begin{equation}\label{rel3o}
i\omega J^s+r\Omega_sU^q=0
\end{equation}
for the stellar SID,
\begin{equation}\label{rel4o}
i\omega\mu^g+\frac{1}{r}\frac{d}{dr}(r\Sigma_0^gU^g)=0\ ,
\end{equation}
\begin{equation}\label{rel5o}
i\omega U^g-2\Omega_g\frac{J^g}{r}
=-\frac{d}{dr}\bigg(\Theta a_g^2\frac{\mu^g}
{\Sigma_0^g}+V^s+\epsilon V^g\bigg)\ ,
\end{equation}
\begin{equation}\label{rel6o}
i\omega J^g+r\Omega_g U^g=0
\end{equation}
for the gaseous isopedic MSID, and
\begin{equation}\label{rel7o}
V^s(r)=\oint d\psi\int_0^{\infty}
\frac{-G\mu^s(r^\prime) r^\prime
dr^\prime}{[r^{\prime2}+r^2-2rr^{\prime}\cos\psi]^{1/2}}\ ,
\end{equation}
\begin{equation}\label{rel8o}
V^g(r)=\oint d\psi\int_0^{\infty}
\frac{-G\mu^g(r^\prime) r^\prime
dr^\prime}{[r^{\prime2}+r^2-2rr^{\prime}\cos\psi]^{1/2}}
\end{equation}
for the two gravitational potential perturbations.

Again, we use equations (\ref{rel2o}) and (\ref{rel3o})
to express $U^s$ and $J^s$ in terms of $\Psi^s\equiv
(a^2_s\mu^s/\Sigma_0^s)+V^s+V^g$ for the stellar SID and
in parallel, we use equations (\ref{rel5o}) and (\ref{rel6o})
to express $U^g$ and $J^g$ in terms of $\Psi^g\equiv\Theta
(a^2_g\mu^g/\Sigma_0^g)+V^s+\epsilon V^g$ for the gaseous
MSID. The resulting expressions are
\begin{equation}\label{Uso}
U^s=\frac{i\omega}{(\omega^2-2\Omega_s^2)}\frac{d\Psi^s}{dr}\ ,
\end{equation}
\begin{equation}\label{Jso}
\frac{J^s}{r}=-\frac{\Omega_s}
{(\omega^2-2\Omega_s^2)}\frac{d\Psi^s}{dr}
\end{equation}
for the stellar SID, and
\begin{equation}\label{Ugo}
U^g=\frac{i\omega}{(\omega^2-2\Omega_g^2)}\frac{d\Psi^g}{dr}\ ,
\end{equation}
\begin{equation}\label{Jgo}
\frac{J^g}{r}=-\frac{\Omega_g}
{(\omega^2-2\Omega_g^2)}\frac{d\Psi^g}{dr}
\end{equation}
for the gaseous MSID, respectively.

A substitution of expressions (\ref{Uso}) and
(\ref{Jso}) into equation (\ref{rel1o}) leads to
\begin{eqnarray}\label{final1o}
0=\omega\mu^s+\frac{1}{r}\frac{d}{dr}
\bigg[\frac{\omega r\Sigma_0^s}
{(\omega^2-2\Omega_s^2)}\frac{d\Psi^s}{dr}\bigg]
\end{eqnarray}
for the stellar SID. Likewise, a substitution of
expressions (\ref{Ugo}) and (\ref{Jgo}) into
equation (\ref{rel4o}) leads to
\begin{eqnarray}\label{final2o}
0=\omega\mu^g+\frac{1}{r}\frac{d}{dr}
\bigg[\frac{\omega r\Sigma_0^g}
{(\omega^2-2\Omega_g^2)}\frac{d\Psi^g}{dr}\bigg]
\end{eqnarray}
for the gaseous isopedic MSID. Ordinary differential
equations (ODEs) (\ref{final1o}) and (\ref{final2o})
are to be solved with Poisson integrals (\ref{rel7o})
and (\ref{rel8o}) simultaneously.


The case of $m=0$ is special. If we set $\omega=0$, all MHD
perturbation equations are satisfied in a trivial manner.
The harmonic factor $\exp{[i(\omega t-m\varphi)]}$ becomes
unity for both $\omega=0$ and $m=0$. In order to construct
stationary configurations, we set $\omega\rightarrow 0$ to
derive limiting solutions.

The two equations above can be cast into
\begin{equation}\label{final3o}
0=\omega\bigg\{\mu^s+\frac{\Sigma_0^s}
{(\omega^2-2\Omega_s^2)}\bigg[\frac{-4\Omega^2_s}
{(\omega^2-2\Omega_s^2)r}
\frac{d}{dr}+\frac{d^2}{dr^2}\bigg]\Psi^s\bigg\}
\end{equation}
for stellar disc and
\begin{equation}\label{final4o}
0=\omega\bigg\{\mu^g+\frac{\Sigma_0^g}
{(\omega^2-2\Omega_g^2)}
\bigg[\frac{-4\Omega^2_g}
{(\omega^2-2\Omega_g^2)r}\frac{d}{dr}+
\frac{d^2}{dr^2}\bigg]\Psi^g\bigg\}
\end{equation}
for the gaseous isopedic MSID.

For small $\omega\neq 0$, we readily obtain
\begin{equation}\label{final5o}
\mu^s+\frac{\Sigma_0^s}{(\omega^2-2\Omega_s^2)}
\bigg[\frac{-4\Omega^2_s}{(\omega^2-2\Omega_s^2)r}
\frac{d}{dr}+\frac{d^2}{dr^2}\bigg]\Psi^s=0
\end{equation}
and
\begin{equation}\label{final6o}
\mu^g+\frac{\Sigma_0^g}{(\omega^2-2\Omega_g^2)}
\bigg[\frac{-4\Omega^2_g}{(\omega^2-2\Omega_g^2)r}
\frac{d}{dr}+\frac{d^2}{dr^2}\bigg]\Psi^g=0\ ,
\end{equation}
by removing $\omega\neq 0$ factor in equations
(\ref{final3o}) and (\ref{final4o}).

For the onset of marginal instability in the limit of
$\omega\rightarrow 0$, we then derive from equations
(\ref{final5o}) and (\ref{final6o})
\begin{equation}\label{final7o}\mu^s
-\frac{\Sigma_0^s}{2\Omega_s^2}
\bigg(\frac{2}{r}\frac{d}{dr}
+\frac{d^2}{dr^2}\bigg)\Psi^s=0
\end{equation}
and
\begin{equation}\label{final8o}
\mu^g-\frac{\Sigma_0^g}{2\Omega_g^2}
\bigg(\frac{2}{r}\frac{d}{dr}
+\frac{d^2}{dr^2}\bigg)\Psi^g=0\ .
\end{equation}

These two resulting equations are exactly the
same as those obtained by setting $m=0$ in
equations (\ref{trans}) and (\ref{trang}),
\begin{equation}\label{transo}
-\mu^s+\frac{\Sigma_0^s}{(m^2-2)\Omega_s^2r}
\bigg(\frac{m^2}{r}-2\frac{d}{dr}
-r\frac{d^2}{dr^2}\bigg)\Psi^s=0
\end{equation}
for the stellar SID, and
\begin{equation}\label{trango}
-\mu^g+\frac{\Sigma_0^g}{(m^2-2)\Omega_g^2r}
\bigg(\frac{m^2}{r}-2\frac{d}{dr}
-r\frac{d^2}{dr^2}\bigg)\Psi^g=0
\end{equation}
for the gaseous isopedic MSID, respectively. This situation
of our composite MSID system parallels the case of Shu et al.
(2000) for a single SID which is isopedically magnetized
as well as the case of Lou \& Shen (2003) for a composite
SID system.

For a background MSID with a coplanar magnetic field,
the two different limiting procedures of $m=0$ and
$\omega\rightarrow 0$ give rise to two different sets
of resulting differential equations (Lou 2002; Lou \&
Zou 2004; Shen, Liu \& Lou 2005).

\section{ALIGNED AND UNALIGNED CASES}

\subsection{Aligned Global Perturbation Configurations}

By definitions, aligned MHD disturbances in a disc mean that maxima
and minima of perturbations occurring at different radii line up in
the azimuth. In contrast, when maxima and minima of perturbations
in a disc shift systematically in the azimuthal angle from one radial
location to the next, the resulting configuration for such disturbances
are referred to as unaligned or spiral patterns (Kalnajs 1973; Binney
\& Tremaine 1987; Shu et al. 2000; Lou \& Shen 2003; Lou \& Zou 2004).

\subsubsection{Axisymmetric Disturbances ($m=0$)}

With $\omega=m=0$, the solution to equations
(\ref{rel1})$-$(\ref{rel8}) takes the form of $U^s=U^g=0$,
$\mu^s=K_1^s/r$, $\mu^g=K_1^g/r$, $J^s=K_2^sr$, $J^g=K_2^gr$,
$V^s=K_3^s\ln r$, and $V^g=K_3^g\ln r$, where $K^{s,g}_i$ for
$i=1,2,3$ are all constants which can be chosen properly such
that equations (\ref{rel2}), (\ref{rel5}), (\ref{rel7}),
(\ref{rel8}) are satisfied. Nevertheless, such a solution
merely represents a rescaling of one axisymmetric
equilibrium to a neighbouring axisymmetric equilibrium.
Such rescaling solutions are allowed but are not of
particular interest here.

\subsubsection{Cases with $|m|\geq1$:
Nonaxisymmetric Disturbances}

In discs with power-law distributions of background physical
variables, we consider aligned perturbation variables that
carry the same power-law dependence as those of the
equilibrium SID do. By this choice, we make use of the
following exact potential-density pair relation, namely
\begin{equation}\label{alignrelation1}
\mu^s=\sigma_s/r\ ,\qquad\qquad \mu^g=\sigma_g/r\ ,
\end{equation}
\begin{equation}\label{alignrelation2}
V^s=-2\pi Gr\mu^s/|m|\ ,\qquad V^g=-2\pi Gr\mu^g/|m|\ ,
\end{equation}
where $\sigma_s$ and $\sigma_g$ are two small constants
for perturbations. A substitution of equations
(\ref{alignrelation1}) and (\ref{alignrelation2}) into
equations (\ref{starel1}) and (\ref{starel2}) then
leads to the following equations:
\begin{equation}\label{mus}
\mu^s=\bigg(\frac{m^2}{r}
-2\frac{d}{dr}-r\frac{d^2}{dr^2}\bigg)(H_1r\mu^s+G_1r\mu^g)\ ,
\end{equation}
\begin{equation}\label{mug}
\mu^g=\bigg(\frac{m^2}{r}-2\frac{d}{dr}
-r\frac{d^2}{dr^2}\bigg)(H_2r\mu^g+G_2r\mu^s)\ ,
\end{equation}
where the four coefficients $H_1, H_2, G_1$
and $G_2$ are defined by
\begin{eqnarray}\label{h1}
H_1\equiv\frac{1}{D^2_s(m^2-2)}\bigg[1-\frac{(D_s^2+1)}{|m|}
\frac{1}{(1+\delta )}\bigg]\ ,\\
\label{h2}
H_2\equiv\frac{1}{D^g_s(m^2-2)}
\bigg[1-\frac{(D_g^2+1)}{|m|}
\frac{\epsilon\delta}{(1+\epsilon\delta )}\bigg]\ ,\\
\label{g1}
G_1\equiv-\frac{(D_s^2+1)}
{D^2_s(m^2-2)|m|}\frac{1}{(1+\delta )}\ ,\\
\label{g2}
G_2\equiv-\frac{(D_g^2+1)}{D^2_g(m^2-2)|m|}
\frac{\delta}{(1+\epsilon\delta )}\ .
\end{eqnarray}
By substituting the forms of $\mu^s$ and $\mu^g$ given
by equation (\ref{alignrelation1}) into equations
(\ref{mus}) and (\ref{mug}), we readily obtain
\begin{eqnarray}\label{hg1}
\qquad\qquad (1-H_1m^2)\mu^s=G_1m^2\mu^g\ ,
\\ \label{hg2}
(1-H_2m^2)\mu^g=G_2m^2\mu^s\ .
\end{eqnarray}
For nontrivial solutions of $\mu^s$
and $\mu^g$, we then derive
\begin{equation}\label{hg3}
(1-H_1m^2)(1-H_2m^2)=G_1G_2m^4\
\end{equation}
for the stationary dispersion relation to
determined $D_s^{2}$ and thus $D_g^{2}$.
Using definitions (\ref{h1})$-$(\ref{g2}), the
solution condition (\ref{hg3}) can be written
explicitly in the form of
\begin{eqnarray}
\bigg[D^2_s(m^2-2)-m^2\bigg(1-\frac{D_s^2+1}{|m|}
\frac{1}{1+\delta}\bigg)\bigg]
\qquad\qquad\nonumber \\
\times\bigg[D^2_g(m^2-2)-m^2\bigg(1
-\frac{D_g^2+1}{|m|}\frac{\epsilon\delta}
{1+\epsilon\delta}\bigg)\bigg]
\qquad\nonumber \\ \label{hg5}
=\frac{(D^2_s+1)(D^2_g+1)m^2\delta}
{(1+\delta)(1+\epsilon\delta)}\ .\qquad\qquad
\end{eqnarray}
In terms of background profiles (\ref{sigmas0})
and (\ref{sigmag0}), we obtain
\begin{equation}\label{DsDg}
a_s^2(1+\epsilon\delta)(1+D_s^2)=\Theta a_g^2(1+\delta)(1+D_g^2)
\end{equation}
and then
\begin{equation}\label{Dg}
D_g^2=\frac{\beta(1+\epsilon\delta)}
{\Theta(1+\delta)}(1+D_s^2)-1\ ,
\end{equation}
where the ratio $\beta\equiv a_s^2/a_g^2$ was introduced earlier.
As is obvious, equation (\ref{hg5}) remains unchanged by replacing
$m$ with $-m$. For simplicity, we use $m$ to represent $|m|$. In
fact, positive and negative $m$ correspond to trailing and leading
spiral patterns relative to the sense of SID rotation, respectively.
After straightforward manipulations and rearrangements, equation
(\ref{hg5}) can be cast into a quadratic equation in terms of
$y\equiv D_s^2$, namely
\begin{equation}\label{quand}
C_2y^2+C_1y+C_0=0\ ,
\end{equation}
where the three coefficients
$C_2$, $C_1$ and $C_0$ are
\begin{eqnarray*}
C_2\equiv\frac{\beta}{\Theta}(|m|+2)(|m|-1)
\qquad\qquad\qquad\qquad\qquad\qquad\qquad  \\
\times\bigg[(m^2-2)\frac{(\epsilon\delta+1)}
{(\delta+1)}+\frac{|m|\delta(\epsilon-1)}
{(\delta+1)^2}\bigg]\ ,\qquad\qquad\\
C_1\equiv\bigg[2(|m|-1)
\qquad\qquad\qquad\qquad\qquad\qquad\qquad\qquad\qquad  \\
\times\frac{(m^2-2\delta-2)(1+\epsilon\delta)
+|m|\delta(\epsilon-2-\epsilon\delta)}{(1+\delta)^2}\bigg]
\frac{\beta}{\Theta}\qquad\\
-\frac{2(m^2-1)(m^2+\delta m^2+|m|-2-2\delta)}{(1+\delta )}\ ,
\qquad\qquad \\
C_0=-(m^2-|m|)
\qquad\qquad\qquad\qquad\qquad\qquad\qquad\qquad\quad \\
\times\frac{m^2(1+\delta)(1+\epsilon\delta)+|m|(2\epsilon\delta^2
+\delta+\epsilon\delta)-2(1+\epsilon\delta)}
{(1+\delta)^2}\frac{\beta}{\Theta}
\quad \\
+\frac{2|m|(m^2-1)(|m|+|m|\delta-1)}{(1+\delta )}\ ,\qquad
\end{eqnarray*}
respectively. From these three coefficient expressions,
the resulting $y$ solutions remain valid for
both positive and negative signs of $m$.
Unless otherwise stated, we shall take $m$
to be positive without loss of generality.

As seen from our MHD perturbation analysis so far, the $\Theta$
parameter always appears together with the $\beta$ parameter;
it is natural and convenient to introduce a new parameter
$\chi\equiv\beta/\Theta$ to simplify notations. Physically,
$\Theta a_g^2$ represents a sort of magnetosonic speed squared
given the isopedic magnetic field geometry; the ratio
$\chi\equiv a_s^2/(\Theta a_g^2)$ thus stands for the square
of the ratio between the stellar velocity dispersion and the
magnetosonic speed.

We now discuss properties of quadratic equation (\ref{quand})
with its determinant $\Delta$ given explicitly by
\begin{eqnarray}
\Delta\equiv C_1^2-4C_0C_2
=\frac{4(m^2-1)^2}{(1+\delta)^2}
\qquad\qquad\qquad\qquad\nonumber\\ \label{deltadeter}\qquad\qquad
\times\big{\{}\big{[}(m+2)
(m-1)(\chi\epsilon\delta-1)
\qquad\qquad\qquad\nonumber\\ \label{deltadeter}\qquad\qquad\qquad
+(m^2-2)(\chi-\delta)\big{]}^2
+4m^2\chi\delta\big{\}}\ .
\end{eqnarray}
Except for the case of $m=1$, the important fact that $\Delta>0$
for $m>1$ means that quadratic equation (\ref{quand}) always have
two real solutions of $y\equiv D_s^2$. The physical meaning is that
for aligned nonaxisymmetric stationary configurations to exist in a
composite MSID system with an isopedic magnetic field, condition
(\ref{hg5}), or equivalently, condition (\ref{quand}) must be
satisfied for appropriate values of $D_s^2$ given a set of specified
parameters $m$, $\delta$, $\epsilon$ and $\chi\equiv\beta/\Theta$.
The two real $y\equiv D_s^2$ solutions may not be necessarily
physical unless the nonnegative requirements of both $D_s^2\geq 0$
and $D_g^2\geq 0$ are met.

We now elaborate several subtle points below.

For the aligned case with $m=1$, it turns out that $C_2=C_1=C_0=0$
by definitions. Equation (\ref{quand}) can therefore be satisfied
for arbitrary values of $D_s^2$. This situation is quite similar
to the $m=1$ cases of a single isopedically magnetized SID studied
by Shu et al. (2000) and of a composite system of two coupled SIDs
analyzed by Lou \& Shen (2003).

When $m\geq 2$, the expressions of coefficient parameters
in the equation can be rearranged into the following forms
by multiplying through a factor of $(1+\delta)^2$, namely
\begin{eqnarray*}
C_2=\chi(m+2)\{(m^2-2)(\epsilon\delta+1)(\delta+1)
\quad\qquad\qquad \\ \qquad\qquad
+m[(\delta\epsilon+1)-(\delta+1)]\}\ ,\qquad
\\
C_1=2\chi\{[m^2-2(\delta+1)](1+\epsilon\delta)
\qquad\qquad\qquad\qquad \\ \qquad
+m[2(\delta\epsilon+1)-(\delta+1)-(\epsilon\delta+1)(\delta+1)]\}
\qquad \\
-2(m+1)(\delta+1)[(\delta+1)(m^2-2)+m]\ ,\quad
\\
C_0=-m\chi\{m^2(1+\delta)(1+\epsilon\delta)
-2(1+\epsilon\delta)
\qquad\qquad \\ \quad
+m[2(\epsilon\delta+1)(\delta+1)-(\delta+1)
-(\epsilon\delta+1)] \}
\\
+2m(\delta+1)(m+1)[m(\delta+1)-1]\ .
\end{eqnarray*}
Since $1+\delta>0$, $1+\epsilon\delta>0$ and $\chi>0$, we
further simplify these three coefficient expressions to
obtain equivalently
\begin{eqnarray}
\label{C2a}C_2&=&(m+2)\bigg[m^2-2-\frac{m}{(\delta\epsilon+1)}
+\frac{m}{(\delta+1)}\bigg]\ ,\quad\qquad\qquad
\\ \nonumber
C_1&=&2\bigg[\frac{m^2}{\delta+1}-2+m\bigg(\frac{2}{\delta+1}
-\frac{1}{\delta\epsilon+1}-1\bigg)\bigg]\quad\qquad\qquad
\\ \label{C1a}
& &-\frac{2(m+1)}{(\epsilon\delta+1)\chi}[(\delta+1)(m^2-2)+m]\
,\quad
\\ \nonumber
C_0&=&-m\bigg[m^2+m\bigg(2-\frac{1}{\delta+1}
-\frac{1}{\epsilon\delta+1}\bigg)-\frac{2}{(\delta+1)}\bigg]
\qquad \\ \label{C0a} &
&+\frac{2m(m+1)}{(\epsilon\delta+1)\chi}[m(\delta+1)-1]\ .
\end{eqnarray}
For the convenience of further analysis, we
introduce two new dimensionless parameters
$\mathcal{B}$ and $X$ defined by
\begin{eqnarray}\label{mathB}
{\mathcal{B}}\equiv\frac{(1+\delta )}
{(1+\epsilon\delta )}\ ,\\
\label{X}\qquad\qquad\qquad
X\equiv m^2-2-\frac{m({\mathcal{B}}-1)}{(1+\delta )}\ .
\end{eqnarray}
It then follows that $C_2$, $C_1$
and $C_0$ can be expressed as
\begin{eqnarray*}
C_2=(m+2)X\ ,
\quad\qquad\qquad\qquad\qquad\qquad\qquad\qquad\qquad\\
C_1=2X-2(m+1)\bigg[\frac{m\delta}{(1+\delta )}
+\frac{{\mathcal{B}}}{\chi}
\bigg(m^2-2+\frac{m}{1+\delta}\bigg)\bigg]\ ,\\
C_0=-m\bigg\{X-2(m+1)
\bigg[\frac{{\mathcal{B}}}{\chi}
\bigg(m-\frac{1}{1+\delta}\bigg)
-\frac{\delta}{(1+\delta )}\bigg]\bigg\}\ .
\end{eqnarray*}

Because there are several parameters in our theoretical model
analysis and our main motivation is for galactic applications,
it would be efficient and more sensible to have rough ranges
to bracket these parameters in order to explore different
regimes for various possible solutions numerically.

For this purpose, we estimate ranges of the parameters $\lambda$,
$\beta$, $\epsilon$ and $\Theta$ for disc galaxies. A typical
magnetic field in a spiral galaxy is about $1\sim 10\mu\hbox{G}$
and higher values of $\sim 30-40\mu\hbox{G}$ may be reached in
central circumnuclear regions by equipartition arguments (e.g.,
Lou et al. 2001). A gas surface mass density is about $\Sigma^g
\sim 2\times 10^{-4}-2\times 10^{-3}\hbox{ g cm}^{-2}$. The
value of mass-to-flux ratio $\Lambda$ is about $20\sim 2000
\hbox{ g cm}^{-2}\hbox{G}^{-1}$. Therefore, the $\lambda$
parameter falls in the range of $0.03\sim 3$. Here, we estimate
the maximum range for every parameter in the galactic context
while keeping in mind the constraint of $1+\epsilon\delta>0$ for
the background magneto-rotational equilibrium. The variation
range of $\epsilon$ may be as large as $-1110\sim 0.9$. Taking
$\delta$ as $0.01$, we choose the $\epsilon$ parameter to fall
within the rough range of $-100\sim 0.9$. In a typical late-type
spiral galaxy, the stellar velocity dispersion $a_s$ is usually
several times higher than the sound speed $a_g$ of the gaseous
disc, such that $\beta>1$ in our model analysis. We have already
shown the $\Theta$ range to be $1<\Theta<2$. As $\Theta$ and
$\beta$ always appear together, we use a single combined parameter
$\chi\equiv\beta/\Theta$ and further take $\chi>1$ in reference to
$\beta >1$. As $1<\Theta<2$, the effect of $\Theta$ variation on
$\chi$ is somewhat limited. For example, one may take
$a_g=7\hbox{ km s}^{-1}$ and $a_s=30\hbox{ km s}^{-1}$ for a typical
late-type spiral galaxy. It is then justifiable to take $\chi=a_s^2/
(\Theta a_g^2)>1$. The introduction of $\mathcal B$ parameter by
definition (\ref{mathB}) is especially useful to estimate effects of
an isopedic magnetic field. It is clear that inequality ${\mathcal B}
\geq 1$ holds valid given the condition $\epsilon\leq 1$ and the
requirement $1+\epsilon\delta>0$ for the background magneto-rotational
equilibrium. With a fixed value of $\delta$, the increase of the
isopedic magnetic field flux will lead to an increase of $\mathcal B$.
It turns out that $\mathcal B$ is a fairly good indicator for the
isopedic magnetic field, although the relationship between the two
is not as simple as a linear one.

Formally, the two real $y\equiv D_s^2$ solutions
of quadratic equation (\ref{quand}) are given by
\begin{equation}\label{solution}
y_{1,2}=\frac{-C_1\pm (C_1^2-4C_0C_2)^{1/2}}{2C_2}\ .
\end{equation}
We now introduce an important parameter $R_a$
(see Appendix A for more details) defined by
\begin{equation}\label{Ra}
R_a\equiv\frac{m[m-1/(1+\delta)]}{m^2-2+m/(1+\delta)}>0\ .
\end{equation}
Based on the mathematical conclusion reached in Appendix A, for
the $y_1$ solution with the plus sign in (\ref{solution}), we
have $y_1>R_a>0$ when $C_2>0$ and $y_1<-1$ when $C_2<0$. For
the $y_2$ solution with the minus sign in (\ref{solution}),
we have $-1<y_2<R_a$.

In the previous analyses (Lou \& Shen 2003; Shen \& Lou 2003,
2004a, b; Lou \& Zou 2004), the two solutions $y_1$ and $y_2$ are
sometimes referred to as supersonic and subsonic rotation solutions
respectively, because as rotational Mach numbers, $y_1$ is almost
always greater than 1 and $y_2$ remains always less than 1. In
comparison, the situation becomes somewhat different in the
presence of an isopedic magnetic field. In particular, $y_1$ may
drop below 1 for some specific values of $m$, $\delta$, $\mathcal B$
and $\chi$; this could also happen when there is no magnetic field.
Furthermore, with the boosting of the magnetic field, $y_2$ may
become greater than 1 even under the condition $\chi\geq 1$. Note
that $y_2$ remains always less than 1 in the absence of magnetic
field. Thus for proper terminologies, we use the {\it F-wave
solution} for $y_1$ and the {\it S-wave solution} for $y_2$,
because when they are both physical, $y_1$ and $y_2$ remain
always faster (F) and slower (S), respectively.

When ${\mathcal B}=1$ for the absence of magnetic field, we have
$\Theta=1$ and hence $\chi=\beta$. This situation is exactly the
same as the problem analyzed by Lou \& Shen (2003) which will be
frequently referred to in our following discussions.

For the F-wave solution $y_1$, there exists an upper limit for the
magnetic field intensity. When the magnetic flux becomes so strong
such that $\mathcal B$ has a sufficiently high value, $C_2$ will
become negative and thus $y_1$ becomes unphysical for being negative.
The physical constraint on $\mathcal B$ for the F-wave solution $y_1$
is therefore
\begin{equation}\label{limbfory1}
1\leq{\mathcal B}<1+\frac{(m^2-2)(1+\delta)}{m}\ ,
\end{equation}
where the lower bound on the left-hand side (LHS) is from
the definition of $\mathcal B$ while the upper bound on the
right-hand side (RHS) is derived by requiring $C_2$ to be
positive. We note that with the increase of both $m$ and
$\delta$, the allowed range of $\mathcal B$ for the existence
of an F-wave solution $y_1$ becomes enlarged. Moreover for
$y_1>0$, $y_1$ decreases with increasing $\chi$.

For the S-wave solution $y_2$, the somewhat loose constraint is
that $-1<y_2<R_a$. For a physical solution, we should require
$y_2>0$. Since $y_1>0$ for $C_2>0$ and $y_1<0$ for $C_2<0$, we
infer from the solution property $y_1y_2=C_0/C_2$ that $y_2>0$
for $C_0>0$ and $y_2<0$ for $C_0<0$, respectively. For a
physical S-wave solution $y_2>0$, the condition $C_0>0$
is simply equivalent to the following inequality
\begin{equation}\label{y2above0}
{\mathcal B}
\bigg[\frac{2(m+1)}{\chi}+\frac{m}{m(1+\delta)-1}\bigg]-(m+2)>0\ .
\end{equation}
When ${\mathcal B}=1$ for the absence of magnetic field, by setting
the LHS of the above inequality (\ref{y2above0}) equal to
zero, we determine a critical value $\chi_c$ for $\chi$ which is the
same as $\beta_c$ parameter of Lou \& Shen (2003); as $\beta$ or $\chi$
is increased to become greater than $\beta_c$ or $\chi_c$, the S-wave
solution $y_2$ will change from positive to negative values. Furthermore,
the increase of $\mathcal B$ leads to an increase of $\chi_c$. For
${\mathcal B}>1$, the critical value $\chi_c$ is given explicitly by
\begin{equation}\label{chic}
\chi_c\equiv\frac{2(m+1)[m(1+\delta)-1]{\mathcal B}}
{(m+2)[m(1+\delta)-1]-m{\mathcal B}}>\frac{2(m+1)
{\mathcal B}}{(m+2)}\ .
\end{equation}
One interesting point to emphasize is that when $\mathcal B$
becomes sufficiently large, there is no limit on $\chi$ and the
S-solution $y_2$ remains always positive. More specifically,
when the following condition (\ref{chin}) is fulfilled
\begin{equation}\label{chin}
{\mathcal B}>1+\frac{(m^2-2)(1+\delta)}{m}
+\frac{2(m+1)\delta}{m}\ ,
\end{equation}
inequality (\ref{y2above0}) will be always satisfied
and thus the S-solution $y_2$ remains always positive.

Without magnetic field, there always exists at least one positive
solution (see Lou \& Shen 2003). In contrast, when an isopedic
magnetic field is anchored in a gaseous SID, there does exist
a combination of parameters such that both $y_1$ and $y_2$
solutions may become negative. Under these circumstances, there
would be no stationary MHD density wave patterns supported by
SID rotation. For this to happen, the range of $\mathcal B$
parameter is given below
\begin{equation}\label{bothnegative}
1+\frac{(m^2-2)(1+\delta)}{m}<{\mathcal B}
<1+\frac{(m^2-2)(1+\delta)}{m}+\frac{2(m+1)\delta}{m}
\end{equation}
[see the RHS of inequality (\ref{limbfory1}) and
inequality (\ref{chin})]; the LHS of inequality
(\ref{bothnegative}) leads to $y_1<0$, while the RHS
of inequality (\ref{bothnegative}) makes $y_2<0$ possible and the
$\chi$ parameter should be adjusted to make $C_0<0$ such that
$y_2<0$. We take parameters $m=2$, $\delta=1$, ${\mathcal B}=4$
and $\chi=20$ as a specific example, and the two negative
solutions are $y_1=-2.8651$ and $y_2=-0.3490$, respectively.

For the phase relationship between the two surface mass
perturbations $\mu^s$ and $\mu^g$, we note that equation
(\ref{hg1}) can be written in the form of
\begin{eqnarray}\nonumber
\frac{\mu^g}{\mu^s}=\frac{1-H_1m^2}{G_1m^2}
=-1-\frac{[D_s^2(m^2-2)-m^2](1+\delta)}
{|m|(D_s^2+1)}\\ \label{phase}
=-1-\frac{(m^2-2)(1+\delta)}{|m|}
+\frac{2(m^2-1)(1+\delta)}{|m|(D_s^2+1)}\\ \nonumber
\label{phase1}=-\frac{[(m^2-2)(1+\delta)+|m|]}{|m|}
\frac{(D_s^2-R_a)}{(D_s^2+1)}\ .\qquad\quad
\end{eqnarray}
According to expression (\ref{phase}) of $\mu^g/\mu^s$, it is fairly
clear that for the F-wave solution (i.e., $y_1>R_a$ with the unphysical
case of $y_1<0$ being ignored), the ratio $\mu^g/\mu^s$ remains always
negative, while for the S-wave solution $y_2$, the ratio $\mu^g/\mu^s$
remains always positive. Physically, when the two density perturbations
are out of phase, the gravity effect is weaker and the MHD density wave
speed is faster, while when the two density perturbations are in phase,
the gravity effect is stronger and the MHD density wave speed is slower
(Lou \& Fan 1998b).

\begin{figure}
\begin{center}
\includegraphics[angle=0,scale=0.45]{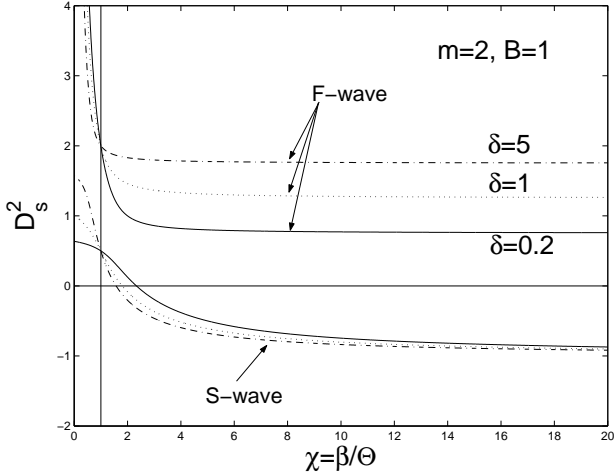}
\caption{\label{f1}Two sets of solution curves for $y\equiv D_s^2$
versus $\chi$ variation for the aligned case with $|m|=2$ and
${\mathcal B}=1$ for the absence of magnetic field. We use $B$
in the figure to correspond to $\mathcal{B}$ in the main text.
The vertical line marks the constraint of $\chi>1$.
We take $\delta=0.2$ (solid lines), 1 (dotted lines) and 5
(dash-dotted lines), respectively. The allowed range for $\chi$ is
$[1,+\infty)$, and we plot these curves within the $\chi$ interval
$(0,20]$. This figure is the same as figure 1 of Lou \& Shen (2003)
and serves as our reference to understand the role and effects of
an isopedic magnetic field. }
\end{center}
\end{figure}

\begin{figure}
\begin{center}
\includegraphics[angle=0,scale=0.45]{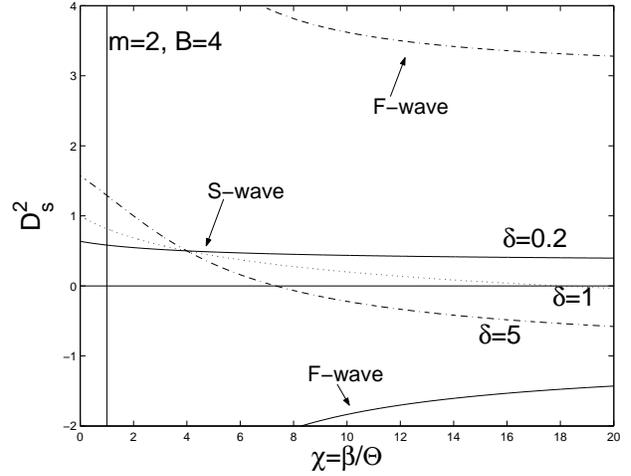}
\caption{\label{f2}Two sets of solution curves of $y\equiv D_s^2$
versus $\chi$ variation for the aligned case with $|m|=2$ and
${\mathcal B}=4$ for the presence of an isopedic magnetic field.
We take $\delta=0.2$ (solid lines), 1 (dotted lines) and 5
(dash-dotted lines), respectively.
Again, we use $B$ in the figure to correspond to $\mathcal B$ in
the main text. The vertical line marks the constraint of $\chi>1$.
Note that the F-wave solution $y_1$ with $\delta=1$ is not shown
as it drops far below zero. In order to give an intuitive
comparison between Fig. 1 and Fig. 2 here, we use the same scale
for the reference frame. It should be noted that only a range of
$\chi$ is permitted when the S-wave solution $y_2$ is positive
due to the constraint of $\chi_s$.}
\end{center}
\end{figure}

In addition to the requirement of $D_s^2>0$, we must also
make sure of $D_g^2>0$ such that the relevant MHD density
wave mode is physically plausible. We now write equation
(\ref{Dg}) in a simple form of
\begin{equation}\label{Dg1}
D_g^2=\frac{\chi}{\mathcal B}(1+D_s^2)-1\ .
\end{equation}
Without magnetic field with ${\mathcal B}=1$ and under the
assumption of $\chi=\beta>1$, it is easy to show that $D_g^2$
remains always positive according to equation (\ref{Dg1}) (Lou
\& Shen 2003; Lou \& Zou 2004). However, this may not be true in
the presence of an isopedic magnetic field with ${\mathcal B}>1$
and $D_g^2$ may become negative even for $\chi>1$. In other words,
one more constraint of $D_g^2>0$ should be checked carefully. On
the basis of Appendix A, we here define a new critical value $\chi_s$
\begin{equation}\label{chis}
\chi_s\equiv\frac{{\mathcal B}}{2(m+1)}
\bigg[m+2-\frac{m\delta}{(m-1)(1+\delta)
+{\mathcal B}}\bigg]
\end{equation}
and only when $\chi>\chi_s$ can inequality $D_g^2>0$
be guaranteed. In the absence of magnetic field (i.e.
${\mathcal B}=1$), we readily see that $\chi_s<1$.
So a more general condition for a physical solution
would be $\chi>\max\{1,\chi_s\}$.
This limit of $\chi_s$ emerges as a constraint
in the presence of an isopedic magnetic field (${\mathcal B}>1$).
With the increase of the magnetic flux (i.e. ${\mathcal B}$
grows), the value of $\chi_s$ increases accordingly. By comparing
the expressions of $\chi_c$ and $\chi_s$, it should be noted that
wherever there exists an upper limit $\chi_c$, this $\chi_c$
remains always larger than $\chi_s$.

We now review several properties of the two solutions of
$y\equiv D_s^2$ as a summary. In the absence of magnetic field
with ${\mathcal B}=1$ (see Figure 1), our analysis here is the
same as those of Lou \& Shen (2003). The F-wave solution $y_1$
remains always positive and the S-wave solution $y_2$ changes
from positive to negative as $\chi$ becomes larger than
$\chi_c$. In the presence of an isopedic magnetic field with
${\mathcal B}>1$, we define a $\chi_s$ instead of $\chi_c$.
For a physical configuration solution, we then require
$\chi>\max\{1,\chi_s \}$.
With the increase of the magnetic flux (i.e., an increase of
${\mathcal B}$), the two solutions $y_1$ and $y_2$ become larger
and the value of $\chi_c$ increases
accordingly. However, as $\mathcal B$ exceeds a certain given value
[see expression (\ref{limbfory1})], the F-wave solution $y_1$
becomes negative and thus unphysical (see Fig. 2). In addition to
the requirement $1+\epsilon\delta>0$ on the magnetic field for the
background magneto-rotational equilibrium, a further condition
(\ref{limbfory1}) is needed for the existence of an F-wave
solution $y_1$. As $\chi_s$ becomes larger than 1, the requirement
of $\chi>\max\{1,\chi_s\}$ means that only if $\chi_s<\chi<\chi_c$,
can the S-wave solution $y_2$ be physically valid. Note that these
constraints on $\chi$ do not apply for the F-wave solution $y_1$.
As the magnetic field strength is increased further [see equation
(\ref{chin})], the upper limit $\chi_c$ for $\chi$ will disappear
and only the lower limit $\chi_s$ exists (see Fig. 3). It is
generally true that when the two solutions $y_1$ and $y_2$ are
both positive and physical, they both increase with the increase
of $\mathcal B$ and decrease with the increase of $\chi$.

Physically, for the F-wave solution with $\mu^g/\mu^s<0$, the surface
mass density perturbations of stellar and gaseous SIDs are completely
out of phase to reduce the effect of gravity. Meanwhile, a faster
azimuthal density wave speed corresponds to a faster SID rotation
(larger $D_s^2$) in order to maintain a stationary MHD configuration.
For the S-wave solution with $\mu^g/\mu^s>0$, the surface mass
perturbations in gaseous and stellar discs are in phase with each
other. As the effect of self-gravity is enhanced, the azimuthal
density wave speed becomes slower corresponding to a slower SID
rotation (smaller $D_s^2$) in order to sustain a stationary MHD
perturbation configuration.

The isopedic magnetic field plays an interesting role. When the
magnetic field grows stronger as $\mathcal B$ becomes larger, the
F-wave solution $y_1$ and the S-wave solution $y_2$ both become
larger. According to expression (\ref{phase}) for the ratio
$\mu^g/\mu^s$, it is easy to demonstrate that $\mu^g/\mu^s$ will
decrease with the increase of $D_s^2$ for both $y_1$ and $y_2$
solutions. For the F-wave solution, the decrease of $\mu^g/\mu^s$
means an increase of the absolute value of $\mu^g/\mu^s$, while
for the S-wave solution, this ratio magnitude becomes smaller. A
physical interpretation is that for a given composite SID system,
an isopedic magnetic field will amplify the contrast of surface
mass density perturbations in the two SIDs for an F-wave mode
and will reduce such contrast of surface mass density
perturbations in the two SIDs for a S-wave mode.

As noted earlier (Lou 2002; Lou \& Fan 2002; Lou \& Shen 2003;
Lou \& Zou 2004), we here emphasize again the physical perspective
that stationary aligned MHD perturbation configurations should be
regarded as purely azimuthal propagation of MHD density waves
counterbalanced by the advection of (M)SID differential rotation.
Together with equation (\ref{Dg}), we may write equation
(\ref{hg5}) in the following form to derive the two $y$ solutions.
\begin{eqnarray}
\bigg(m^2-2+\frac{m}{1+\delta}\bigg)
\big(D^2_s-R_a\big)
\qquad\qquad\qquad\qquad\nonumber \\
\times\bigg[D^2_g\bigg(\frac{C_2}{m+2}
+\frac{m\delta}{1+\delta}
\bigg)-m\bigg(m-1+\frac{\mathcal B}
{1+\delta}\bigg)\bigg]\nonumber
\\\label{hg6}=(D^2_s+1)(D^2_g+1)
\frac{{\mathcal B}m^2\delta}{(1+\delta)^2}\ ,
\end{eqnarray}
where $C_2$ and $R_a$ are defined by expressions (\ref{C2a}) and
(\ref{Ra}), respectively. For the F-wave solution $y_1$, we have
$D_s^2>R_a$ when $D_s^2>0$ which is guaranteed by $C_2>0$. Moreover
from equations (\ref{hg6}) and (\ref{Dg1}), we can show that for a
positive F-wave solution $y_1$, the corresponding $D_g^2$ is also
positive such that the overall solution is physically plausible.

We now write solution condition (\ref{hg5}) for stationary MHD
configurations of a composite SID system in a physically more
suggestive form of
\begin{eqnarray}\nonumber
\quad\big{[}\Omega_s^2(m^2-2)-m^2a_s^2/r^2+2\pi
G\Sigma_0^s|m|/r\big{]}
\quad\qquad \\ \quad \nonumber
\times\big{[}\Omega_g^2(m^2-2)-m^2\Theta a_g^2/r^2
+2\pi\epsilon G\Sigma_0^g|m|/r\big{]}\\
\label{disperall} =4\pi^2G^2\Sigma_0^s\Sigma_0^gm^2/r^2\ .
\end{eqnarray}
The RHS of equation (\ref{disperall}) represents
the mutual gravitational coupling between the `fluid' stellar
SID and the `magnetofluid' of a gaseous isopedic MSID. Without
this gravitational coupling, the two factors on the LHS would
emerge as two separate solution conditions for
stationary aligned perturbation configurations with $|m|\geq 2$
of stellar SID and of gaseous isopedic MSID, respectively. For
the stellar SID alone, the condition would be
\begin{equation}\label{disperste}
m^2\Omega_s^2=\kappa_s^2+m^2a_s^2/r^2
-2\pi G\Sigma_0^s|m|/r
\end{equation}
and for the gaseous MSID alone with an isopedic
magnetic field, the condition would be
\begin{equation}\label{dispergas}
m^2\Omega_g^2=\kappa_g^2+m^2\Theta
a_g^2/r^2-2\pi\epsilon G\Sigma_0^g|m|/r\ .
\end{equation}
According to the well-known dispersion relation of density waves
first derived under the tight-winding or WKBJ approximation [Lin
\& Shu 1964, 1966; Lin 1987; or equation (39) of Shu et al. 2000],
dispersion relations (\ref{disperste}) and (\ref{dispergas}) can
be readily recovered by replacing the radial wavenumber $|k|$ with
the azimuthal wavenumber $|m|/r$ and setting $\omega=0$ in an
inertial frame of reference as noted by Lou (2002) in the study of
stationary MHD perturbation configurations of a single MSID with a
coplanar magnetic field. Along this line of reasoning, we know that
dispersion relations (\ref{disperste}) and (\ref{dispergas})
separately describe azimuthal propagations of hydrodynamic density
waves in a stellar SID and of MHD density waves in a gaseous MSID
with an isopedic magnetic field.

\begin{figure}
\begin{center}
\includegraphics[angle=0,scale=0.45]{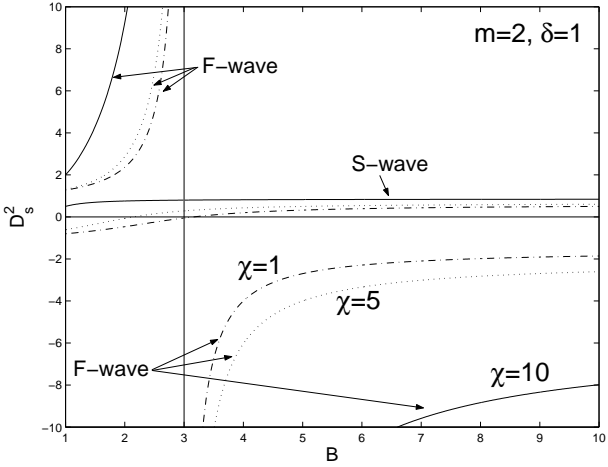}
\caption{\label{f3}Two sets of solution curves
for $D_s^2$ versus $\mathcal B$ variation.
Here we use $B$ in the figure to correspond to $\mathcal B$
in the main text. The vertical line represents the diverging
point of $y_1$ for the aligned case with $|m|=2$, $\delta=1$
and $\chi=1$, 5 and 10, respectively.  }
\end{center}
\end{figure}

\subsubsection{Secular Barlike Instabilities}

We have constructed analytically stationary configurations for
aligned nonaxisymmetric MHD perturbations in a composite MSID
system. Whether these perturbation configurations are stable or
just represent transition states from axisymmetric equilibria to
non-axisymmetric configurations still remains an open question
(see discussions of Shu et al. 2000 and extensive references
therein).

It was tentatively suggested (Shu et al. 2000; Galli et al. 2001)
that in the case of a single (M)SID, these stationary solutions
signal onsets of bifurcations from an axisymmetric (M)SID to
non-axisymmetric configurations in (M)SIDs.


Following the earlier procedures (Shu et al. 2000; Lou 2002; Lou
\& Shen 2003), we explore parameter regimes of these stationarity
conditions in reference to the disc stability criterion first
hypothesized by Ostriker \& Peebles (1973) for the onsets of
bar-type instabilities. Judging similarities and differences
between a single (M)SID and a composite (M)SID system, we examine
stability properties for stationary configurations of aligned and
unaligned perturbations with an isopedic magnetic field in
reference to the analysis of Shu et al. (2000).

The criteria for (secular and dynamic) barlike instabilities
were sometimes expressed in terms of the ratio of the rotational
kinetic energy ${\mathcal{T}}$ to the absolute value of the
gravitational potential energy ${\mathcal{W}}$ (Ostriker \&
Peebles 1973; Binney \& Tremaine 1987; Shu et al. 2000; Lou \&
Shen 2003; Lou \& Zou 2004). To obtain the scalar virial theorem
in the present case of a composite MSID system, we start from
the radial force-balance equations of the equilibrium state,
\begin{equation}\label{equs1}
\Sigma_0^s\Omega^2_sr=\frac{d}{d r}
{(a^2_s\Sigma_0^s)}+\Sigma_0^s\frac{d}{d r}(\phi^s+\phi^g)\ ,
\end{equation}
\begin{equation}\label{equg1}
\Sigma_0^g\Omega^2_gr
=\frac{d}{d r}{(\Theta a^2_g\Sigma_0^g)}
+\Sigma_0^g\frac{d}{d r}(\phi^s+\epsilon\phi^g)\ ,
\end{equation}
where the axisymmetric dark matter
halo potential is not included yet.
An addition of equations (\ref{equs1}) and (\ref{equg1}) gives
\begin{eqnarray}\nonumber
\qquad\quad (\Sigma_0^g\Omega^2_g+\Sigma_0^s\Omega^2_s)r
=\frac{d}{dr}{(\Theta a^2_g\Sigma_0^g+a^2_s\Sigma_0^s)}
\qquad\\ \label{all}\qquad\quad
+\Sigma_0^g\frac{d}{d r}(\phi^s+\epsilon\phi^g)
+\Sigma_0^s\frac{d}{d r}(\phi^s+\phi^g)\ .
\end{eqnarray}
Multiplying equation (\ref{all}) by a ring area element
$2\pi r^2dr$ and integrating from 0 to a finite radius
$L$, we obtain
\begin{equation}\label{secular}
2({\mathcal{T}}+{\mathcal{U}})+{\mathcal{W}}=2\pi L^2
[\Theta a_g^2\Sigma_0^g(L)+a_s^2\Sigma_0^s(L)]\ ,
\end{equation}
where
\begin{eqnarray}\label{T}
{\mathcal{T}}\equiv \int_0^L\frac{1}{2}\Sigma_0^g(r\Omega_g)^22\pi
r dr+\int_0^L\frac{1}{2}\Sigma_0^s(r\Omega_s)^22\pi r dr\ ,\\
\label{U} {\mathcal{U}}\equiv \int_0^L(\Theta a_g^2\Sigma_0^g
+a_s^2\Sigma_0^s)2\pi r dr\ ,
\qquad\qquad\qquad\qquad\\ \nonumber {\mathcal{W}}\equiv-\int_0^L
r\Sigma_0^g\frac{d(\phi^s+\epsilon\phi^g)}{dr}2\pi rdr
\qquad\qquad\qquad\qquad\\ \label{W} \qquad\qquad
-\int_0^L r\Sigma_0^s\frac{d(\phi^s+\phi^g)}{dr}2\pi rdr\ .\qquad
\end{eqnarray}
Here $\mathcal{T}$ is the rotational kinetic energy in the composite
SID system, $\mathcal{U}$ is the equivalent `thermal energy' contained
in the composite MSID system including the effect of magnetic pressure,
and $\mathcal{W}$ is the equivalent gravitational-work integral
including the effect of magnetic tension.

Using $\delta\equiv\Sigma^g_0/\Sigma^s_0$, $\Omega_s=a_sD_s/r$,
$\Omega_g=\Theta^{\frac{1}{2}}a_gD_g/r$ together with equations
(\ref{sigmas0}) and (\ref{sigmag0}), we cast these integrals in
the following forms of
\begin{eqnarray}
\label{T1}
{\mathcal{T}}=\frac{a_s^4(1+D_s^2)}{2G(1+\delta)}
\bigg[D_s^2+\frac{\delta(1+\epsilon\delta)}{(1+\delta)}
(1+D^2_s)-\frac{\Theta\delta}{\beta}\bigg]L\ ,
\\
\label{U1}
{\mathcal{U}}=\frac{a_s^4(1+D_s^2)}{G(1+\delta)}
\bigg(\frac{\Theta\delta}{\beta}+1\bigg)L\ ,
\quad\qquad\qquad\qquad\qquad\\
\label{W1}
{\mathcal{W}}=-\frac{a_s^4(1+D_s^2)^2}{G(1+\delta)^2}
(1+2\delta+\epsilon\delta^2)L\ .\qquad\qquad\qquad
\end{eqnarray}

For a composite MSID system of an infinite radial extent, all three
integrals (\ref{T1})$-$(\ref{W1}) above diverge as $L\rightarrow+
\infty$ but their mutual ratios remain finite. For example, the
ratio of the kinetic energy of disc rotation to the absolute
value of the gravitational potential energy is
\begin{eqnarray}\nonumber
\frac{\mathcal{T}}{|\mathcal{W}|}
=\frac{(1+D_s^2)(1+2\delta+\epsilon\delta^2)
-(1+\delta)[1+(\Theta\delta/\beta)]}
{2(1+D_s^2)(1+2\delta+\epsilon\delta^2)}
\\\label{TW}
=\frac{1}{2}-\frac{1+(\delta/\chi )}
{2(1+D^2_s)(1+\delta/{\mathcal{B}})}\ .
\ \ \qquad\qquad\qquad\qquad
\end{eqnarray}
This ratio can also be arranged into the following form
\begin{equation}\label{TW1}
\frac{\mathcal{T}}{|\mathcal{W}|}=\frac{1}{2}
\frac{a_s^2\Sigma_0^sD_s^2+\Theta
a_g^2\Sigma_0^gD_g^2}{[a_s^2\Sigma_0^s(1+D_s^2)
+\Theta a_g^2\Sigma_0^g(1+D_g^2)]}
\end{equation}
which is explicitly symmetrized with respect
to physical parameters of the two SIDs.
Here, $\epsilon$ does not appear explicitly in
the final expression.

Note that the ratio ${\mathcal T}/|\mathcal W|$ falls between 0
and 0.5 as usual (Binney \& Tremaine 1987; Lou \& Shen 2003;
Lou \& Zou 2004) and increases with increasing $D^2_s$. For
stationary configurations of aligned MHD perturbations in a
composite system of two coupled SIDs, the two possible values
of $D_s^2$ give rise to two different values of
${\mathcal T}/|\mathcal W|$ ratio; the larger and smaller
values of $D^2_s$ correspond to larger and smaller values
of the ${\mathcal T}/|\mathcal W|$ ratio, respectively.

Furthermore, as the magnetic flux increases (i.e., an increase of
$\mathcal B$), the ${\mathcal T}/|\mathcal W|$ ratio decreases.
This simply means that the effect of an isopedic magnetic field
tends to make the composite SID system more stable (Lou \& Shen
2003; Shen \& Lou 2003, 2004a, b; Lou \& Zou 2004, 2005).

On the basis of an extensive numerical exploration for $m=2$ to
$m=10$ and so forth, one can show the existence of a dividing line
in the ratio ${\mathcal T}/|\mathcal W|$ for global stationary
configurations of aligned MHD perturbations. More specifically,
for F-wave solutions $y_1$, the ratio ${\mathcal T}/|\mathcal W|$
ranges from $m/(4m+4)$ to $1/2$, while for S-wave solutions $y_2$,
the ratio ${\mathcal T}/|\mathcal W|$ ranges from 0 to $m/(4m+4)$
when $m\geq2$. Based on $N-$body numerical simulation experiments
involving only $\sim 300$ particles for a stability analysis of a
rotating disc under self-gravity, Ostriker \& Peebles (1973)
suggested an empirical criterion that
${\mathcal T}/|{\mathcal W}|\leq0.14\pm 0.02$ is necessary
but not sufficient against bar-type instabilities (see also
earlier numerical simulation results of Miller et al. 1970; Hohl
1971). When ${\mathcal T}/|\mathcal W|\geq 0.14\pm 0.02$, a disc
system would evolve rapidly into bar-type configurations (e.g.
Binney \& Tremaine 1987). In the current problem of a composite
MSID system, the lowest value of ${\mathcal T}/|\mathcal W|$
ratio for F-wave solutions $y_1$ is $\sim 0.1667$,
while for S-wave solutions $y_2$, the ${\mathcal T}/|\mathcal W|$
ratio can be lower than $0.14\pm 0.02$.

In the above analysis, we have not included the effect of
a massive dark matter halo. One obvious consequence of a
massive dark matter halo is to increase $|\mathcal W|$ and
thus decrease ${\mathcal T}/|\mathcal W|$ ratio, and for
both classes of global perturbation configurations, this
implies the tendency towards a stability. At any rate, a
comparison of the above two criteria suggests that given
a background dark matter halo, F-wave configurations tend
to be more stable than S-wave configurations.


By numerical experiments for $m=2$ up to $m=10$ and so
forth, we found that when ${\mathcal B}=\chi$, the value
of ${\mathcal T}/|\mathcal W|$ ratio for S-wave solutions
$y_2$ is always $m/(4m+4)$.

\begin{figure}
\begin{center}
\includegraphics[angle=0,scale=0.45]{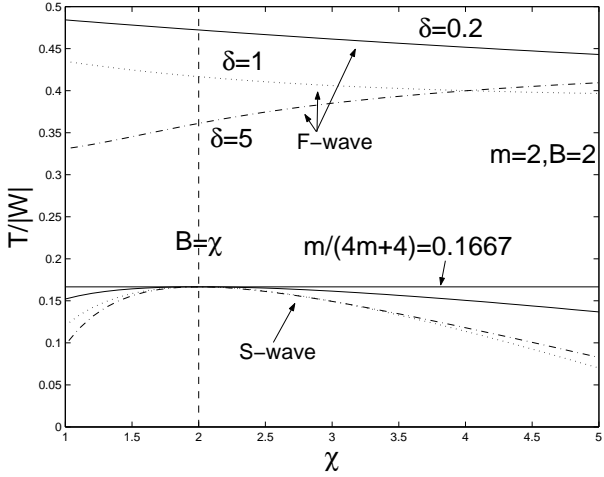}
\caption{\label{f4}The ${\mathcal T}/|\mathcal W|$ ratio
versus $\chi$ variation for the aligned cases with $m=2$,
${\mathcal B}=1$ and $\delta=0.2$ (solid lines), 1
(dotted lines) and 5 (dash-dotted lines), respectively.
Here, we use $B$, $W$ and $T$ in the figure to correspond to
$\mathcal B$, $\mathcal W$ and $\mathcal T$ in the main text,
respectively. The horizontal line $m/(4m+4)=0.1667$ divides the
ratios ${\mathcal T}/|\mathcal W|$ for the S-wave solutions
$y_2$ and for the F-wave solutions $y_1$. Through numerical
explorations for $m=2$ up to $m=10$ and so forth, the ratio
${\mathcal T}/|\mathcal W|$ of the S-wave solution $y_2$
reaches a maximum value of $m/(4m+4)$ at $\chi={\mathcal B}$.}
\end{center}
\end{figure}

\subsection{Unaligned Logarithmic Spiral Disturbances}

For global stationary configurations of unaligned or logarithmic
spiral MHD disturbances, we take on the following set of exact
density-potential pair relations that satisfy the Poisson
integrals, namely
\begin{equation}\label{us1}
\mu^l=\sigma^lr^{-3/2}\exp(\rm{i}\alpha\ln r)\ ,
\end{equation}
\begin{equation}\label{vs1}
V^l=v^lr^{-1/2}\exp(\rm{i}\alpha\ln r)\ ,
\end{equation}
where $\sigma^l$ and $v^l$ are small constants for
perturbations, $\alpha$ is a parameter to characterize
the radial variation; the two small coefficients $v^l$
and $\sigma^l$ are algebraically related by
\begin{equation}\label{vs2}
v^l=-2\pi G{\mathcal{N}}_m(\alpha)\sigma^l\ ,
\end{equation}
with ${\mathcal{N}}_m(\alpha)\equiv K(\alpha,m)$ being the Kalnajs
function (Kalnajs 1971) and superscript $l=s, g$. For logarithmic
spiral perturbations in a composite MSID system, the above
description is a sensible extension of earlier model studies.
The radial scaling parameter $\alpha$ (closely related to the
radial wavenumber) is naturally taken to be the same for
perturbations in both stellar SID and gaseous isopedic MSID.

In the following analysis and computations, we use two useful
formulae of ${\mathcal{N}}_m(\alpha)$. One is the recursion
relation in $m$ of ${\mathcal{N}}_m(\alpha)$ for a fixed
$\alpha$ value (Kalnajs 1971)
\begin{equation}\label{Nm1}
{\mathcal{N}}_{m+1}(\alpha){\mathcal{N}}_m(\alpha)
=[(m+1/2)^2+\alpha^2]^{-1}\ ,
\end{equation}
and the other is the asymptotic expression
of ${\mathcal{N}}_m(\alpha)$
\begin{equation}\label{Nm2}
{\mathcal{N}}_m(\alpha)\approx(m^2+\alpha^2+1/4)^{-1/2}
\end{equation}
in the regime of $m^2+\alpha^2\gg 1$ (e.g., Shu et al. 2000).
When the requirement for accuracy is not so stringent in some
quantitative analyses, expression (\ref{Nm2}) may even be
used to compute logarithmic spiral solutions with $|m|=1$.

For $|m|>0$, we now proceed to solve equations
(\ref{starel1}) and (\ref{starel2}) that can
be cast into more compact forms of
\begin{equation}\label{mus1}
\mu^s=\bigg{(}\frac{m^2}{r}
-2\frac{d}{dr}-r\frac{d^2}{dr^2}\bigg{)}(H_1r\mu^s+G_1r\mu^g)\ ,
\end{equation}
\begin{equation}\label{mug1}
\mu^g=\bigg(\frac{m^2}{r}-2\frac{d}{dr}
-r\frac{d^2}{dr^2}\bigg)(H_2r\mu^g+G_2r\mu^s)\ ,
\end{equation}
where the four coefficients $H_1$, $H_2$, $G_1$
and $G_2$ are
\begin{eqnarray}\label{h1s}
\qquad H_1\equiv\frac{1}{D^2_s(m^2-2)}
\bigg[1-\frac{(D_s^2+1){\mathcal{N}}_m(\alpha)}
{(1+\delta )}\bigg]\ ,\\
\label{h2s}
\qquad H_2\equiv\frac{1}{D^2_g(m^2-2)}
\bigg[1-\frac{(D_g^2+1)
\epsilon\delta{\mathcal{N}}_m(\alpha)}
{(1+\epsilon\delta)}\bigg]\ ,\\
\label{g1s}
G_1\equiv-\frac{(D_s^2+1)}{D^2_s(m^2-2)}
\frac{{\mathcal{N}}_m(\alpha)}{(1+\delta)}\ ,\\
\label{g2s}
G_2\equiv-\frac{(D_g^2+1)}{D^2_g(m^2-2)}
\frac{\delta{\mathcal{N}}_m(\alpha)}
{(1+\epsilon\delta )}\ .
\end{eqnarray}
Substituting $\mu^s$ and $\mu^g$ in the form of expression
(\ref{us1})  into equations (\ref{mus1}) and (\ref{mug1}), we
readily obtain
\begin{eqnarray}\label{HGU1}
[1-H_1(m^2+\alpha^2+1/4)]\mu^s=G_1(m^2+\alpha^2+1/4)\mu^g\ ,\\
\label{HGU2}
[1-H_2(m^2+\alpha^2+1/4)]\mu^g=G_2(m^2+\alpha^2+1/4)\mu^s\ .
\end{eqnarray}
Multiplying both sides of the above two
equations and removing $\mu^s\mu^g\neq 0$,
we derive the solution condition as
\begin{eqnarray}\nonumber
\quad [1-H_1(m^2+\alpha^2+1/4)][1-H_2(m^2+\alpha^2+1/4)]\\
\label{HGU} =G_1G_2(m^2+\alpha^2+1/4)^2\ .
\end{eqnarray}
Using definitions (\ref{h1s})$-$(\ref{g2s}),
condition (\ref{HGU}) then becomes
\begin{eqnarray}\nonumber
\bigg\{1-\frac{1}{D^2_s(m^2-2)}\bigg[1-\frac{(D_s^2+1)
{\mathcal{N}}_m(\alpha)}{(1+\delta )}\bigg]
\bigg(m^2+\alpha^2+\frac{1}{4}\bigg)
\bigg\}\times\\\nonumber
\bigg\{1-\frac{1}{D^2_g(m^2-2)}\bigg[1-\frac{(D_g^2+1)
\epsilon\delta{\mathcal{N}}_m(\alpha)}{(1+\epsilon\delta )}\bigg]
\bigg(m^2+\alpha^2+\frac{1}{4}\bigg)\bigg\}\\
\nonumber=\frac{(D_s^2+1)(D_g^2+1)}{D^2_sD^2_g(m^2-2)^2}
\frac{\delta{\mathcal{N}}_m^2(\alpha)}{(1+\delta)
(1+\epsilon\delta)}\bigg(m^2+\alpha^2+\frac{1}{4}\bigg)^2\ .
\end{eqnarray}
This equation may be rearranged into the form of
\begin{eqnarray}\nonumber
\bigg\{D^2_s(m^2-2)
-\bigg[1-\frac{(D_s^2+1){\mathcal{N}}_m(\alpha)}{(1+\delta)}\bigg]
\big(m^2+\alpha^2+1/4\big)\bigg\}
\quad\\\nonumber \times
\bigg\{D^2_g(m^2-2)
-\bigg[1-\frac{(D_g^2+1)
\epsilon\delta{\mathcal{N}}_m(\alpha)}{(1+\epsilon\delta )}\bigg]
\big(m^2+\alpha^2+1/4\big)\bigg\}\\ \nonumber
=\frac{(D_s^2+1)(D_g^2+1)\delta{\mathcal{N}}_m^2(\alpha)}
{(1+\delta)(1+\epsilon\delta)}(m^2+\alpha^2+1/4)^2\ .
\end{eqnarray}
Using expression (\ref{Dg}) for $D_g^2$,
we readily obtain a quadratic equation
in terms of $y\equiv D_s^2$,
\begin{equation}\label{quand1}
C_2y^2+C_1y+C_0=0\ ,
\end{equation}
where the three coefficients $C_2$, $C_1$ and $C_0$ are
\begin{eqnarray*}
C_2\equiv\chi\bigg\{\bigg[(m^2-2)
+\frac{{\mathcal{N}}_m(\alpha)}
{(1+\delta )}\bigg(m^2+\alpha^2+\frac{1}{4}\bigg)\bigg]
\quad\quad\\
\qquad \times\bigg[\frac{(1+\epsilon\delta )}{(1+\delta )}
(m^2-2)+\frac{\epsilon\delta{\mathcal{N}}_m(\alpha)}
{(1+\delta)}\bigg(m^2+\alpha^2+\frac{1}{4}\bigg)\bigg]
\\ \qquad\qquad -\frac{\delta{\mathcal{N}}_m^2(\alpha)}
{(1+\delta)^2}\bigg(m^2+\alpha^2+\frac{1}{4}\bigg)^2\bigg\}\ ,\\
\end{eqnarray*}
\begin{eqnarray*}
C_1\equiv
\chi\bigg\{\bigg[\frac{(1+\epsilon\delta )}
{(1+\delta )}(m^2-2)+\frac{\epsilon\delta{\mathcal{N}}_m(\alpha)}
{(1+\delta )}\bigg(m^2+\alpha^2+\frac{1}{4}\bigg)\bigg]
\\
\times\bigg[\frac{2{\mathcal{N}}_m(\alpha)}
{(1+\delta)}\bigg(m^2+\alpha^2+\frac{1}{4}\bigg)
-\bigg(\alpha^2+\frac{9}{4}\bigg)\bigg]\\
-\frac{2\delta{\mathcal{N}}_m^2(\alpha)}{(1+\delta)^2}
\bigg(m^2+\alpha^2+\frac{1}{4}\bigg)^2\bigg\}
\qquad\qquad\\
-\bigg[(m^2-2)+\frac{{\mathcal{N}}_m(\alpha)}
{(1+\delta)}\bigg(m^2+\alpha^2+\frac{1}{4}\bigg)\bigg]
\bigg(2m^2+\alpha^2-\frac{7}{4}\bigg)\ ,
\end{eqnarray*}
\begin{eqnarray*}
C_0\equiv\chi\bigg(m^2+\alpha^2+\frac{1}{4}\bigg)
\bigg\{\bigg[
\frac{(1+\epsilon\delta )}{(1+\delta)}(m^2-2)
\qquad\qquad\qquad \\
+\frac{\epsilon\delta{\mathcal{N}}_m(\alpha)}
{(1+\delta )}\bigg(m^2+\alpha^2+\frac{1}{4}\bigg)\bigg]
\qquad\\ \qquad
\times\bigg[\frac{{\mathcal{N}}_m(\alpha)}
{(1+\delta )}-1\bigg]-
\frac{\delta{\mathcal{N}}_m^2(\alpha)}
{(1+\delta)^2}\bigg(m^2+\alpha^2+\frac{1}{4}\bigg)\bigg\}
\qquad\\ \qquad\quad
-\bigg(2m^2+\alpha^2-\frac{7}{4}\bigg)
\bigg(m^2+\alpha^2+\frac{1}{4}\bigg)
\bigg[\frac{{\mathcal{N}}_m(\alpha)}{(1+\delta )}-1\bigg]\ .
\end{eqnarray*}
In the following, we shall remove some nonzero common factors
of $C_2$, $C_1$ and $C_0$ to simplify these expressions. For
this purpose, we introduce a new parameter
\begin{equation}\label{M}
{\mathcal{M}}\equiv\frac{(m^2-2)}{(m^2+\alpha^2+1/4)}\ .
\end{equation}
Together with parameter ${\mathcal B}$ defined by
equation (\ref{mathB}) as in the aligned case, we obtain
\begin{eqnarray*}
C_2\equiv\big{[}{\mathcal{M}}+{\mathcal{N}}_m(\alpha)\big{]}
\bigg[\frac{1}{\mathcal{B}}\bigg({\mathcal{M}}
+\frac{{\mathcal{N}}_m(\alpha)}{1+\delta}\bigg)
-\frac{{\mathcal{N}}_m(\alpha)}{(1+\delta )}\bigg]\ ,\\
C_1\equiv\bigg\{\frac{1}{\mathcal{B}}\big{[}{\mathcal{M}}
+{\mathcal{N}}_m(\alpha)\big{]}
\bigg[{\mathcal{M}}+\frac{2{\mathcal{N}}_m(\alpha)}
{(1+\delta )}-1\bigg]
\qquad\qquad\\
-\frac{{\mathcal{N}}_m(\alpha)}{(1+\delta )}
\big{[}{\mathcal{M}}+2{\mathcal{N}}_m(\alpha)-1\big{]}\bigg\}
-\frac{({\mathcal{M}}+1)}{\chi}\bigg[{\mathcal{M}}
+\frac{{\mathcal{N}}_m(\alpha)}{(1+\delta )}\bigg]\ ,
\\
C_0\equiv\bigg{\{}\frac{1}{\mathcal{B}}\big{[}{\mathcal{M}}
+{\mathcal{N}}_m(\alpha)\big{]}
\bigg[\frac{{\mathcal{N}}_m(\alpha)}{(1+\delta )}-1\bigg]
\quad\qquad\qquad\qquad\\
-\frac{{\mathcal{N}}_m(\alpha)}{(1+\delta )}
\big{[}{\mathcal{N}}_m(\alpha)-1\big{]}\bigg{\}}
-\frac{({\mathcal{M}}+1)}{\chi}\bigg[
\frac{{\mathcal{N}}_m(\alpha)}{(1+\delta )}-1\bigg]\ .\quad
\end{eqnarray*}
Consequently, we have the determinant $\Delta$
of quadratic equation (\ref{quand1}) above as
\begin{eqnarray*}
\Delta\equiv C_1^2-4C_0C_2=({\mathcal{M}}+1)^2\bigg{\{}\Big{\{}
\frac{1}{\mathcal{B}}[{\mathcal{M}}+{\mathcal{N}}_m(\alpha)]
\qquad\\ \quad
-\frac{{\mathcal{N}}_m(\alpha)}{(1+\delta )}-\frac{1}{\chi}
\bigg[{\mathcal{M}}+\frac{{\mathcal{N}}_m(\alpha)}{(1+\delta )}\bigg]
\Big{\}}^2+\frac{4\delta{\mathcal{N}}_m^2(\alpha)}
{(1+\delta)^2\chi}\bigg{\}}\geq 0\ .
\end{eqnarray*}
Formally, there are two different real
solutions $y_1$ and $y_2$ of quadratic
equation (\ref{quand1}) in the forms of
\begin{equation}\label{solutionspiral}
y_{1,2}=\frac{-C_1\pm (C_1^2-4C_0C_2)^{1/2}}{2C_2}\
\end{equation}
for $\Delta>0$.

Based on our earlier discussion for the axisymmetric situation
of $m=0$ in equations (\ref{final5o}) and (\ref{final6o}) for
both aligned and unaligned cases, the above results remain
valid for the axisymmetric case of $m=0$ in the unaligned
case (i.e. perturbations with radial oscillations). Note that
$\Delta$ remains always positive for $m\geq1$. The rare case of
$\Delta=0$ can only happen when $m=0$ and ${\mathcal M}+1=0$.
In general, there are therefore two different real solutions
$y_1$ and $y_2$ and these two solutions may become the same
only if $m=0$ and ${\mathcal M}+1=0$.

We now proceed to discuss the three situations $m\geq2$,
$m=1$ and $m=0$ in three separate subsections below.

\subsubsection{Logarithmic Spiral Cases with $m\geq2$}

When $m\geq2$, we take the following approximate
form of ${\mathcal{N}}_m(\alpha)$, namely
$$
{\mathcal{N}}_m(\alpha)
\cong\frac{1}{(m^2+\alpha^2+1/4)^{1/2}}\ .
$$
As in the aligned case, we introduce a similar
$R_s$ parameter
\begin{equation}
R_s\equiv\frac{[1-{\mathcal{N}}_m(\alpha)/(1+\delta)]}
{[\mathcal{M}+{\mathcal{N}}_m(\alpha)/(1+\delta)]}
\end{equation}
(see Appendix B for more details).

On the basis of the mathematical analysis in Appendix B, for the
$y_1$ solution, we have $y_1>R_s>0$ when $C_2>0$ while $y_1<-1$
when $C_2<0$, while for the $y_2$ solution, we have $-1<y_2<R_s$
independent of the sign of $C_2$. Here, the role of this $R_s$
parameter for the unaligned case is similar to the role of $R_a$
parameter for the aligned case.
Note that when $m\geq 2$,
parameter $R_s$ remains always positive. As in the aligned case,
we still refer to $y_1$ as the F-wave solution and $y_2$ as the
S-wave solution for the unaligned case, respectively.

To analyze the phase relationship between the two surface mass
perturbations $\mu^s$ and $\mu^g$ in the coupled stellar SID and
isopedic MSID, we recast equation (\ref{HGU1}) in the form of
\begin{eqnarray}\nonumber
\qquad\frac{\mu^g}{\mu^s}&=&\frac{[1-H_1(m^2+\alpha^2+1/4)]}
{G_1(m^2+\alpha^2+1/4)}\\
\label{phasespiral1}&=&-1-\frac{{\mathcal{M}}(1+\delta)}
{{\mathcal{N}}_m(\alpha)}
+\frac{({\mathcal{M}}+1)(1+\delta)}
{{\mathcal{N}}_m(\alpha)(D_s^2+1)}\\
\label{phasespiral2}&=&
-\frac{[{\mathcal{N}}_m(\alpha)+{\mathcal{M}}(1+\delta)]}
{{\mathcal{N}}_m(\alpha)}\frac{(D_s^2-R_s)}{(D_s^2+1)}\ .
\end{eqnarray}
From expression (\ref{phasespiral2}) of ratio $\mu^g/\mu^s$, it
is clear that for the F-wave solution $y_1$ (i.e., $y_1>R_s$
with the unphysical situation of $y_1<0$ being ignored), ratio
$\mu^g/\mu^s$ is always negative, while for the S-wave solution
$y_2$, ratio $\mu^g/\mu^s$ is always positive. Note that these
results are qualitatively the same as those in the aligned case with
$m\geq2$. The same as the aligned case, for the F-wave solution
$y_1$, ratio $\mu^g/\mu^s<0$ means that surface mass density
perturbations in stellar SID and gaseous MSID are out of phase
while for the S-wave solution $y_2$, ratio $\mu^g/\mu^s>0$ means
that the surface mass density disturbance in the gaseous MSID is
in phase with the surface mass density disturbance in the stellar
SID. It becomes transparent that in-phase and out-of-phase surface
mass density perturbations are intrinsic characters for the two
solutions $y_1$ and $y_2$ when $m\geq 2$. Physically, both aligned
and unaligned perturbations represent MHD density wave propagations
in a composite MSID system. As MHD density wave patterns with
out-of-phase surface mass density perturbations travel faster,
a higher rotational Mach number $D_s$ is required in order to
strike a global stationary configuration. In contrast, MHD density
wave patterns with in-phase surface mass density perturbations
travel slower, a lower rotational Mach number $D_s$ is needed in
order to strike a global stationary configuration. This phase
relationship for the two surface mass density perturbations is
retained even in the presence of an isopedic magnetic field (Lou
\& Fan 1998; Lou \& Shen 2003; Lou \& Zou 2004).

The solution $y$ stands for $D_s^2$ which is the square of the
rotational Mach number $D_s$ of the stellar SID. The two solutions
$y_1$ and $y_2$ correspond to the square of two different rotational
Mach number. One obvious physical requirement is that both $y_1$ and
$y_2$ must be positive. Meanwhile, we also need to require $D_g^2$
to be positive in order to establish a plausible composite MSID
system. According to equation (\ref{Dg1}), we have the
correspondence between the condition $D_g^2>0$ and the condition
\begin{equation}
D^2_s=y>\frac{\mathcal B}{\chi}-1\ .
\end{equation}
We discuss in the following these two constraints for F-wave
solutions $y_1$ and S-wave solutions $y_2$, separately. The
two constraints must be satisfied simultaneously to limit
the parameter space, otherwise there are no physical $y$
solutions.

For F-wave solutions $y_1$ to exist, we have the following
mathematical conclusions from Appendix B, namely
\begin{eqnarray*}
\qquad\qquad y_1>0 \ \ \quad\mbox{ and }\quad \ \
y_1>\frac{\mathcal B}{\chi}-1
\end{eqnarray*}
are guaranteed by $C_2>0$.
As a result, we can set a limit on the strength of an
isopedic magnetic field by requiring $C_2>0$. For the
$\mathcal B$ parameter, we thus have inequalities
\begin{equation}\label{spiraly1b}
1\leq{\mathcal B}<1+\frac{{\mathcal M}(1+\delta)}
{{\mathcal N}_m(\alpha)}
\end{equation}
for the existence of a physical F-wave solution $y_1$. As in
the aligned case, this constraint means that too strong an
isopedic magnetic field would be impossible for sustaining
a stationary F-wave solution $y_1$. By increasing both $m$
and $\delta$ and decreasing $\alpha$, the allowed range of
$\mathcal B$ for sustaining F-wave solution $y_1$ becomes
enlarged. Once the F-wave solution $y_1$ exists, we always have
a decreasing $y_1$ as both $\chi$ and $\mathcal B$ increase
(see Appendix B for details). Through our analysis, we also
find that for ${\mathcal B}>1$ and $\alpha\rightarrow\infty$,
the $C_2$ coefficient will become negative and thus lead to
a negative $y_1$. Therefore, there exists a critical value
$\alpha_c$ of $\alpha$ such that $y_1>0$ exists only when
$\alpha<\alpha_c$; this critical $\alpha_c$ is
$$
\alpha_c\equiv\bigg[\frac{(m^2-2)^2(1+\delta)^2}
{({\mathcal B}-1)^2}-\bigg(m^2+\frac{1}{4}\bigg)\bigg]^{1/2}\ .
$$
For the S-wave solution $y_2>0$ to exist, we establish
the following two correspondences between
\begin{eqnarray*}
\qquad\qquad  y_2>0 \ \ \quad\mbox{ and }
\quad \ \ C_0>0\ ,
\end{eqnarray*}
and between
\begin{eqnarray*}
\qquad\qquad y_2>\frac{\mathcal B}{\chi}-1
\ \qquad\qquad \mbox{and} \qquad\qquad\qquad\qquad\qquad\ \  \\
\qquad C_2\bigg(\frac{{\mathcal B}}{\chi}-1\bigg)^2
+C_1\bigg(\frac{{\mathcal B}}{\chi}-1\bigg)
+C_0
>0\ ,
\end{eqnarray*}
respectively.
The requirement of $C_0>0$ for a positive $y_2$ leads to a
constraint on $\chi$ parameter, that is, $\chi<\chi_c$ where
the critical value $\chi_c$ for $\chi$ is explicitly defined by
\begin{eqnarray}\label{chics}
\nonumber
\chi_c\equiv\frac{{\mathcal B}({\mathcal M}+1)
[1+\delta-{\mathcal N}_m(\alpha)]}
{[1+\delta-{\mathcal N}_m(\alpha)][{\mathcal M}
+{\mathcal N}_m(\alpha)]-{\mathcal B}{\mathcal N}_m(\alpha)
[1-{\mathcal N}_m(\alpha)]}\\
>\frac{{\mathcal B}({\mathcal M}+1)}
{[{\mathcal M}+{\mathcal N}_m(\alpha)]}\ .
\qquad\qquad\qquad
\end{eqnarray}
We further note that when $\mathcal B$ parameter becomes
sufficiently large to satisfy the following inequality
\begin{equation}
{\mathcal B}>1+\frac{{\mathcal M}(1+\delta)}
{{\mathcal N}_m(\alpha)}+\frac{(1+{\mathcal M})\delta}
{[1-{\mathcal N}_m(\alpha)]}\ ,
\end{equation}
the $\chi$ parameter will be free from the constraint of
$\chi<\chi_c$ because such $\chi_c$ no longer exists.

From the inequality
$$
C_2\bigg(\frac{{\mathcal B}}{\chi}-1\bigg)^2
+C_1\bigg(\frac{{\mathcal B}}{\chi}-1\bigg)+C_0>0\ ,
$$
we derive another constraint on $\chi$, that is,
$\chi>\chi_s$ where $\chi_s$ is explicitly defined by
\begin{eqnarray} \nonumber
\chi_s\equiv \frac{{\mathcal B}}{({\mathcal
M}+1)}\bigg\lbrace{\mathcal M} +{\mathcal N}_m(\alpha)
\qquad\qquad\qquad\qquad \\ \label{chiss} \qquad\qquad\qquad
-\frac{\delta{\mathcal N}_m(\alpha) [1-{\mathcal
N}_m(\alpha)]}{{\mathcal B}{\mathcal N}_m(\alpha)
+(1+\delta)[1-{\mathcal N}_m(\alpha)]}\bigg\rbrace\ .
\end{eqnarray}
We can readily demonstrate that $\chi_s<\chi_c$ such that physical
solution $y_2>0$ could exist for $\chi_s<\chi<\chi_c$. However,
we show presently that $\chi_s>\chi_c$ may happen in certain
situations and the two requirements are incompatible with each
other; there is thus no physical solution for a positive $y_2$.

The allowed range for $\chi$ parameter is $\chi>1$. When
${\mathcal B}=1$ for the absence of an isopedic magnetic
field, the critical $\chi_s$ remains always less than 1
such that the second requirement will be satisfied
automatically. As the strength of an isopedic magnetic
field gets stronger, the $\chi>\chi_s$ constraint becomes
indispensable for a positive $y_2$.

Our analysis above bears a strong resemblance to that for the
aligned case. In fact, should we set the parameter $\alpha=0$
for a purely azimuthal propagation of MHD density waves, we
can approximately obtain nearly the same results as those of
the aligned case. This naturally suggests that the aligned case
should be only regarded as a special case of the unaligned cases
(Lou 2002; Lou \& Fan 2002; Lou \& Shen 2003; Lou \& Zou 2004).

Through extensive numerical explorations, we can show empirically
that both the F-wave solution $y_1$ and the S-wave solution $y_2$
increase with increasing $\alpha$ (see Figure 5). From the
perspective of global stationary MHD density waves, this is
a physically sensible result.

\begin{figure}
\begin{center}
\includegraphics[angle=0,scale=0.45]{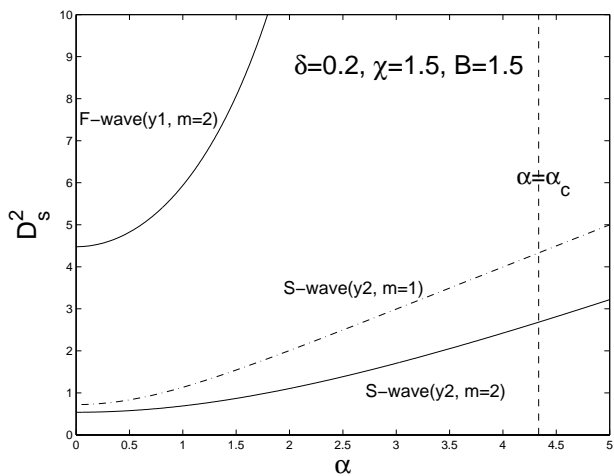}
\caption{For $D_s^2$ versus $\alpha$ of the unaligned case with
specified $\delta=0.2$, $\chi=1.5$ and ${\mathcal B}=1.5$, the
two solid solution curves $y_1$ and $y_2$ are for $m=2$ and
the one dash-dotted solution curve $y_2$ is for $m=1$.
Here, we use $B$ in the figure to correspond to $\mathcal B$ in the
main text. All these solutions increase with increasing $\alpha$.
The dashed vertical line represents $\alpha=\alpha_c$. When $\alpha$
becomes greater than $\alpha_c$, the branch of $y_1$ ($m=2$) solution
flips and becomes negative. In comparison, we see that $y_2$ with
$m=1$ is greater than $y_2$ with $m=2$.  }
\end{center}
\end{figure}

\subsubsection{The $m=1$ Case}

For $m=1$, definition (\ref{M}) and recurrence
relation (\ref{Nm1}) together give
\begin{displaymath}
{\mathcal{M}}=-\frac{1}{(\alpha^2+5/4 )}
\ \quad\ \hbox{and}\ \quad
{\mathcal{N}}_1(\alpha)=\frac{(\alpha^2+17/4)^{1/2}}
{(\alpha^2+9/4 )}\ .
\end{displaymath}
By numerical experiments (see Appendix B), we have $C_2<0$ and
only $y_2$ solution being positive. The ratio $\mu^g/\mu^s$ of
$y_2$ solution remains always positive with in-phase surface
mass density perturbations. We also know that $y_2$ increases
with decreasing $\chi$ and increasing $\mathcal B$. Numerically,
we find that $y_2$ solution increases with increasing $\alpha$.

The two constraints for sensible
$y_2$ solutions may be expressed as
\begin{eqnarray*}
\qquad y_2>0 \ \qquad\qquad \hbox{and} \qquad\qquad
\ y_2>\frac{\mathcal B}{\chi}-1
\end{eqnarray*}
corresponding to
\begin{eqnarray*}
C_0>0 \qquad  \hbox{and} \qquad  C_2\bigg(\frac{{\mathcal B}}
{\chi}-1\bigg)^2+C_1\bigg(\frac{{\mathcal B}}{\chi}
-1\bigg)+C_0
>0\ ,
\end{eqnarray*}
respectively (see Appendix B2).

The mathematical forms of these constraints are too involved to be
shown here. Note that properties of $y_2$ solution with $m=1$ are
just like those of $y_2$ in the cases of $m\geq 2$ (see Fig. 6).
As the $y_2$ solutions of $m=2$ and $m=1$ are fairly similar to
each other, the $y_2$ solution changes smoothly as $m$ increases
sequentially.
%

\subsubsection{The $m=0$ Case: Marginal Stability
\newline of Axisymmetric MHD Disturbances}

For $m=0$ in definition (\ref{M}) and
recurrence relation (\ref{Nm1}), we have
\begin{eqnarray}\nonumber
{\mathcal{M}}=-\frac{2}{(\alpha^2+1/4)}\
\\ \hbox{ and }
\qquad\qquad\qquad\qquad\qquad\qquad\qquad\nonumber
\qquad\qquad \\ {\mathcal{N}}_0(\alpha)
=\frac{(\alpha^2+9/4)}{(\alpha^2+1/4)
(\alpha^2+17/4)^{1/2}}\ .
\end{eqnarray}
On the basis of extensive numerical experiments,
we found that the five expressions below
\begin{displaymath}
1-\frac{{\mathcal{N}}_0(\alpha)}{(1+\delta )}\ ,\ \ \
1-{\mathcal{N}}_0(\alpha)\ ,\ \ \ 1+{\mathcal{M}}\ ,\ \ \
{\mathcal{M}}+{\mathcal{N}}_0(\alpha)\ ,\ \ \
{\mathcal{M}}+\frac{{\mathcal{N}}_0(\alpha)}{(1+\delta)}
\end{displaymath}
will become negative when $\alpha$ is sufficiently small. As
$\alpha$ is increased across several different critical values,
they will in turn become positive. We now proceed to discuss
consequences of this $\alpha$ variation in different ranges
separately.

The analysis is fairly complicated and we
summarize the basic results here. By imposing the two
constraints of $y>0$ and $y>({\mathcal B}/\chi)-1$, we
found that when $\alpha<1.113$ and $\alpha>1.793$, there
is only one $y$ solution that is physically valid under
certain conditions. For $1.113\leq\alpha\leq1.793$, there
is no physically valid solution at all. We then find that
the valid $y$ solution is characterized by in-phase surface
mass density perturbations (i.e. a positive $\mu^g/\mu^s$
ratio). As before, the $y$ solution increases
with decreasing $\chi$ and increasing $\mathcal B$ in both
ranges of $\alpha<1.113$ and $\alpha>1.793$. Numerically, we
again find that when $\alpha<1.113$, the solution of $D_s^2$
decreases with increasing $\alpha$. In comparison, when
$\alpha>1.793$, $D_s^2$ first decreases with increasing
$\alpha$ and upon crossing a certain point, $D_s^2$ increases
with increasing $\alpha$.

For the positive portions of solution $D^2_s$ without going
into the extreme, the basic profile of $D_s^2$ versus $\alpha$
is qualitatively similar to the results of the single SID in
Shu et al. (2000) and of the composite SIDs in Lou \& Shen
(2003). In fact, our model analysis is more general to include
the results of Shu et al. (2000) and of Lou \& Shen (2003).

From the dispersion relation of MHD density waves in the WKBJ
or tight-winding approximation (Shen \& Lou 2003, 2004b; Lou \&
Zou 2004), we see that the solution curve in the $D_s^2$ versus
$\alpha$ profile represents the marginal stability curve that
separates the regime of stable oscillations from the regime of
unstable oscillations. As noted by Shu et al. (2000), when
$\alpha<1.113$, the parameter regime under the curve in the
lower-left corner represents the rotation and magnetic field
modified Jeans collapse regime. As $\alpha$ stands for the
radial wavenumber and $m$ stands for the azimuthal wavenumber,
a small $\alpha$ corresponds to a long-radial wavelength MHD
density wave disturbance. When $\alpha>1.793$, the parameter
regime above the solution curve represents the ring
fragmentation regime for large $\alpha$ corresponding to
short-radial wavelength MHD density wave (see Fig. 7).
\begin{figure}
\begin{center}
\includegraphics[angle=0,scale=0.45]{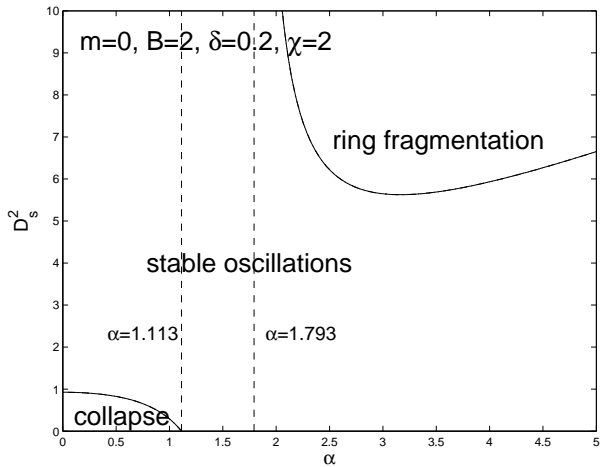}
\caption{The marginal stability curve of $D_s^2$ versus $\alpha$
when $m=0$, ${\mathcal B}=2.0$, $\delta=0.2$ and $\chi=2.0$.
Here, we use $B$ in the figure to correspond to $\mathcal B$
in the main text. The solution curve decreases with increasing
$\alpha$ when $\alpha<1.113$. When $\alpha>1.793$, the solution
curve first decreases and then increases with increasing $\alpha$.
There exists a minimum for the ring fragmentation curve.}
\end{center}
\end{figure}

As $\chi$ parameter increases, the $y$ solution decreases. Hence
the collapse regime shrinks and the ring fragmentation regime grows
with increasing $\chi$ (see Fig. 8). As $\mathcal B$ parameter
increases, the $y$ solution increases. Hence the collapse regime
grows and the ring fragmentation regime shrinks with increasing
$\mathcal B$ (see Fig. 9).

\begin{figure}
\begin{center}
\includegraphics[angle=0,scale=0.45]{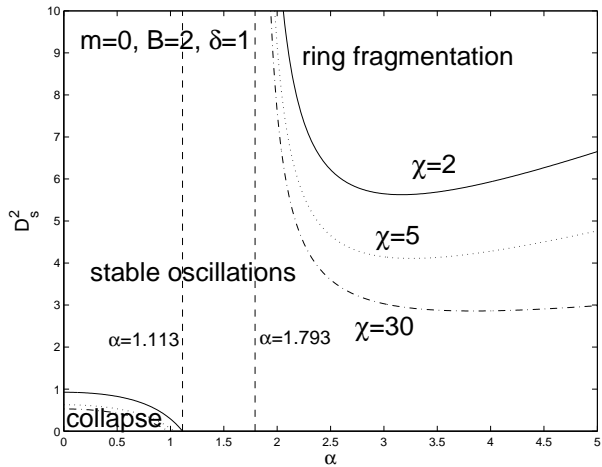}
\caption{The marginal stability curves of $D_s^2$ versus
$\alpha$ with different values of $\chi=2$, 5, 30,
respectively, for specified values of $m=0$,
${\mathcal B}=2$, and $\delta=1$.
Here, we use $B$ in the figure to correspond to $\mathcal B$
in the main text. The collapse regime shrinks and the ring
fragmentation regime grows with increasing $\chi$.   }
\end{center}
\end{figure}

\begin{figure}
\begin{center}
\includegraphics[angle=0,scale=0.45]{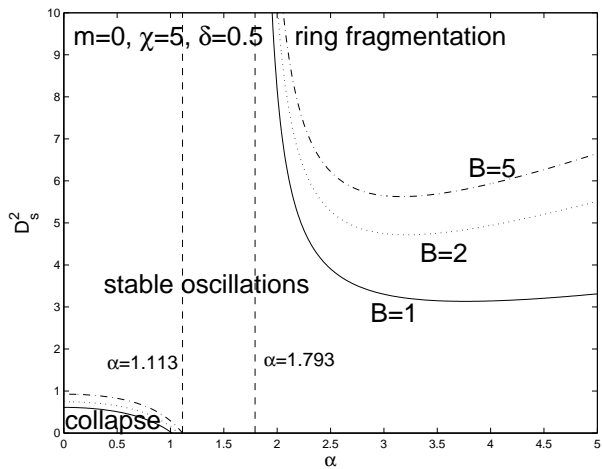}
\caption{The marginal stability curves of $D_s^2$ versus
$\alpha$ with different values of ${\mathcal B}=1$, 2, 5,
respectively, for specified values of $m=0$, $\chi=5$,
and $\delta=0.5$.
Here, we use $B$ in the figure to correspond to $\mathcal B$ in
the main text. The collapse regime grows and the ring-fragmentation
regime shrinks with increasing $\mathcal B$.     }
\end{center}
\end{figure}

Physically, $\chi$ parameter stands for the ratio in the
effective temperatures of the two SIDs and $\mathcal B$ represents
the strength of the isopedic magnetic field. The above results
mean that when the `fluid' stellar SID becomes much `hotter' than
the gaseous isopedic MSID, the stellar SID tends to be more stable
against collapse for large-scale disturbances but less stable
against ring-like fragmentation for small-scale disturbances. For a
stronger isopedic magnetic field, the tendency of collapse becomes
easier while the ring-fragmentation becomes more difficult.
%
Here the $\mathcal{B}$ parameter involves the background equilibrium
parameters. While we seem to vary only the background magnetic field
strength, the background magneto-rotational equilibrium actually
changes accordingly.

\section{Conclusions, Summary and Discussions }


For modelling an actual spiral galaxy, our treatment of a composite
system of MSIDs, while highly idealized still, contains several
physically realistic key ingredients. The formulation and results
of our analysis provide the basic rationale for developing useful
astrophysical concepts. The two theorems of Shu \& Li (1997) have
now been generalized to a composite MSID system with an isopedic
magnetic field. Instead of an assumption, the isopedic relation
is shown here to be an initial condition maintained by ideal MHD
equations. In particular, we have taken into account of long-range
effects of gravity in full without the usual local WKBJ
approximation. We have succeeded in constructing various global
perturbation configurations, such as bars, spirals, barred spirals
etc. By analogies, we proposed specific transition criteria among
different stationary perturbation configurations. With the isopedic
magnetic field geometry, we have set the stage to further study
global patterns of circumnuclear and spiral arm galactic winds.
We now discuss plausible consequences and possible applications
to spiral galaxies from several perspectives.

On much smaller scales, physical processes involve various
star formation activities and their interactions with the
surrounding ISM. As luminous massive OB star formation
tends to be more numerous in regions of relatively high gas
concentration and of stronger magnetic field, the collection
of numerous young bright stars will then outline
a spiral pattern embedded in relatively high density spiral
arms. Theoretical studies on the large-scale dynamics of
galactic spiral structure are closely tied to a chain of
star formation processes and to the physical manifestation
of spiral MHD density waves.
%
For the global star formation rate in a disc galaxy (e.g.,
Kennicutt 1989; Elmegreen 1994; Silk 1997), magnetic field plays
an important role in the gaseous disc in the context of large-scale
instabilities (e.g., Lou \& Fan 2000; Lou et al. 2001).
We would emphasize that should large-scale galactic disc
instabilities be indeed responsible for the global star formation
rate, then analyses and simulations of a composite disc system
involving magnetic field would be more realistic and physically
meaningful. The oft-used axisymmetric stability criterion involving
the $Q$ parameter is significantly modified in a composite disc
system with an isopedic magnetic field (Shen \& Lou 2003; Lou
\& Zou 2004).

Galaxies form and evolve on large scales. To provide a dynamic
basis for the classification of galactic morphologies is one
major goal of galactic research (e.g., Bertin et al. 1989a, b;
Bertin \& Lin 1996). The modal approach to this problem has
gained remarkable progress in explaining various configurations
of spiral, barred-spiral galaxies with observational constraints.
For example, a barred galaxy is expected to be associated with a
relatively heavy disc whose halo mass inside the optical radius
is small. The potential of a rotating bar can give rise to quasi
periodic perturbations in the disc and hence affect the entire
configuration of a disc. While limited by the stationarity
requirement in our combined analytical and numerical solution
procedure, we are able to take into account of long-range effects
of gravitation in full for MHD perturbations in a composite MSID
system without the deficiency of the usual local WKBJ approximation.
For example for $m=2$, we have constructed separately barred
configurations (aligned) and logarithmic spiral configurations
(unaligned). By the principle of linear superposition, we might be
able to also construct barred logarithmic spiral configurations in
a composite MSID system by choosing appropriate sets of parameters.
In addition to theoretically modelling galactic configurations,
these solutions are valuable in testing, initializing and
benchmarking hydrodynamic and MHD numerical simulation codes for
further nonlinear studies.

As important extension and generalization of the earlier theoretical
analyses of Shu et al. (2000) on zero-frequency (i.e. stationary)
aligned and unaligned MHD perturbation configurations of an
isopedically magnetized SID and of Lou \& Shen (2003) on stationary
aligned and unaligned perturbation structures in a composite system
of two-fluid SIDs, we have constructed analytically stationary
configurations of aligned and unaligned MHD perturbations in a
composite system of a `fluid' stellar SID and a `magnetofluid'
gaseous MSID. Other closely relevant work include Galli et al.
(2001), Lou (2002), Lou \& Fan (2002), Lou \& Shen (2003), Shen
\& Lou (2004a, b), Chakrabarti et al. (2003), Shen, Liu \& Lou
(2005), Lou \& Zou (2004, 2005). While this composite model of
one SID and one MSID is highly idealized in many aspects, it
does contain several necessary and more realistic elements that
are crucial to understand large-scale structures and dynamics
of spiral galaxies.

From the basic fluid-magnetofluid equations for a composite
system of a stellar SID and a gaseous MSID, we begin with an
axisymmetric background in a magneto-rotational equilibrium. The
necessary condition of $1+\epsilon\delta>0$ arises here in order
to maintain an equilibrium with a positive gas surface mass
density $\Sigma^g$ (eqn. \ref{sigmag0}) which in turn puts
an upper limit on the magnetic flux under the isopedic condition
of constant $\Sigma^g/B_z$. One important conclusion of our
analysis is that a constant $\Sigma^g/B_z$ is an initial condition
sustained by ideal MHD equations during disc evolution. Physically,
the magnetic tension force cannot be too strong to make an
equilibrium of the gaseous MSID impossible (eqn \ref{backg}) in the
first place. Massive dark-matter halos tend to stabilize disclike
galaxies as expected (Ostriker \& Peeble 1973; Hohl 1976; Miller
1978). Although quantitative results for a partial composite MSID
system are not detailed here, we suggest that the potential of a
dark matter halo will certainly stabilize a composite MSID system
and allow for a stronger flux of isopedic magnetic field based on
our working experience. The $\mathcal B$ parameter [see definition
(\ref{mathB})] contains this constraint explicitly.

For MHD perturbations, we derive two sets of linearized equations
in the stellar SID and the gaseous MSID. Setting $\omega=0$, we
derive the stationary MHD dispersion relation with the gravitational
coupling. By properly choosing different potential-density pairs to
satisfy the Poisson integral, we examine the two classes of aligned
and unaligned logarithmic spiral MHD configurations in parallel. As
a result of the gravitational coupling between the two SIDs, we
obtain a quadratic equation in terms of $y\equiv D_s^2$ to construct
stationary MHD perturbation configurations of the composite MSID
system, namely, equation (\ref{quand}) for the aligned case and
equation (\ref{quand1}) for the unaligned logarithmic spiral case.
The two necessary physical requirements are that both $D_s^2$ and
$D_g^2$ should be positive.

On the basis of rigorous mathematical derivation and analysis,
we have reached several conclusions and results. Meanwhile,
we have explored parameter regimes numerically to reveal a
few trends of parameter variations empirically.

Let us first summarize the basic results of the aligned case.
For axisymmetric MHD perturbations with $m=0$, the resulting
configuration is simply a rescaling of one axisymmetric
equilibrium to a neighbouring axisymmetric equilibrium which
is certainly allowed but trivial for our purpose. For
nonaxisymmetric MHD perturbations with $m=1$, the resulting
quadratic equation of $y\equiv D_s^2$ is automatically
satisfied for arbitrary values of $y$ (Shu et al. 2000; Lou
\& Shen 2003). For nonaxisymmetric MHD perturbations with
$m\geq 2$, there are two sets of $y\equiv D_s^2$ solutions
representing two different classes of stationary global MHD
perturbation configurations. For the phase relationships
between the surface mass density perturbations in the two SIDs,
the lower $D_s$ solutions are featured by an in-phase relation
while the higher $D_s$ solutions are featured by an out-of-phase
relation. For a weak isopedic magnetic field, the out-of-phase
solution always exists ($D_s^2>0$ and $D_g^2>0$ simultaneously),
while for the existence of the in-phase solution, the condition
$\chi_s<\chi<\chi_c$ must be met with $\chi_s$ and $\chi_c$
defined by equations (\ref{chis}) and (\ref{chic}) for the
aligned case and by equations (\ref{chics}) and (\ref{chiss})
for the unaligned case. As $\chi_s<1$ may happen as the isopedic
magnetic field becomes weak, the requirement $\chi>\chi_s$ will
be automatically satisfied by the physical constraint $\chi>1$.
As the isopedic magnetic flux increases across a certain value,
the out-of-phase solution vanishes while the constraint of the
in-phase solution becomes $\chi>\chi_s$ with $\chi_s>1$ and with
the constraint $\chi<\chi_c$ being irrelevant. Moreover, we
emphasize that in order to have $D_g^2>0$, the condition
$\chi_s<\chi$ must be satisfied for both aligned and unaligned
cases and for both F-wave and S-wave solutions; the constraint
$\chi_s<\chi$ for F-wave solutions is satisfied automatically
once other requirements are met.

In the framework of the density wave theory, the case of aligned
perturbations represents a purely azimuthal propagation of density
wave (Lou 2002) and the stationarity is sustained by the advection
of a counter SID rotation. In general, the class of unaligned
logarithmic spiral perturbations represents MHD density waves with
both radial and azimuthal propagations where parameter $\alpha$ is an
effective radial wavenumber. Another special situation is a purely
radial propagation with $m=0$ in the `unaligned' class. It is thus
not surprising that with $\alpha\neq 0$ in the unaligned class, we
obtain fairly similar results in parallel with the aligned class for
$m\geq 2$. For example, by setting $\alpha=0$ in the unaligned class,
we arrive at essentially the same results as in the aligned class for
$m\geq 2$. This justifies our physical interpretation that the aligned
class is merely a special case to the unaligned class. For the
unaligned class, we obtain two solutions of $D_s^2$ characterized
by out-of-phase and in-phase surface mass density perturbations,
respectively. Other solution properties of the unaligned class
remain qualitatively the same as those of the aligned class.

The unaligned cases with $m=1$ and $m=0$ need more special
considerations. For $m=1$, only the $D_s^2$ solution with in-phase
surface mass density perturbations could be valid under certain
constraints. For $m=0$, only the $D_s^2$ solution with in-phase
surface mass density perturbations could still exist; when
$1.113\leq\alpha\leq1.793$, there is no physical solution though.
From the local WKBJ dispersion relation, the profile $D_s^2$ versus
$\alpha$ represents the marginal stability curve (Lou \& Shen 2003;
Shen \& Lou 2003, 2004a, b; Lou \& Zou 2005; Shen, Liu \& Lou 2005).
In general, when $D_s^2$ and $\alpha$ are both sufficiently small,
rotation and magnetic field modified gravitational instabilities
for Jeans collapse occur and when $D_s^2$ and $\alpha$ are both
sufficiently large, MHD ring fragmentation instabilities appear.
Between these two unstable regimes is the stable regime of SID
rotation for axisymmetric oscillations. By analytical analyses,
we find that all physically valid $D_s^2$ solutions increase with
decreasing $\chi$ and increasing isopedic magnetic flux. This is a
quite general conclusion for both aligned and unaligned cases. The
substantial mathematical procedures for these conclusions can be
found in Appendices A and B.

Finally, we discuss several possible implications and applications
of our model analysis. After the theoretical prediction of the solar
wind by Parker (1958, 1963) and the subsequent confirmation by spacecraft
observations in the early 1960s, the more general concept of stellar
winds and the relevant theories have been developed in parallel (e.g.,
Burke 1968; Holzer \& Axford 1970; Lamers \& Cassinelli 1999).
Meanwhile, the concept of galactic winds has gradually emerged (e.g.,
Johnson \& Axford 1971). Intrigued by the almost absence of ISM
materials in most elliptical galaxies, Mathews \& Baker (1971) also
came to the initial concept of galactic winds. It is now widely
believed that a galactic wind (sometimes also referred to as a
galactic superwind) is powered by starburst activities and supernova
explosions (e.g., Chevalier \& Clegg 1985; Heckman et al. 1990).
Because of various observational limitations, the detection of a
galactic wind has remained a challenge for many years. Along with
the advance of observational technologies, evidence for galactic
winds have gradually surfaced in the multi-wavelength studies.
Using the {\it ROSAT} data for X-ray emissions, Strickland et al.
(1997) presented a galactic wind picture for the nearby galaxy M82.
Matthews \& de Grijs (2004) found the evidence for a galactic wind
in the edge-on Sbc galaxy UGC 10043 by optical imaging and
spectroscopy. Melo et al. (2003) reported the detection of
supergalactic winds in the edge-on starburst galaxy NGC 4631. Together
with galactic-scale outflows, a relevant concept and phenomenon of
superbubbles have been proposed and observed (e.g., Tomisaka, Ikeuchi
\& Habe 1981; Ferri\`ere 2001). Both phenomena are the results of
supernovae, hypernovae, violent burst activities of massive star
formation in the densest regions of host galaxies. Galactic winds
are important processes to channel ISM materials along open magnetic
fields into the intergalactic medium, while superbubbles fail to get
out of the reign of the host galaxies (e.g., Tenorio-Tagle, Silich \&
Mu\~noz-Tu\~n\'on 2003). The physical reasons why superbubbles cannot
become galactic winds remain unclear.

Based on our model analysis, we here propose that the global structure
and geometry of galactic magnetic field should be mainly responsible
for the coexistence of large-scale outflows and superbubbles in disc
galaxies. In strong analogy to similar phenomena such as global coronal
mass ejections, violent solar flares from active regions with closed
magnetic field over the solar surface and fast solar wind streams
coming out of solar coronal hole regions with open magnetic fields, we
suspect that bubbles and superbubbles may involve closed magnetic fields
anchored at the galactic disc plane, while large-scale galactic winds
pumped by energetic sources must be guided by open magnetic field lines
to reach outer regions.

The model analysis here and those of Lou \& Zou (2004, 2005) and
of Shen, Liu \& Lou (2005) are complementary to each other in terms
of overall magnetic field geometries associated with discs. In the
case of a coplanar magnetic field, Parker instabilities inflated by
galactic cosmic rays and supernova explosions can lead to formation
of bubbles and superbubbles. In the case of an isopedic magnetic
field, Velikhov-Chandrasekhar-Balbus-Hawley instabilities, starburst
activities and the cosmic-ray pressure together pump galactic outflows
from circumnuclear regions and along spiral arms. We have separately
prepared MHD stages in an idealized manner for further research on
pertinent physical processes. In a real disc galaxy, these two
different magnetic field geometries are intermingled and randomly
distributed all over the disc. Intuitively, it is natural to imagine
that outflows will be stronger at places where open magnetic field
flux is high. As large-scale dense regions in a galactic plane, the
spiral arms in a galaxy are expected to carry more magnetic field
flux because of the frozen-in condition for magnetic flux in the gas
and of small-scale MHD dynamo processes. Comparing other regions in
the disc (except for the nucleus and circumnuclear regions), stronger
outflows are expected to emerge from spiral arm regions. We refer to
this scenario as Spiral Arm Galactic Winds.
%
%

\section*{Acknowledgments}
This research has been supported in part by the ASCI Center for
Astrophysical Thermonuclear Flashes at the University of Chicago
under Department of Energy contract B341495, by the Special Funds
for Major State Basic Science Research Projects of China, by
Tsinghua Centre d'Astrophysique, by the Collaborative Research
Fund from the National Natural Science Foundation of China (NSFC)
for Young Outstanding Overseas Chinese Scholars (NSFC 10028306) at
the National Astronomical Observatory, Chinese Academy of Sciences,
by the NSFC grant 10373009 at the Tsinghua University, and by the
Yangtze Endowment from the Ministry of Education through Tsinghua
University. The hospitality and support of the Mullard Space
Science Laboratory at University College London, U.K. and of
Centre de Physique des Particules de Marseille (CPPM/IN2P3/CNRS)
+ Universit\'e de la M\'editerran\'ee Aix-Marseille II, France
are also gratefully acknowledged. Affiliated institutions of
Y.Q.L. share this contribution.

\begin{appendix}
\section[]{}
In the aligned cases of $m\geq 2$, we defined two parameters
\begin{eqnarray}
\qquad\qquad
{\mathcal{B}}\equiv\frac{(1+\delta)}{(1+\epsilon\delta)}\geq 1\ ,\\
\qquad\qquad
X\equiv m^2-2-\frac{m({\mathcal{B}}-1)}{(1+\delta)}\ .
\end{eqnarray}
In the quadratic equation of $y\equiv D_s^2$ for a
stationary global MHD perturbation configuration
\begin{equation}\label{quadr}
C_2y^2+C_1y+C_0=0\ ,
\end{equation}
the three coefficients $C_2$,
$C_1$ and $C_0$ are defined by
\begin{eqnarray*}
C_2&\equiv &(m+2)X\ ,\\
C_1&\equiv &
2X-2(m+1)\bigg{\{}\frac{m\delta}{1+\delta }
+\frac{{\mathcal{B}}}{\chi}
\bigg[m^2-2+\frac{m}{1+\delta }\bigg]\bigg{\}}\ ,\\
C_0&\equiv &-m\bigg{\{}X-2(m+1)
\bigg[\frac{{\mathcal{B}}}{\chi}
\bigg(m-\frac{1}{1+\delta}\bigg)
-\frac{\delta}{1+\delta }\bigg]\bigg{\}}\ .
\end{eqnarray*}
From quadratic equation (\ref{quadr}), one can
readily derive the expression of the determinant
$\Delta$ and show that
\begin{eqnarray*}
\Delta\equiv C_1^2-4C_0C_2=4(m+1)^2
\bigg{\{}\frac{4m^2{\mathcal{B}}^2\delta}{(1+\delta)^2\chi}
\qquad\qquad\\  \qquad\qquad
+\bigg[X+\frac{m\delta}{1+\delta}
-\frac{\mathcal{B}}{\chi}
\bigg(m^2-2+\frac{m}{1+\delta}\bigg)\bigg]^2\bigg{\}}>0\ .
\end{eqnarray*}
There are thus two real $y\equiv D_s^2$
solutions for equation (\ref{quadr}).

For the convenience of analysis, we introduce
a useful and important $R_a$ parameter,
\begin{equation}\label{A4}
R_a\equiv\frac{m[m-1/(1+\delta)]}
{m^2-2+m/(1+\delta)}>0\ .
\end{equation}
The two following inequalities hold, namely
\begin{eqnarray*}
C_2R_a^2+C_1R_a+C_0=-\frac{4{\mathcal{B}}m^2(m-1)(m+1)^2\delta}
{[(m^2-2)(1+\delta)+m]^2}<0\ ,\\
C_2(-1)^2+C_1(-1)+C_0=\frac{4(m-1)(m+1)^2{\mathcal B}}{\chi}>0\ .
\end{eqnarray*}
According to the Viete theorem
(e.g., John \& Horst 1998) for a quadratic equation,
we know that $y_1>R_a>y_2>-1$ for $C_2>0$, while $R_a>y_2>-1>y_1$
for $C_2<0$. Note that when $C_2<0$ (i.e., $X<0$), one must
also have $C_1<0$. It is then clear that $y_1$ remains
positive and negative for $C_2>0$ and $C_2<0$, respectively.

To determine the phase relationship between the two surface
mass density perturbations $\mu^g$ and $\mu^s$, we derive
\begin{eqnarray*}
\frac{\mu^g}{\mu^s}=\frac{1-H_1m^2}{G_1m^2}
=-\frac{(m^2-2)(1+\delta)+|m|}{|m|}
\frac{(D_s^2-R_a)}{(D_s^2+1)}\ ,
\end{eqnarray*}
where $H_1$ and $G_1$ are separately defined by expressions
(\ref{h1}) and (\ref{g1}). Here, we are primarily interested
in a $y$ solution which is at least positive; a negative $y$
solution is unphysical. As the case of $y_1>0$ always corresponds to
$y_1>R_a>0$, it follows that $\mu^g/\mu^s$ corresponding to $y_1$
solution remains always negative for $\mu^g$ and $\mu^s$ being
out of phase. As $y_2<R_a$ for any values of $C_2$, it follows
immediately that the ratio of $\mu^g/\mu^s$ corresponding to
$y_2$ solution remains always positive for $\mu^g$ and $\mu^s$
being in phase.

In addition to the physical requirement of $y\equiv D_s^2>0$, we
must also impose $D_g^2>0$. In the purely hydrodynamic case of
Lou \& Shen (2003) without an isopedic magnetic field as studied
here, $D_g^2>0$ follows from $D_s^2>0$ automatically. According
to the background magneto-rotational equilibrium condition of
radial force balance, we readily derive the inequality
\begin{displaymath}
D_g^2=\frac{\chi}{\mathcal B}(D_s^2+1)-1>0
\end{displaymath}
which is equivalent to the following inequality
\begin{displaymath}
D_s^2>\frac{\mathcal B}{\chi}-1\ .
\end{displaymath}
In other words, besides $D_s^2>0$, we should also
require $D_s^2>{\mathcal B}/\chi-1$ in the case
of a positive RHS.

We now explore the relations between the two real
$y$ solutions and their variations with specific
parameters such as $\chi$ and ${\mathcal{B}}$.
For a chosen parameter, we apply the procedure of
taking the derivative of the relevant $y$ solution
with respect to this specific parameter, namely
\begin{equation}
2C_2y'y+C_2'y^2+C_1'y+C_1y'+C_0'=0\ ,
\end{equation}
where the prime $'$ indicates a derivative
with respect to this chosen parameter. It
follows immediately that
\begin{equation}
y'=-\frac{(C_2'y^2+C_1'y+C_0')}{(2C_2y+C_1)}\ ,
\end{equation}
where $y$ stands for either $y_1$
or $y_2$ solution given by
\begin{eqnarray}
\qquad\qquad y_{1,2}=\frac{-C_1\pm(C_1^2-4C_0C_2)^{1/2}}{2C_2}\ .
\end{eqnarray}
By simple manipulations, we immediately arrive at
\begin{equation}\label{A9}
y_1'=-\frac{C_2'y_1^2+C_1'y_1+C_0'}{(C_1^2-4C_0C_2)^{1/2}}
=-\frac{C_2'y_1^2+C_1'y_1+C_0'}{\Delta^{1/2}}\ ,\\
\end{equation}
\begin{equation}\label{A10}
y_2'=\frac{C_2'y_2^2+C_1'y_2+C_0'}{(C_1^2-4C_0C_2)^{1/2}}
=\frac{C_2'y_2^2+C_1'y_2+C_0'}{\Delta^{1/2}}\ ,\quad
\end{equation}
where $\Delta$ is the determinant of quadratic equation
(\ref{quadr}). By these expressions, we examine the
relations between the two $y$ solutions and their
dependence on relevant parameters.

We first discuss solution properties of $y_1$ and $y_2$
by varying parameter $\chi$ and thus have explicitly
\begin{eqnarray*}
C_2'\equiv\frac{\partial C_2}{\partial\chi}\ ,
\ \ \
C_1'\equiv\frac{\partial C_1}{\partial\chi}\ ,
\ \ \
C_0'\equiv\frac{\partial C_0}{\partial\chi}\ ,
\ \ \
y'\equiv\frac{\partial y}{\partial\chi}\ .
\end{eqnarray*}
It follows immediately that
\begin{eqnarray*}
C_2'&=&0\ , \\
C_1'&=&\frac{2{\mathcal{B}}}{\chi^2}
(m+1)\bigg(m^2-2+\frac{m}{1+\delta}\bigg)\ ,\\
C_0'&=&-\frac{2{\mathcal{B}}}{\chi^2}m(m+1)
\bigg(m-\frac{1}{1+\delta}\bigg)\ .
\end{eqnarray*}

By expressions (\ref{A9}) and (\ref{A10}), the first
derivatives $y_1'$ and $y_2'$ then take the forms of
\begin{eqnarray}
\qquad y_1'=-\frac{C_1'(y_1-R_a)}{(C_1^2-4C_0C_2)^{1/2}}\ ,\\
\qquad y_2'=-\frac{C_1'(R_a-y_2)}{(C_1^2-4C_0C_2)^{1/2}}\ .
\end{eqnarray}

On the basis of our analysis following equation (\ref{A4}),
we have $y_1'<0$ and $y_2'<0$ for $C_2>0$, and $y_1'>0$ and
$y_2'<0$ for $C_2<0$. The basic conclusion is that when $y_1$
and $y_2$ are both positive and thus physical, we have both
$$
\frac{\partial y_1}{\partial \chi}<0
\ \qquad\hbox{ and } \ \qquad
\frac{\partial y_2}{\partial \chi}<0\ .
$$

We next examine the solution properties of $y_1$ and $y_2$
as parameter $\mathcal B$ varies and define explicitly
\begin{eqnarray*}
C_2'\equiv\frac{\partial C_2}{\partial {\mathcal{B}}}\ ,
\ \ \
C_1'\equiv\frac{\partial C_1}{\partial {\mathcal{B}}}\ ,
\ \ \
C_0'\equiv\frac{\partial C_0}{\partial {\mathcal{B}}}\ ,
\ \ \
y'\equiv\frac{\partial y}{\partial {\mathcal{B}}}\ .
\end{eqnarray*}
It then follows immediately that
\begin{eqnarray*}
\quad C_2'&=&-\frac{m(m+2)}{(1+\delta)}\ ,
\\
\quad C_1'&=&-\frac{2m}{(1+\delta)}-\frac{2(m+1)}{\chi}
\bigg[m^2-2+\frac{m}{(1+\delta )}\bigg]\ ,
\\
\quad C_0'&=&\frac{m^2}{(1+\delta)}
+\frac{2m(m+1)}{\chi}\bigg[m-\frac{1}{(1+\delta )}\bigg]\ .
\end{eqnarray*}
By rearrangements and manipulations, it is clear that
\begin{eqnarray}\label{A13}\nonumber
C_2y^2+C_1y+C_0=
\quad\qquad\qquad\qquad\qquad\qquad
\\\label{alignappe1}
\ \ {\mathcal{B}}(C_2'y^2+C_1'y+C_0')+D_2y^2+D_1y+D_0=0\ ,
\end{eqnarray}
where the three coefficients $D_2$,
$D_1$ and $D_0$ are defined by
\begin{eqnarray*}
\quad D_2&\equiv&(m+2)\bigg[m^2-2+\frac{m}{(1+\delta)}\bigg]\ ,\\
\quad D_1&\equiv&2(m+2)\bigg[\frac{m}{(1+\delta)}-1\bigg]\ ,\\
\quad D_0&\equiv&-m(m+2)\bigg[m-\frac{1}{(1+\delta)}\bigg]\ .
\end{eqnarray*}
Now in the second line of equation (\ref{A13}), we find
\begin{eqnarray*}
\frac{D_2y^2+D_1y+D_0}{(m+2)}=\bigg(m^2
-2+\frac{m}{1+\delta}\bigg)(y+1)(y-R_a)\ .
\end{eqnarray*}
By the properties of the two real solutions $y_1$ and
$y_2$ and the expression above, we can readily show that
$D_2y_1^2+D_1y_1+D_0>0$ and $D_2y_2^2+D_1y_2+D_0<0$,
respectively.

With the expression below from the second
line of equation ~(\ref{alignappe1})
\begin{equation}
C_2'y^2+C_1'y+C_0'=-\frac{(D_2y^2+D_1y+D_0)}{\mathcal{B}}\ ,
\end{equation}
we immediately conclude that
\begin{eqnarray*}
\frac{\partial y_1}{\partial \mathcal B}
&=&-\frac{(C_2'y_1^2+C_1'y_1+C_0')}{\Delta}>0\ ,\\
\frac{\partial y_2}{\partial \mathcal B}
&=&\frac{(C_2'y_2^2+C_1'y_2+C_0')}{\Delta}>0\ .
\end{eqnarray*}
We note that for a given value of $\delta$, parameter
$\mathcal B$ is a function of $\epsilon$, while
parameter $\chi\equiv\beta/\Theta$ is yet another
function of $\epsilon$ according to the definition of
$\Theta$. It is then possible to write $\Theta$ as a
function of $\mathcal B$ in the explicit form of
\begin{eqnarray}
\Theta=1+\frac{(1+\eta^2)({\mathcal{B}}-1)}
{{\mathcal{B}}(1+\delta^{-1})+\eta^2({\mathcal{B}}-1)}\ .
\end{eqnarray}
As we have
\begin{eqnarray*}
\frac{d\chi}{d{\mathcal{B}}}=-\frac{\beta(1+\eta^2)
\delta(1+\delta)}{(2\eta^2{\mathcal{B}}\delta
-2\eta^2\delta +{\mathcal{B}}+2{\mathcal{B}}
\delta-\delta)^2}<0
\end{eqnarray*}
and, by the chain rule of taking derivatives,
\begin{eqnarray*}
\frac{d
y[{\mathcal{B}},\chi({\mathcal{B}})]}{d{\mathcal{B}}}
=\frac{\partial y}{\partial {\mathcal{B}}}
+\frac{\partial y}{\partial\chi}\frac{d\chi}{d{\mathcal{B}}}\ ,
\end{eqnarray*}
together with inequalities derived earlier
\begin{eqnarray*}
\frac{\partial y_1}{\partial \chi}<0\ (\hbox{for } C_2>0)\ ,
\ \frac{\partial y_2}{\partial \chi}<0
\ \ \hbox{ and }\ \
\frac{\partial y_{1,2}}{\partial {\mathcal{B}}}>0\ ,
\end{eqnarray*}
we can clearly demonstrate that
\begin{eqnarray*}
\qquad\quad \frac{d y_1}{d{\mathcal{B}}}>0\ \quad
(\hbox{ for } C_2>0)
\quad \ \ \hbox{ and }\ \ \
\frac{d y_2}{d{\mathcal{B}}}>0\ .
\end{eqnarray*}
Note that for $C_2<0$ and $y_1<0$, the case is not considered
here because of an unphysical negative $y=D_s^2$.

Finally, we proceed to derive analytically the relevant
constraints of requiring $D^2_s>0$ and $D^2_g>0$, which are
equivalent to $y>0$ and $y>({\mathcal B}/\chi-1)$, respectively.

Let us consider $y_1$ first. For $C_2>0$, we can readily
demonstrate that $y_1>0$ and $y_1>{\mathcal B}/\chi-1$.

We then consider $y_2$. As we know $y_1>{\mathcal B}/\chi-1$
(equivalent to $D_g^2>0$) and the properties of a quadratic
equation, we therefore have the following two correspondences
$$
y_2>\frac{{\mathcal B}}{\chi}-1
\ \Leftrightarrow\
C_2\bigg(\frac{{\mathcal B}}{\chi}-1\bigg)^2
+C_1\bigg(\frac{{\mathcal B}}{\chi}-1\bigg)+C_0>0
$$
and
$$
y_2>0\qquad\Leftrightarrow\qquad C_0>0\ ,
$$
separately.
%
For the following two inequalities
\begin{eqnarray*}
C_2\bigg(\frac{{\mathcal B}}{\chi}-1\bigg)^2
+C_1\bigg(\frac{{\mathcal B}}{\chi}-1\bigg)+C_0=
\qquad\qquad\qquad
\\
\frac{m{\mathcal B}}{\chi(1+\delta)} \bigg{\{}
m(m+m\delta+{\mathcal B}-1)
\quad\qquad\qquad\qquad\qquad
\\
-\bigg(\frac{{\mathcal B}}{\chi}-1\bigg)
\big[(m+2)(m+{\mathcal B}-1)+\delta(m^2-2)\big]
\bigg{\}}>0
\end{eqnarray*}
and
\begin{eqnarray*}
C_0=m\bigg{\{}m(m+m\delta+{\mathcal B}-1)
\qquad\qquad\qquad
\\ \qquad
+2(m+1)(m+m\delta-1)
\bigg(\frac{{\mathcal B}}{\chi}-1\bigg)\bigg{\}}>0
\end{eqnarray*}
to be valid, we clearly need to require that
\begin{equation}
\frac{{\mathcal B}}{\chi}-1<\frac{m(m+m\delta
+{\mathcal B}-1)}{(m+2)(m+{\mathcal B}-1)
+\delta(m^2-2)}
\end{equation}
and
\begin{equation}
\frac{{\mathcal B}}{\chi}-1>-\frac{m(m+m\delta
+{\mathcal B}-1)}{2(m+1)(m+m\delta-1)}\ ,
\end{equation}
respectively.


\section[]{}
For the unaligned cases or the logarithmic spiral cases,
we first introduce a new parameter $\mathcal{M}$ as
already defined in the main text, namely
\begin{eqnarray*}
\qquad {\mathcal{M}}\equiv\frac{m^2-2}{(m^2+\alpha^2+1/4)}\ .
\end{eqnarray*}
In the quadratic equation
\begin{equation}\label{quandric}
C_2y^2+C_1y+C_0=0\ ,
\end{equation}
we have the following forms of $C_2$, $C_1$ and $C_0$ coefficients
\begin{eqnarray*}
C_2\equiv\big{[}{\mathcal{M}}+{\mathcal{N}}_m(\alpha)\big{]}
\bigg\lbrace\frac{1}{\mathcal{B}}\bigg[{\mathcal{M}}
+\frac{{\mathcal{N}}_m(\alpha)}{(1+\delta)}\bigg]
-\frac{{\mathcal{N}}_m(\alpha)}{(1+\delta)}\bigg\rbrace\ ,
\qquad\qquad\ \\
C_1\equiv\bigg{\{}\frac{1}{\mathcal{B}}\big({\mathcal{M}}
+{\mathcal{N}}_m(\alpha)\big)
\bigg[{\mathcal{M}}+\frac{2{\mathcal{N}}_m(\alpha)}
{(1+\delta)}-1\bigg]
\qquad\qquad\qquad\qquad \\
-\frac{{\mathcal{N}}_m(\alpha)}{(1+\delta)}
\big{[}{\mathcal{M}}+2{\mathcal{N}}_m(\alpha)-1\big{]}\bigg{\}}
-\frac{({\mathcal{M}}+1)}{\chi}\bigg[{\mathcal{M}}
+\frac{{\mathcal{N}}_m(\alpha)}{(1+\delta)}\bigg]\ ,
\qquad\\
C_0\equiv\bigg{\{}\frac{1}{\mathcal{B}}\big{[}{\mathcal{M}}
+{\mathcal{N}}_m(\alpha)\big{]}
\bigg[\frac{{\mathcal{N}}_m(\alpha)}{(1+\delta)}-1\bigg]
\quad\qquad\qquad\qquad\qquad\qquad \\
-\frac{{\mathcal{N}}_m(\alpha)}{(1+\delta)}
\big{[}{\mathcal{N}}_m(\alpha)-1\big{]}\bigg{\}}
-\frac{({\mathcal{M}}+1)}{\chi}
\bigg[\frac{{\mathcal{N}}_m(\alpha)}{(1+\delta)}-1\bigg]\ .
\quad\qquad
\end{eqnarray*}
The determinant $\Delta$ of quadratic equation
(\ref{quandric}) can then be explicitly shown as
\begin{eqnarray*}
\Delta\equiv C_1^2-4C_0C_2=({\mathcal{M}}+1)^2
\bigg[\!\!\bigg[
\bigg\{
\frac{1}{\mathcal{B}}[{\mathcal{M}}+{\mathcal{N}}_m(\alpha)]
\qquad\\
-\frac{{\mathcal{N}}_m(\alpha)}{(1+\delta)}-\frac{1}{\chi}
\bigg[{\mathcal{M}}+\frac{{\mathcal{N}}_m(\alpha)}{(1+\delta)}\bigg]
\bigg\}^2+\frac{4\delta{\mathcal{N}}_m^2(\alpha)}{(1+\delta)^2\chi}
\bigg]\!\!\bigg]\geq 0\ ,
\end{eqnarray*}
so that two real $y$ solutions are guaranteed.

In parallel to what we have done for the aligned case in
Appendix A, we would like to find out the relationships
between the $y$ solutions and their variations with
relevant parameters. Again, we use $y_1$ and $y_2$
to represent the two real $y$ solutions defined by
\begin{eqnarray}
y_{1,2}=\frac{-C_1\pm(C_1^2-4C_0C_2)^{1/2}}{2C_2}\ .
\end{eqnarray}
Following the same procedure of Appendix A, we derive
\begin{eqnarray}
y_1'=-\frac{C_2'y_1^2+C_1'y_1+C_0'}{(C_1^2-4C_0C_2)^{1/2}}
=-\frac{C_2'y_1^2+C_1'y_1+C_0'}{\Delta^{1/2}}\ ,\\
y_2'=\frac{C_2'y_2^2+C_1'y_2+C_0'}{(C_1^2-4C_0C_2)^{1/2}}
=\frac{C_2'y_2^2+C_1'y_2+C_0'}{\Delta^{1/2}}\ ,
\end{eqnarray}
where the prime $'$ denotes the first derivative of
a quantity with respect to a specific parameter.

We first discuss the $y$ solution properties with respect
to the variation of parameter $\chi$ and define explicitly
\begin{eqnarray*}
\quad C_2'\equiv\frac{\partial C_2}{\partial\chi}\ ,
\ \ \
C_1'\equiv\frac{\partial C_1}{\partial\chi}\ ,
\ \ \
C_0'\equiv\frac{\partial C_0}{\partial\chi}\ ,
\ \ \
y'\equiv\frac{\partial y}{\partial\chi}\ .
\end{eqnarray*}
It follows immediately that
\begin{eqnarray*}
\qquad C_2'&=&0\ ,\\
\qquad C_1'&=&\frac{({\mathcal{M}}+1)}{\chi^2}\bigg[{\mathcal{M}}
+\frac{{\mathcal{N}}_m(\alpha)}{(1+\delta)}\bigg]\ ,\\
\qquad C_0'&=&\frac{({\mathcal{M}}+1)}{\chi^2}
\bigg[\frac{{\mathcal{N}}_m(\alpha)}{(1+\delta)}-1\bigg]\ .
\end{eqnarray*}
In parallel with the analysis of the aligned case, we
here introduce a similar parameter $R_s$ in the form of
\begin{eqnarray*}
\qquad  R_s\equiv\frac{1-{\mathcal{N}}_m(\alpha)/(1+\delta)}
{{\mathcal{M}}+{\mathcal{N}}_m(\alpha)/(1+\delta)}\ .
\end{eqnarray*}
It is then straightforward to demonstrate that
\begin{eqnarray*}
\frac{\partial y_1}{\partial\chi}
=-\frac{({\mathcal{M}}+1)[{\mathcal{M}}
+{\mathcal{N}}_m(\alpha)/(1+\delta)]}
{\chi^2\Delta^{1/2}}(y_1-R_s)\ ,\\
\frac{\partial y_2}{\partial\chi}
=-\frac{({\mathcal{M}}+1)[{\mathcal{M}}
+{\mathcal{N}}_m(\alpha)/(1+\delta)]}
{\chi^2\Delta^{1/2}}(R_s-y_2)\ .
\end{eqnarray*}
We now examine behaviours of $y$ solutions by
varying parameter $\mathcal B$. Following the
same procedure, we write
\begin{eqnarray*}
\quad C_2'\equiv\frac{\partial C_2}{\partial {\mathcal{B}}}\ ,
\ \ \
C_1'\equiv\frac{\partial C_1}{\partial {\mathcal{B}}}\ ,
\ \ \
C_0'\equiv\frac{\partial C_0}{\partial {\mathcal{B}}}\ ,
\ \ \
y'\equiv\frac{\partial y}{\partial {\mathcal{B}}}\ ,
\end{eqnarray*}
and derive explicitly
\begin{eqnarray*}
\quad C_2'&=&-\frac{1}{{\mathcal{B}}^2}\big{[}{\mathcal{M}}
+{\mathcal{N}}_m(\alpha)\big{]}\bigg[{\mathcal{M}}
+\frac{{\mathcal{N}}_m(\alpha)}{(1+\delta)}\bigg]\ ,
\\
C_1'&=&-\frac{1}{{\mathcal{B}}^2}\big{[}{\mathcal{M}}
+{\mathcal{N}}_m(\alpha)\big{]}\bigg[{\mathcal{M}}
+\frac{2{\mathcal{N}}_m(\alpha)}{(1+\delta)}-1\bigg]\ ,
\\
C_0'&=&-\frac{1}{{\mathcal{B}}^2}\big{[}{\mathcal{M}}
+{\mathcal{N}}_m(\alpha)\big{]}\bigg[
\frac{{\mathcal{N}}_m(\alpha)}{(1+\delta)}-1\bigg]\ .
\end{eqnarray*}
It is straightforward to demonstrate that
\begin{eqnarray*}
\frac{\partial y_1}{\partial {\mathcal{B}}}
=\frac{[{\mathcal{M}}+{\mathcal{N}}_m(\alpha)]
[{\mathcal{M}}
+{\mathcal{N}}_m(\alpha)/(1+\delta)]}
{{\mathcal{B}}^2\Delta^{1/2}}(y_1+1)(y_1-R_s)\ ,
\\
\frac{\partial y_2}{\partial{\mathcal{B}}}
=\frac{[{\mathcal{M}}+{\mathcal{N}}_m(\alpha)]
[{\mathcal{M}}
+{\mathcal{N}}_m(\alpha)/(1+\delta)]}
{{\mathcal{B}}^2\Delta^{1/2}}(y_2+1)(R_s-y_2)\ ,
\end{eqnarray*}
respectively.

Two important inequalities are summarized below
\begin{displaymath}
C_2R_s^2+C_1R_s+C_0=-\frac{(1+{\mathcal{M}})^2
{\mathcal{N}}_m^2(\alpha)\delta}{[{\mathcal{M}}
+{\mathcal{M}}\delta+{\mathcal{N}}_m(\alpha)]^2}<0
\end{displaymath}
and
\begin{displaymath}
C_2(-1)^2+C_1(-1)+C_0=\frac{(1+{\mathcal{M}})^2}{\chi}>0\ .
\end{displaymath}
The phase relationship between the two surface mass density
perturbations $\mu^g$ and $\mu^s$ is then given by
\begin{eqnarray*}
\qquad \frac{\mu^g}{\mu^s}&=&\frac{1-H_1(m^2+\alpha^2+1/4)}
{G_1(m^2+\alpha^2+1/4)}
\\&=&-\frac{{\mathcal{N}}_m(\alpha)+{\mathcal{M}}(1+\delta)}
{{\mathcal{N}}_m(\alpha)}\frac{(D_s^2-R_s)}{(D_s^2+1)}\ ,
\end{eqnarray*}
where $H_1$ and $G_1$ are separately defined by (\ref{h1s})
and (\ref{g1s}). In order to have both $D_s^2$ and $D_g^2$
being positive, we still have the same two requirements of
$D_s^2=y>0$ and $D_s^2=y>{\mathcal B}/\chi-1$ simultaneously
as in the aligned case.

In the following, we analyze the three situations
of $m\geq2$, $m=1$, and $m=0$, separately.

\subsection{The Cases of $m\geq2$ }

For $m\geq2$, the approximation ${\mathcal{N}}_m(\alpha)
\cong 1/(m^2+\alpha^2+1/4)^{1/2}$ is valid (Shu et al.
2000) and we have
\begin{displaymath}
\frac{1}{2}>{\mathcal{N}}_m(\alpha)>0\ ,\quad\qquad
1>{\mathcal{M}}>0\ ,\quad\qquad R_s>0\ .
\end{displaymath}
According to the basic theory of a quadratic equation, we infer
that $y_1>R_s>y_2>-1$ for $C_2>0$, while $R_s>y_2>-1>y_1$ for
$C_2<0$. This inference is entirely similar to that of the
aligned case. And we can get the similar condition for $y_2$
that $y_2>0\Leftrightarrow C_0>0$. Note also that $R_s$
increases with increasing $\delta$ and $\alpha$.

By these analyses, we reach the following conclusions. For
$y_1$ and $y_2$ being positive and thus physical, we have
\begin{displaymath}
\frac{\partial y_1}{\partial\chi}<0\ ,
\ \ \ \
\frac{\partial y_2}{\partial \chi}<0\ ,
\ \ \ \
\frac{\partial y_1}{\partial {\mathcal{B}}}>0\ ,
\ \ \ \
\frac{\partial y_2}{\partial {\mathcal{B}}}>0\ .
\end{displaymath}
The perturbation mass density ratio $\mu^g/\mu^s$ corresponding to
$y_1$ remains negative with $\mu^g$ and $\mu^s$ being out of phase,
while $\mu^g/\mu^s$ corresponding to $y_2$ remains positive with
$\mu^g$ and $\mu^s$ being in phase. The result obtained and the
procedure taken here are just the same as those for the aligned
case. Moreover, by setting $\alpha=0$ in the unaligned case, the
equation for the unaligned case can be simplified to the same
equation of the aligned case, implying that the aligned case
is just a special example of the unaligned case.

We now indicate consequences of the two physical
requirements $y>0$ and $y>{\mathcal B}/\chi-1$. By
analytical derivations for the $y_1$ solution, the
two physical requirements
\begin{displaymath}
y_1>0 \qquad\qquad \mbox{ and } \qquad\qquad y_1
>\frac{\mathcal B}{\chi}-1
\end{displaymath}
are simply equivalent to $C_2>0$.
Meanwhile for the $y_2$ solution,
we find that the condition
\begin{equation}
y_2>0
\end{equation}
is equivalent to
\begin{equation}
C_0>0\ ,
\end{equation}
while the condition
\begin{equation}
y_2>\frac{\mathcal B}{\chi}-1
\end{equation}
is equivalent to
\begin{equation}
C_2\bigg(\frac{\mathcal B}{\chi}-1\bigg)^2
+C_1\bigg(\frac{\mathcal B}{\chi}-1\bigg)+C_0>0\ .
\end{equation}

\subsection{The Case of $m=1$ }

For $m=1$, we have inequalities
\begin{displaymath}
-\frac{4}{5}<{\mathcal{M}}=-\frac{1}{(\alpha^2+5/4)}<0\
\end{displaymath}
and
\begin{displaymath}
0<{\mathcal{N}}_1(\alpha)=\frac{(\alpha^2+17/4)^{1/2}}
{(\alpha^2+9/4)}<1\ .
\end{displaymath}
By straightforward algebraic manipulations, we find that
${\mathcal{M}}+{\mathcal{N}}_1(\alpha)>0$ and ${\mathcal{M}}+1>0$.
As ${\mathcal B}\geq1$, $C_2$ remains always less than 0 and there
must be a negative $y$ solution (in fact this $y$ solution must
also be less than $-1$) that corresponds to $y_1$. In this case,
only the $y_2$ solution may be positive in order to be physical.

Because of $y_1y_2=C_0/C_2$ and the solution property of $y_1$
and $C_2$ derived earlier, the physical requirement of
\begin{displaymath}
y_2>0
\end{displaymath}
is clearly equivalent to
\begin{displaymath}
C_0>0\ .
\end{displaymath}

One can readily show the equivalence
between the two inequalities $R_s>0$ and
${\mathcal{M}}+{\mathcal{N}}_1(\alpha)/(1+\delta)>0$. When
$\delta<0.145307118$, $R_s$ remains always positive for
arbitrary values of $\alpha$. Furthermore, we know that
$R_s>y_2>-1$ for $R_s>0$, while $y_2>-1>R_s$ for $R_s<0$.
It then follows that
\begin{displaymath}
\frac{\partial y_2}{\partial\chi}<0
\qquad\hbox{ and }\qquad
\frac{\partial y_2}{\partial{\mathcal B}}>0\ ;
\end{displaymath}
and the ratio of $\mu^g/\mu^s$ corresponding to $y_2$
remains always positive with $\mu^g$ and $\mu^s$ being in
phase. This phase relationship is just the same as that
of $y_2$ solution in the case of $m\geq2$. Here, we have
again the following two equivalent inequalities between
\begin{eqnarray*}
\qquad y_2>0
\qquad\qquad\hbox{ and }\qquad
\qquad C_0>0\ ,
\end{eqnarray*}
and between
\begin{eqnarray*}
\qquad y_2>\frac{\mathcal B}{\chi}-1
\end{eqnarray*}
and
\begin{eqnarray*}
\qquad C_2\bigg(\frac{\mathcal B}{\chi}-1\bigg)^2
+C_1\bigg(\frac{\mathcal B}{\chi}-1\bigg)+C_0>0\ ,
\end{eqnarray*}
respectively.

\subsection{The Case of $m=0$ }

For $m=0$, we have
\begin{displaymath}
-8\leq{\mathcal{M}}=-\frac{2}{(\alpha^2+1/4)}<0\ ,
\end{displaymath}
\begin{displaymath}
{\mathcal{N}}_0(\alpha)
=\frac{(\alpha^2+9/4)}{(\alpha^2+1/4)(\alpha^2+17/4)^{1/2}}>0\ .
\end{displaymath}
The situation becomes somewhat involved. First, we
identify the following five expressions, namely
\begin{displaymath}
1-\frac{{\mathcal{N}}_0(\alpha)}{1+\delta}\ ,\ \ \
1-{\mathcal{N}}_0(\alpha),\ \ \ \ 1+{\mathcal{M}}\ ,\ \ \
{\mathcal{M}}+{\mathcal{N}}_0(\alpha)\ ,\ \ \
{\mathcal{M}}+\frac{{\mathcal{N}}_0(\alpha)}{1+\delta}\ .
\end{displaymath}
By extensive numerical computations using the two
expressions of ${\mathcal{M}}$ and ${\mathcal{N}}_0$, we
infer several basic results. All these expressions are
negative when $\alpha$ is small enough. When $\alpha$
becomes larger than some critical point, which is different
for each expression, these expressions become positive.

For the convenience of analysis, we denote the five
critical points by $\alpha_1$, $\alpha_2$, $\alpha_3$,
$\alpha_4$, and $\alpha_5$ satisfying the conditions
\begin{eqnarray*}
1-\frac{{\mathcal{N}}_0(\alpha_1)}{(1+\delta)}=0\ ,\ \ \
1-{\mathcal{N}}_0(\alpha_2)=0\ ,\ \ \
1+{\mathcal{M}}(\alpha_3)=0\ ,\ \ \ \\
{\mathcal{M}}(\alpha_4)+{\mathcal{N}}_0(\alpha_4)=0\ ,
\ \hbox{ and } \
{\mathcal{M}}(\alpha_5)
+\frac{{\mathcal{N}}_0(\alpha_5)}{(1+\delta)}=0\ ,
\end{eqnarray*}
respectively. More specifically, numerical calculations
indicate that $\alpha_2=1.113$, $\alpha_3=1.323$ and
$\alpha_4=1.793$. It is found that $\alpha_1<\alpha_2$ and
$\alpha_4<\alpha_5$. Actually, $\alpha_1$ and $\alpha_5$
depend on the value of parameter $\delta$. For example,
when $\delta>3.4$, expression
$1-{\mathcal{N}}_0(\alpha)/(1+\delta)$ remains always
positive and $\alpha_1$ does not exist. In general,
our analysis indicates the following ordering for the
five $\alpha_i$ parameters,
\begin{displaymath}
\alpha_1<\alpha_2<\alpha_3<\alpha_4<\alpha_5\ .
\end{displaymath}
We now proceed to discuss each $\alpha-$range separately.

\subsubsection{The Range of $\ 0\leq\alpha<\alpha_1$}

For this $\alpha-$range to exist, we must require $\delta <3.4$.
In this $\alpha-$range, we have the following five inequalities
\begin{eqnarray*}
1-\frac{{\mathcal{N}}_0(\alpha)}{(1+\delta)}<0\ ,\ \ \
1-{\mathcal{N}}_0(\alpha)<0\ ,\ \ \
1+{\mathcal{M}}<0\ ,\ \ \ \\
{\mathcal{M}}+{\mathcal{N}}_0(\alpha)<0\ ,\ \ \
{\mathcal{M}}+\frac{{\mathcal{N}}_0(\alpha)}{(1+\delta)}<0\ .
\end{eqnarray*}
In this case with $C_2>0$ and $0.93>R_s>0$, we have $y_1>R_s>y_2>-1$.
The surface mass density ratio of $\mu^g/\mu^s$ corresponding to
$y_1$ remains positive with $\mu^g$ and $\mu^s$ being in phase and
the ratio of $\mu^g/\mu^s$ corresponding to $y_2$ remains negative
with $\mu^g$ and $\mu^s$ being out of phase. We further infer
\begin{displaymath}
\frac{\partial y_1}{\partial\chi}<0\ ,\ \ \ \
\frac{\partial y_2}{\partial\chi}<0\ ,\ \ \ \
\frac{\partial y_1}{\partial{\mathcal B}}>0\ ,\ \ \ \
\frac{\partial y_2}{\partial{\mathcal B}}>0\ .
\end{displaymath}
For the $y_1$ solution, we only need to think of the requirement
$y_1>({\mathcal B}/\chi)-1$. After some analysis, we can
demonstrate the equivalence of two inequalities
\begin{displaymath}
y_1>\frac{\mathcal B}{\chi}-1
\end{displaymath}
and
\begin{displaymath}
C_2\bigg(\frac{{\mathcal B}}{\chi}-1\bigg)^2
+C_1\bigg(\frac{{\mathcal B}}{\chi}-1\bigg)+C_0<0
\end{displaymath}
for $({\mathcal B}/\chi)-1>R_s$.
For the $y_2$ solution, we found by analytical derivation that
the two inequalities $y_2>0$ and $y_2>({\mathcal B}/\chi)-1$
are incompatible with each other. Thus in this $\alpha-$range,
there is no physical $y_2$ solution.

We note that $R_s=0$ for the special case of $\alpha=\alpha_1$,
so that $y_1>R_s=0>y_2>-1$. Here, only the $y_1$ solution is
positive and thus physical with the ratio of $\mu^g/\mu^s$
being positive for in-phase $\mu^g$ and $\mu^s$.

\subsubsection{The Range of $\ \alpha_1<\alpha<\alpha_2$}

In this $\alpha-$range, we have inequalities
\begin{eqnarray*}
1-\frac{{\mathcal{N}}_0(\alpha)}{(1+\delta)}>0\ ,\ \ \
1-{\mathcal{N}}_0(\alpha)<0\ ,\ \ \
1+{\mathcal{M}}<0\ ,\ \ \
\\
{\mathcal{M}}+{\mathcal{N}}_0(\alpha)<0\ ,\ \ \
{\mathcal{M}}+\frac{{\mathcal{N}}_0(\alpha)}{(1+\delta)}<0\ .
\end{eqnarray*}
In this case with $C_2>0$ and $-0.7444<R_s<0$, we have
$y_1>R_s>y_2>-1$, such that $y_2$ is negative and unphysical.
The surface mass density ratio of $\mu^g/\mu^s$ of $y_1$
solution is positive with $\mu^g$ and $\mu^s$ being in
phase. We further infer
\begin{displaymath}
\frac{\partial y_1}{\partial\chi}<0
\qquad \ \ \hbox{ and } \qquad \ \
\frac{\partial y_1}{\partial {\mathcal B}}>0\ .
\end{displaymath}
Through an analytical analysis, we can demonstrate
the equivalence of two pairs of inequalities between
\begin{equation}
y_1>\frac{\mathcal B}{\chi}-1
\end{equation}
and
\begin{equation}
C_2\bigg(\frac{{\mathcal B}}{\chi}-1\bigg)^2
+C_1\bigg(\frac{{\mathcal B}}{\chi}-1\bigg)+C_0<0
\end{equation}
for $({\mathcal B}/\chi)-1>R_s$, and between
\begin{equation}
y_1>0
\end{equation}
and
\begin{equation}
C_0<0\ .
\end{equation}
For the special case of $\alpha=\alpha_2$, we still have a
positive $y_1$ solution. However, the conditions $y_1>0$
and $y_1>({\mathcal B}/\chi)-1$ are incompatible with
each other and there is then no physical $y$ solution.

\subsubsection{The Range of $\ \alpha_2<\alpha<\alpha_3$}

In this $\alpha-$range, we have the
following set of inequalities
\begin{eqnarray*}
1-\frac{{\mathcal{N}}_0(\alpha)}{(1+\delta)}>0\ ,\ \ \
1-{\mathcal{N}}_0(\alpha)>0\ ,\ \ \
1+{\mathcal{M}}<0\ ,\ \ \
\\
{\mathcal{M}}+{\mathcal{N}}_0(\alpha)<0\ ,\ \ \
{\mathcal{M}}+\frac{{\mathcal{N}}_0(\alpha)}{(1+\delta)}<0\ .
\end{eqnarray*}
In this case with $C_2>0$ and $-1<R_s<-0.7444$, we have
$y_1>R_s>y_2>-1$. The $y_2$ solution remains negative and
thus unphysical. The perturbation surface mass density
ratio of $\mu^g/\mu^s$ for the $y_1$ solution is positive
with $\mu^g$ and $\mu^s$ being in phase. We further
infer the following inequalities
\begin{displaymath}
\frac{\partial y_1}{\partial \chi}<0
\qquad\hbox{ and }\qquad
\frac{\partial y_1}{\partial {\mathcal B}}>0\ .
\end{displaymath}
The two inequalities $y_1>0$ and $y_1>({\mathcal B}/\chi)-1$
are incompatible and there is no physical $y_1$ solution in
this $\alpha-$range.

When $\alpha=\alpha_3$, we have ${\mathcal M}=-1$ and
$C_2>0$. It then follows that $y_1=y_2=-1$ and there
is thus no physical $y$ solution for $\alpha=\alpha_3$.

\subsubsection{The Range of $\ \alpha_3<\alpha<\alpha_4$}

In this $\alpha-$range, we have the following inequalities
\begin{eqnarray*}
1-\frac{{\mathcal{N}}_0(\alpha)}{(1+\delta)}>0\ ,\ \ \
1-{\mathcal{N}}_0(\alpha)>0\ ,\ \ \
1+{\mathcal{M}}>0\ ,\ \ \
\\
{\mathcal{M}}+{\mathcal{N}}_0(\alpha)<0\ ,\ \ \
{\mathcal{M}}+\frac{{\mathcal{N}}_0(\alpha)}{(1+\delta)}<0\ .
\end{eqnarray*}
With $C_2>0$ and $R_s<-1$, we have $-1>y_1>R_s>y_2$. Both
$y_1$ and $y_2$ solutions are negative and thus unphysical.

For $\alpha=\alpha_4$, we have $C_2=0$, corresponding
to the divergence point of the quadratic equation; and
only one negative $y$ solution exists.

\subsubsection{The Range of $\ \alpha_4<\alpha<\alpha_5$ }

In this $\alpha-$range, we have the following inequalities
\begin{eqnarray*}
\ 1-\frac{{\mathcal{N}}_0(\alpha)}{(1+\delta)}>0\ ,\ \ \
1-{\mathcal{N}}_0(\alpha)>0\ ,\ \ \ \
1+{\mathcal{M}}>0\ ,\ \ \
\\
{\mathcal{M}}+{\mathcal{N}}_0(\alpha)>0\ ,\ \ \
{\mathcal{M}}+\frac{{\mathcal{N}}_0(\alpha)}{(1+\delta)}<0\ .
\end{eqnarray*}
With $C_2<0$ and $R_s<-1$, we have $y_2>-1>y_1>R_s$.
The negative $y_1$ solution is unphysical. The perturbation
surface mass density ratio $\mu^g/\mu^s$ of $y_2$ solution
is positive with $\mu^g$ and $\mu^s$ being in phase. We
further infer
\begin{displaymath}
\frac{\partial y_2}{\partial\chi}<0
\qquad\hbox{ and }\qquad
\frac{\partial y_2}{\partial {\mathcal B}}>0\ .
\end{displaymath}
By an analytical analysis, we can demonstrate
the equivalence of the following two sets of
inequalities between
\begin{equation}
y_2>\frac{\mathcal B}{\chi}-1
\end{equation}
and
\begin{equation}
C_2\bigg(\frac{{\mathcal B}}{\chi}-1\bigg)^2
+C_1\bigg(\frac{{\mathcal B}}{\chi}-1\bigg)+C_0>0\ ,
\end{equation}
and between
\begin{equation}
y_2>0
\end{equation}
and
\begin{equation}
C_0>0\ ,
\end{equation}
respectively.

For $\alpha=\alpha_5$, we have $C_2<0$ and thus a negative $y_1$
solution. Only the $y_2$ solution has the possibility of being
positive. The ratio $\mu^g/\mu^s$ of $y_2$ solution for
$\alpha=\alpha_5$ is positive with $\mu^g$ and $\mu^s$ being in phase.

\subsubsection{The Range of $\ \alpha>\alpha_5$}

In this $\alpha-$range, we have
the following five inequalities
\begin{eqnarray*}
1-\frac{{\mathcal{N}}_0(\alpha)}{(1+\delta)}>0\ ,\ \ \
1-{\mathcal{N}}_0(\alpha)>0\ ,\ \ \
1+{\mathcal{M}}>0\ ,\ \ \
\\
{\mathcal{M}}+{\mathcal{N}}_0(\alpha)>0\ ,\ \ \
{\mathcal{M}}+\frac{{\mathcal{N}}_0(\alpha)}{(1+\delta)}>0\ .
\end{eqnarray*}
With $C_2<0$ and $R_s>0$, we have $R_s>y_2>-1>y_1$. The negative
$y_1$ solution is unphysical. The perturbation surface mass
density ratio $\mu^g/\mu^s$ of the $y_2$ solution is positive
with $\mu^g$ and $\mu^s$ being in phase. We further infer
\begin{displaymath}
\frac{\partial y_2}{\partial\chi}<0
\qquad\hbox{ and }\qquad
\frac{\partial y_2}{\partial {\mathcal B}}>0\ .
\end{displaymath}
By an analytical analysis, we can demonstrate
the equivalence of the following two sets of
inequalities between
\begin{equation}
y_2>\frac{\mathcal B}{\chi}-1
\end{equation}
and
\begin{equation}
C_2\bigg(\frac{{\mathcal B}}{\chi}-1\bigg)^2
+C_1\bigg(\frac{{\mathcal B}}{\chi}-1\bigg)+C_0>0\ ,
\end{equation}
and between
\begin{equation}
y_2>0
\end{equation}
and
\begin{equation}
C_0>0\ ,
\end{equation}
respectively.

Finally, we give a complete summary for the $m=0$ case with
different ranges of $\alpha$ values. Physically, we could
only have one real positive $y$ solution with surface mass
density perturbations $\mu^g$ and $\mu^s$ being in phase.
For $0<\alpha<\alpha_2=1.113$ and some constraints, $y_1$
is the only physical solution.
For $1.113=\alpha_2\leq\alpha\leq\alpha_4=1.793$, no
physical solution exists.
For $\alpha>\alpha_4=1.793$ and some constraints, $y_2$
is the only physical solution. All the physical $y$
solutions increase with increasing $\mathcal B$ and
decreasing $\chi$.
\end{appendix}

\end{document}